\documentclass[symmetry,article,accept,pdftex,moreauthors]{Definitions/mdpi}

\newcommand{\ms}{\scriptscriptstyle}

\def\be{\begin{equation}}
\def\ee{\end{equation}}
\def\ba{\begin{eqnarray}}
\def\ea{\end{eqnarray}}

\def\sp{\kern +3pt}
\def\sm{\kern -3pt}
\def\spQ{\kern +6pt}

\newcommand{\ket}[1]{|#1\rangle}

\def\sfrac#1#2{{\textstyle \frac{#1}{#2}}}

\firstpage{1} 
\pubvolume{1}
\issuenum{1}
\articlenumber{0}
\pubyear{2025}
\copyrightyear{2025}
\externaleditor{Firstname Lastname} 
\datereceived{21 March 2025} 
\daterevised{13 April 2025} 
\dateaccepted{18 April 2025} 
\datepublished{ } 
\hreflink{https://doi.org/} 

\Title{Electroweak
 Form Factors of Baryons in Dense Nuclear Matter}

\TitleCitation{Electroweak Form Factors of Baryons in Dense Nuclear Matter}

\Author{G.~Ramalho 
 $^{1,}$*\orcidA{},  K.
~Tsushima $^{2}$\orcidB{} and Myung-Ki Cheoun $^{1}$\orcidC{}$^\dagger$}

\AuthorNames{G. Ramalho, K. Tsushima and Myung-Ki Cheoun}

        \AuthorCitation{Ramalho, G.; Tsushima, K.; Cheoun, M.-K.}
\address{%
$^{1}$ \quad Department of Physics and OMEG Institute, 
  Soongsil University, Seoul 06978, Republic of Korea;  
\\
$^{2}$ \quad Laborat\'orio 
 de 
F\'{i}sica Te\'orica e Computacional---LFTC,
Programa de Posgradua\c{c}\~ao em Astrof\'{i}sica e F\'{i}sica Computacional,  
  Universidade Cidade de  S\~ao Paulo,   S\~ao Paulo 01506-000, SP, Brazil; kazuo.tsushima@gmail.com, kazuo.tsushima@cruzeirodosul.edu.br}

\corres{Correspondence: gilberto.ramalho2013@gmail.com}

\firstnote{cheoun@ssu.ac.kr}

\abstract{There is evidence that the properties of hadrons are modified in a nuclear medium.
  Information about the medium modifications of the internal structure of  hadrons is fundamental for the study of dense nuclear matter and high-energy processes, including heavy-ion and nucleus--nucleus collisions.
  {At the moment, however,  empirical information about  medium modifications of hadrons is limited; therefore, theoretical studies are essential for progress in the field.}
  {In the present work, we review theoretical studies of the electromagnetic and axial form factors of octet baryons in symmetric nuclear matter.}
  The calculations are based on a model that takes into account the degrees of freedom revealed in experimental studies of low and intermediate square transfer momentum $q^2=-Q^2$: valence quarks and meson cloud excitations of baryon cores.
  The formalism combines a covariant constituent quark model, developed for a free space (vacuum) with the quark--meson coupling model for extension to the nuclear medium.
  We conclude that the nuclear medium modifies the baryon properties differently according to the flavor content of the baryons and the medium density.
  The effects of the medium increase with  density and are stronger (quenched or enhanced) for light baryons than for heavy baryons. 
  In particular, the in-medium neutrino--nucleon and antineutrino--nucleon cross-sections are reduced compared to the values in free space.
  The proposed formalism can be extended to densities above the normal nuclear density and applied to neutrino--hyperon and antineutrino--hyperon scattering in dense nuclear matter.
}


\begin{document}


\section{Introduction \label{secIntro}}

The structure of  hadrons and their interactions are
described by Quantum Chromodynamics (QCD) in terms of quarks
and gluons, the QCD degrees of freedom.
It is then natural to assume that, in the presence of a strong mean field pervading nuclear matter, the motion of quarks and gluons inside hadrons is modified.
Modifications due to the medium are expected to be reflected
on the electromagnetic and weak structure functions of the nucleon
and other other baryons when compared with the structure functions 
in free space~\cite{BrownRho,QMCReview,JLabbook}.

Information about the medium properties of hadrons is
important for understanding
environments with dense nuclear matter
from high-energy nucleus--nucleus collisions
to the cores of compact stars~\cite{QMCReview,GA-Medium1,Miyatsu15a}.
At the moment, however, due to the complexity of experiments
with nuclear matter, the experimental information about the structure form factors
of baryons in nuclear medium is limited, and
theoretical studies are fundamental for the progress in the field.

Modifications of the properties of  baryons
in the nucleus have been observed regarding the European Muon Collaboration
(EMC effect) in the deep inelastic structure functions of the
nucleus~\cite{EMC1,EMC2,QMCReview,Hen17a}.
In the present work, however, we focus on the electromagnetic
and weak form factors of  octet baryons 
and study  their dependence on transfer momentum
and medium density.

Evidence of the modification of the electromagnetic properties
of baryons in a nuclear matter has been reported for the proton
based on measurements in polarized $(\vec{e}, e' \vec{p})$ scattering
on different targets at MAMI and Jefferson Lab (JLab)~\cite{Dieterich01,Paolone10,Strauch03a,JLab-data,Malace11a,Malov00,MAMI2021,Kolar23a}.
The measurements indicate an enhancement of the ratio between
the electric and magnetic form factors of the bound proton
when compared with the free space ratio,
a clear manifestation of the in-medium modifications
of the electromagnetic structure of the proton.
This interpretation is consistent with theoretical
calculations based on different frameworks~\cite{Octet2,Octet3,Frank96,Lu99a,Miller02,Smith04,Cloet09a,deAraujo18,Yakhshiev03,QMCEMFFMedium2,QMCEMFFMedium3,QMCEMFFMedium4,Christov95a}.
At the moment, the experimental study of the in-medium modifications
of baryons other than the proton is more challenging,
but there are some proposals for measurements on the
in-medium neutron electromagnetic form factors~\cite{PAC35}.
Model estimations for the proton predict a suppression (quenching)
for the ratio between the electric
and magnetic form factors~\cite{Octet2,Lu99a,Cloet09a,deAraujo18,QMCEMFFMedium2,QMCEMFFMedium3,QMCEMFFMedium4}.
For the remaining members of the baryon octet, 
one can expect enhancement or quenching
of the ratio depending on the flavor content of the baryon~\cite{Octet3}.

The experimental information about the in-medium modifications of the axial-vector
form factor of baryons in general and the nucleon in particular is very scarce.
Experiments related to the beta decay of heavy
nuclei suggest the quenching of the nucleon axial-vector coupling constant~\cite{Brown85a,Gysbers19a}.
Model calculations based on the MIT bag model~\cite{Thomas84} and
quark--meson coupling (QMC) model~\cite{QMCReview,Lu01a} and
Skyrme and soliton models~\cite{Meissner87a,Meier97a,Gysbers19a,Meissner89a,Rakhimov98a,Christov95a,Christov96a}
are consistent with the reduction in the 
nucleon axial-vector form factor in nuclear matter.

There is then a strong motivation to develop formalisms that can be used to 
estimate structure functions of baryons at relatively high densities
and moderate and large $Q^2$ based on the degrees of freedom
manifest in the vacuum: valence quarks and meson cloud excitations of baryon cores.
Model calculations based on Effective Field Theory, Skyrme and quark--soliton models,
and the QMC model that take into account the effective medium modification
of the quarks and hadron masses are particularly appropriate
for the study of the properties of  baryons immersed in a nuclear medium~\cite{Lu99a,Meissner87a,Meier97a,Christov96a}.
In the present work, we review recent calculations of the electromagnetic
and axial form factors for the baryon octet in a symmetric nuclear matter
based on a covariant quark model combined with the QMC model~\cite{Octet2,Octet3,GA-Medium1}.
We assume that the nuclear matter is distributed by a large volume, and that the interactions between baryons with the medium can be simulated by mean-field one-body currents, and also that the final state interactions are small~\cite{QMCReview,Octet2,GA-Medium1}.
The octet baryon form factors are calculated in
terms of $Q^2$ and the medium density $\rho$
in the interval between $\rho=0$ (free space) and
$\rho= \rho_0$, where $\rho_0$ is the normal nuclear matter density,
characterized by $\rho_0 \simeq 0.15$ fm$^{-3}$.   

In the free space, there is an extensive body of literature
related to the electromagnetic and axial structure of baryons.
For a review, see Refs.~\cite{Ramalho-Pena23,NSTAR,Aznauryan12,Compton,AxialFF,Gaillard84,Bernard02,Gorringe04,Schindler07a}.
In the study of the electromagnetic and axial structure of  octet baryons,
we consider an extension of
the covariant spectator quark model~\cite{Spectator-Review,Nucleon,Nucleon2,Omega,NDeltaSL1},
developed for the study of baryons in free space~\cite{Octet,OctetDecuplet1,Octet4,AxialFF}.
In the covariant spectator quark model, the electroweak
interactions with the baryon systems are described in relativistic impulse approximation
using parametrizations of the quark currents
and radial wave functions, determined in previous studies
of the nucleon, octet baryons, and decuplet baryons~\cite{Nucleon,Nucleon2,Omega,Octet,Sigma0Lambda,Octet2,AxialFF,OctetDecuplet1,OctetDecuplet2,HyperonFF1,HyperonFF2,HyperonFF3,OctetDecupletD1,OctetDecupletD2,DeltaFF,Octet4,DeltaFF2,DeltaFF3,Omega2,Delta-shape}.
We also consider  effective parametrizations
of the meson cloud contributions motivated by theoretical principles
for a better description of the low-$Q^2$ region~\cite{Octet,Octet2,OctetDecuplet2,OctetDecupletD2,AxialFF}.
The extension of the formalism to the nuclear medium takes into account
the modifications of the properties of hadrons in the medium (masses and coupling constants),
as determined by the QMC model~\cite{GA-Medium1,Octet2,QMCReview,Lu99a,QMCEMFFMedium2,QMCEMFFMedium5}.
The combination of formalisms is justified
because there is a correspondence between the
quark effective electromagnetic structure
of the MIT bag model/cloudy bag model (CBM)~\cite{Thomas84,Thomas83,Kubodera85,Yamaguchi89,Tsushima88,Miller02,Lu01b,Lu98x,Lu01a} and the
covariant spectator quark model~\cite{OctetDecuplet1,OctetDecuplet2,OctetDecupletD1,OctetDecupletD2}.

In the covariant spectator quark model, the baryon cores are described as systems of three quarks that interact with electroweak probes under the relativistic impulse approximation (additive quark model).
The quarks have their own internal structure (constituent quarks)
that simulates  dressing by gluons and quark--antiquark pairs,
consistently with chiral symmetry in the confining limit~\cite{Nucleon}.
This complex structure is taken into account considering
a vector meson dominance structure for electroweak quark currents.
This representation of the quark structure is consistent with the
picture discussed within the Dyson--Schwinger framework
and visualized in lattice QCD simulations,
associated with  dynamical chiral symmetry breaking
that leads to massive extended quarks at low energies~\cite{Eichmann16a,Cloet14a,Maris03a}.
The quark internal structure and quark masses can then be regraded
as a manifestation of the dynamical chiral symmetry breaking.

Using the proposed formalism, we calculate the 
elastic electromagnetic octet baryon form factors
(electric $G_E$ and magnetic $G_M$)
and the axial form factors (axial-vector $G_A$
and induced pseudoscalar $G_P$)
associated with the allowed transitions between
octet baryon states and discuss the
impact of the medium effects in terms of the nuclear medium density.
The study of the transitions includes $|\Delta I|=1$ transitions
(conversion $u \leftrightarrow d$), $|\Delta S|=1$ transitions
(conversion $s \leftrightarrow d$), and neutral current transitions
($\Delta I= \Delta S=0$).
The impact of the medium on the electromagnetic form factors depends
on the flavor content (charged baryons, neutral baryons, and number of strange quarks).
In general, $G_A$ and $G_P$ are suppressed in a nuclear medium,
but the magnitude of the suppression depends on the type
of the transition ($|\Delta I|=1$ and $|\Delta S|=1$) and on the mass of the baryons.

We use the numerical results to
 study reactions of neutrinos/antineutrinos
with nuclei and reactions of neutrinos/antineutrinos with hyperons
in dense nuclear matter for different densities.
{
The methods developed here for the octet baryons
can be extended in the future to  other baryon systems, like decuplet baryons and 
transitions between octet baryons and decuplet baryons,
as well as for densities larger than the normal nuclear matter.
There is also the possibility of  generalization
to asymmetric nuclear matter.}

The present article is organized as follows:
In the next section, we introduce the structure functions used
in the study of  electromagnetic and axial interactions
with  octet baryons and transitions between octet baryon states.
In Section~\ref{secModel}, we explain the theoretical formalism
used in the calculation of the octet baryon electromagnetic and axial form factors in free space and in the nuclear medium.
Numerical calculations of octet baryon electroweak form factors
in the nuclear medium for different nuclear matter densities
are presented and discussed in Section~\ref{secResults1}.
In Section~\ref{secResults2}, we discuss the applications
of the numerical calculations for nucleons bound to nuclei,
the calculation of  neutrino--nucleon
and antineutrino--nucleon cross-sections for bound nucleons,
and the extension of the formalism for
nuclear matter densities above the normal nuclear matter ($\rho > \rho_0$).
{
It is the first time that octet baryon form
factors are being calculated for densities larger than
the normal nuclear matter within our framework.}
The limitations and possible improvements of
the formalism are discussed in Section~\ref{secDiscussion}.
The outlook and conclusions are provided in Section~\ref{secConclusions}.

\section{Octet Baryon Electromagnetic and Axial Form Factors \label{secDefs}}

We focus our attention now on octet baryons.
The octet baryons are characterized by spin $J= \frac{1}{2}$
and positive parity ($P=+$): $J^P=\frac{1}{2}^+$ in a compact notation.
The baryon octet includes the nucleons $N$ (proton $p$ and neutron $n$),
the $\Lambda$, the $\Sigma$ baryons ($\Sigma^+$, $\Sigma^0$, and $\Sigma^-$),
and the $\Xi$ baryons ($\Xi^-$ and $\Xi^0$).

We will use the label $B$ to represent properties associated with the baryon $B$
(masses, charges, magnetic moments, etc.).
The elastic form factors associated with the interaction with an octet baryon member
are also labeled by the index $B$.

As for the axial transitions, related to the weak interaction,
we need to consider different labels ($B$ and $B'$) for the $B' \to B$ transitions.
For simplicity, we avoid the representation
$B' \to B x$, where $x =  \ell \bar \nu_{\ell}$ for beta decays,
or $x = \bar \ell \nu_{\ell}$ for inverse beta decays,
with \mbox{$\ell =e, \mu, \tau$} (electron, muon, or tau), 
since the properties of the current are independent of the \mbox{lepton family.}

The axial transitions can be divided into three different types:
change in isospin (like the $n \to p$ beta decay),
change in strangeness (like $\Sigma^- \to p$),  and neutral current transitions.
The first two types (charged transitions) are mediated by the bosons $W^\pm$; the neutral current transitions are mediated by the boson $Z^0$.
The list of possible transitions is included in Table~\ref{tab-Axial-Transitions}.

These transitions are classified according to the kinematic of the free space.
The decays of $\Sigma^-$ to $\Sigma^0$ and $\Sigma^0$ to $\Sigma^+$ are kinetically allowed due
to the magnitude of their masses.
Notice that the order of the decays can be inverted when
the baryons are excited by interaction with the nuclear medium.
In nuclear medium, one can have for instance the transitions
$p \to n$, $\Sigma^0 \to \Sigma^-$ and $\Sigma^+ \to \Sigma^0$.

\begin{table}[H]
\caption{Axial transitions between octet baryon members:
    $|\Delta I|=1$, $|\Delta S|=1$ and neutral current transitions~\cite{AxialFF}.
    There are no contributions to the $\Lambda \to \Lambda$ transition.
    The axial flavor operators $X$ associated with  $|\Delta I|=1$ ($I_\pm$),
    $|\Delta S|=1$ ($V_\pm$) and $\Delta I= \Delta S=0$ ($I_0$)
    are defined in Appendix~\ref{appGM-matrices}
    and discussed in the next section.
    \label{tab-Axial-Transitions}}

  \begin{tabularx}{\textwidth}{CCC}
\toprule
      & \boldmath{$B \to B'$} & \boldmath{$X$} \\
\midrule
$|\Delta I|=1$  & $ n \to p$ & $I_+$ \\
                &  $\Sigma^\pm \to \Lambda$ & $I_\mp$ \\
                   & $\Sigma^- \to \Sigma^0$ &  $I_+$ \\
                           & $\Sigma^0 \to \Sigma^+$  & $I_+$ \\
                         & $\Xi^- \to \Xi^0$  &  $I_+$ \\
               &                   &  \\
$|\Delta S| = 1$                       &  $\Sigma^{-} \to n$ & $V_+$  \\
                                    &  $\Sigma^0 \to p$  &  $V_+$    \\
                                  &  $\Xi^- \to \Lambda$    & $V_+$\\
                            & $\Xi^- \to \Sigma^0$ & $V_+$\\
                                  & $\Xi^0 \to \Sigma^+$  & $V_+$ \\
                  &                   &  \\
$\Delta I = 0$    &    $N \to N$  & $I_0$\\
$\Delta S = 0$   &     $\Sigma \to \Sigma $  & $I_0$ \\
                  &    $\Xi \to \Xi$  & $I_0$\\
\bottomrule
\end{tabularx}
\end{table}

We discuss now the formalism associated with
the electromagnetic transitions and weak axial transitions
 between octet baryon members.

For the studies in nuclear medium, we
use * to label the properties in medium
(masses $M_B^\ast$, form factors, etc.).

\subsection{Electromagnetic Transitions \label{secEMFFdef}}

We consider first the electromagnetic
interactions with an octet baryon:  $\gamma^\ast B (P) \to B(P')$,
where $q=P'-P$ is the transfer momentum.
The electromagnetic current associated with
the $\gamma^\ast B \to B$ transition can be written as~\cite{Octet2,Octet,Octet4}
\ba
J_B^\mu = \bar u_B(P') \left[
  F_{1B} (Q^2) \gamma^\mu + F_{2B}  (Q^2) \frac{i \sigma^{\mu \nu} q_\nu}{2 M_B} \right] u_B(P),
\label{eqJem}
\ea
where $Q^2= -q^2$, $u_B(P')$ and  $u_B(P)$ are
the final and initial Dirac spinors,  respectively.
The functions $F_{1B}$ and $F_{2B}$ are the Dirac and Pauli form factors, respectively.

The current $J_B^\mu$ defines the baryon $B$
elastic form factors.
In this discussion, we exclude the kinetically
allowed transition between octet baryon members:
$\gamma^\ast \Lambda \to \Sigma^0$ transition~\cite{Sigma0Lambda}.
This transition has properties that differ from
the octet baryon elastic transitions
and has similarities with the inelastic
$\gamma^\ast N \to N^\ast$ transitions~\cite{Ramalho-Pena23}.

Using  $F_{1B}$ and $F_{2B}$, we define the Sachs form factors,
electric and magnetic~\cite{Octet,HyperonFF1},
\ba
G_{EB} (Q^2) &= & F_{1B} (Q^2) - \frac{Q^2}{4 M_B^2}   F_{2B} (Q^2),  \\
G_{MB} (Q^2) &= & F_{1B} (Q^2) + F_{2B} (Q^2).
\ea
At $Q^2=0$, one has $F_{1B} (0)= e_B$,
where $e_B$ is the baryon charge, and $F_{2B} (0)= \kappa_B$
is the baryon $B$ anomalous magnetic moment.
The function  $G_{MB} (0)$ represents the magnetic moment
in natural units $\frac{e}{2 M_B}$,
where $e$ is the elementary electric charge
\ba
\mu_B = G_{MB} (0) \frac{e}{2 M_B}.
\ea

The comparison of magnetic moments is usually conducted
in units of nuclear magneton ($\hat \mu_N$), corresponding to
the nucleon natural units in the free space, $\hat \mu_N \equiv \frac{e}{2 M_N}$,
where $M_N$ is the physical nucleon mass.
We can then write~\cite{Octet2,Octet3}
\ba
\mu_B = G_{MB} (0) \frac{M_N}{M_B} \hat \mu_N,
\label{eqMu-nucleon}
\ea
where $ G_{MB} (0) \frac{M_N}{M_B}$ is the numerical value
of $\mu_B $ in nuclear magneton.
We can also represent the anomalous magnetic moment $F_{2B}(0)$
in nuclear magneton.

  From the expression for $G_{EB}(Q^2)$, we can conclude 
  that, for neutral baryons, $G_{EB}$ should vanish for $Q^2=0$ since
  the charge is zero.
  This condition is expressed by $F_{1B} \propto Q^2$,
  leading also to the analytic relation $G_{EB} \propto Q^2$, near $Q^2=0$.

In the nuclear medium, we redefine
the Dirac and Pauli form factors as $F_{1B}^\ast$ and $F_{2B}^\ast$, respectively.
As a consequence, the electric and the magnetic form factors
take the form
\ba
G_{EB}^\ast (Q^2) &= & F_{1B}^\ast (Q^2) - \frac{Q^2}{4 M_{B}^{\ast 2}}   F_{2B}^\ast (Q^2), \label{eqGEdef} \\
G_{MB}^\ast (Q^2) &= & F_{1B}^\ast (Q^2) + F_{2B}^\ast (Q^2),
\label{eqGMdef}
\ea
where $M_B^\ast$ is the in-medium baryon mass.

In the nuclear medium, the baryon masses are modified (to $M_B^\ast$).
In these conditions, the comparison with the magnetic moments
in free space must take into account the units in vacuum.
Taking the nucleon as an example, in the nuclear medium,
the nucleon mass is modified to $M_N^\ast < M_N$, 
and the magnetic moment in the nuclear medium ($B=N$) is
\ba
\mu_N^\ast = G_{MN}^\ast (0) \left( \frac{e}{2 M_N^{\ast}} \right) =
G_{MN}^\ast (0) \frac{M_N}{M_N^{\ast}} \left(  \frac{e}{2 M_N}   \right).
\ea
To obtain the magnetic form factor in units of the nuclear magneton, we need then to multiply $G_{MN}^\ast (0)$ (the magnetic form factor in natural units) by the factor $ \frac{M_N}{M_N^{\ast}}$.
In other words, in units of the nuclear magneton, the magnetic moment is corrected by the factor $\frac{M_N}{M_N^\ast} > 1$.


\subsection{Weak Transitions and Axial Form Factors}

We consider now the transition $B^\prime (P') \leftrightarrow B(P) x(q)$,
where $x$ represents the weak transition mediator ($W^\pm$ or $Z^0$ bosons).
We include the double direction arrow to take into account all possible transitions,
including decays (beta or muon) and inverse decays~\cite{Gaillard84,GA-Medium1}.

The $B \to B'$  weak transition current can be represented as~\cite{Bernard02,Gorringe04}
\ba
( J_5^\mu)_{B'B}
= \frac{1}{2}  \bar u_{B'}(P')  \left[ G_A(Q^2) \gamma^\mu + G_P (Q^2) \frac{q^\mu}{2 M_{BB'}}  
  \right] \gamma_5 u_B(P),
\label{eqJaxial}
\ea
and defines the axial-vector $G_A$ and the induced pseudoscalar $G_P$ form factors, 
and $M_{BB'} = \frac{1}{2}(M_{B'} + M_B)$ is the average between
the mass of the initial state ($M_B$) and the final state ($M_{B'}$), 
and $q=P' -P$. As before, $Q^2=-q^2$.
The factor $1/2$ is included to be consistent with the nucleon case
($n \to p$ beta decay).
For simplicity, we omitted the indexes $B$ and $B'$ from the representation
of the form factors.

The expression (\ref{eqJaxial}) is obtained after a proper projection in
the flavor space.
The axial flavor operators can be represented
in terms of the Gell--Mann matrices $\lambda_i$ (\mbox{$i=1,\ldots,8$})~\cite{Gaillard84}, 
as given in Appendix~\ref{appGM-matrices}.  
{The allowed transitions correspond to the operators: $I_0= \lambda_3$  (neutral current transitions), $I_\pm$ (increases/decreases the isospin projection), and $V_\pm$ (decreases/increases the number of strange quarks).}

In the nuclear medium, the current  (\ref{eqJaxial})
is replaced by in-medium equivalents, including the in-medium effective masses
$M_B^\ast$, $M_{B'}^\ast$ that define $M_{BB'}^\ast$,
and the axial form factors $G_A^\ast$ and $G_P^\ast$.


\section{Covariant Quark Model for
  the Free Space and Nuclear Medium \label{secModel}}

In the present section, we review the formalism
associated with the covariant spectator quark model
used in the description of the electromagnetic
and axial structure of baryons in the free space
and discuss how the formalism can be extended to the nuclear medium.
The extension is obtained from a consistent combination of
the covariant spectator quark model with the CBM and with the QMC model~\cite{OctetDecuplet1,OctetDecuplet2,OctetDecupletD1,OctetDecupletD2}.

We start by discussing the model at the microscopic level, how the baryon wave functions are defined
in terms of the quark flavor and spin properties,
and how the electroweak interactions with the quarks are described.
In the covariant spectator quark model, the quarks
have an internal structure that is a consequence
of the quark dressing (gluons and quark--antiquark pairs).
In the interaction with electroweak probes,  we can reduce the probe--quark interactions
to a coupling with a single quark in a quark--diquark system.
In this formalism, the wave functions of the baryons are not determined
by a dynamical wave equation but are instead
built in terms of the quark spin-flavor structure
with radial wave functions determined phenomenologically
for a few ground state baryons.
The motivation of the formalism is not to
determine the baryon mass spectrum but 
to describe the structure functions of the baryons. 
In the numerical calculations, we consider then the
experimental masses of the baryons.

Within the covariant spectator quark model, the particular form used for the quark--diquark radial wave functions and for the quark currents is employed in a first stage, in the extension of the formalism to a lattice QCD regime associated with a defined pion mass, and in a second stage
in the extension to the nuclear medium.
These extensions are \mbox{discussed below.}

It is worth mentioning, however, that the consideration of
the quark degrees of freedom exclusively is not sufficient to describe
the electromagnetic and axial structure of the baryons
and transitions between baryon states at low $Q^2$.
In that regime, there are excitations that cannot be described
only in terms of valence quark effects, and excitations
related to quark--antiquark or meson states may have relevant contributions.
Those processes can be regarded as a meson cloud dressing
at the hadronic level
and can be interpreted as interactions between different quarks
inside the baryon (not quark self-dressing)~\cite{Octet4,Octet,Octet2,OctetDecuplet2,OctetDecupletD1,OctetDecupletD2}.
Since the meson cloud dressing at the hadronic level differs from the
quark self-dressing, there is no double counting.

\begin{figure}[H]
  \includegraphics[width=2.9in]{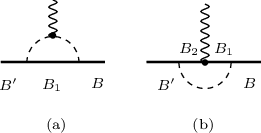}
  \caption{Electroweak interactions with a baryon $B$ within the one-meson-loop level.
      (a) Interaction with meson. (b) Interaction with intermediate baryons.
      $B'$ differs from $B$ in the inelastic transition (weak axial transitions).
      The states $B_1$ and $B_2$ represent generic intermediate states.
\label{fig-Meson-Cloud}}
\end{figure}

We consider then that, in the electroweak interactions, the elastic and transition form factors can be decomposed into a valence quark contribution and a contribution associated with the electromagnetic and axial interaction with the bare core dressed by meson cloud, hereafter mentioned as the meson cloud contribution (see Figure~\ref{fig-Meson-Cloud}).
The separation between valence and meson cloud contributions
is naturally model-dependent. The model dependence is a consequence of the differences 
in the calibration of the background and in
the identification of the bare states~\cite{Bernard98a,Hammer04a,Meissner07a,Pascalutsa07a,Ronchenn13a}.
It is expected that the impact of the model dependence is reduced in our formalism
since the bare parameters of the model are determined by
lattice QCD simulations associated with large pion masses.
Although lattice QCD simulations
include some meson cloud effects, their effects
are reduced when the pion mass is large~\cite{Pascalutsa07a,Alexandrou08a}.

The masses of the octet baryons are not calculated within
the quark model formalism.
In the calculations in the free space,
we consider the averages of the experimental values
of the masses of the different octet baryon families ($N$, $\Lambda$, $\Sigma$, and $\Xi$).
The values for the effective masses in medium are discussed next
(see Section~\ref{secMedium1}).

The covariant spectator quark model has been used successfully in the study of
the $\gamma^\ast N \to N^\ast$ transitions
for the nucleon resonances $N^\ast= \Delta(1232)$, $N(1440)$, $N(1520)$,
and $N(1535)$~\cite{Spectator-Review,NDeltaSL1,NDeltaSL2,LatticeD,Lattice,Roper1,Roper2,SemiRel,N1535SL1,N1535SL2,N1520SL,Siegert1,Siegert2}, among others~\cite{Delta1600,SQTM,LambdaStar}.
The formalism has also been applied to the Dalitz decay
of nucleon resonances and other baryons
($B^\ast \to e^+ e^- B$)~\cite{OctetDecupletD1,OctetDecupletD2,NDeltaTL3,NDeltaTL4,N1535TL,N1520TL}, 
to the study of the elastic electromagnetic form factors
of baryons in the spacelike region~\mbox{\cite{Octet,Omega,Omega2,DeltaFF,DeltaFF2,DeltaFF3}}, in the timelike
region~\cite{HyperonFF1,HyperonFF2,HyperonFF3}, 
and to the nucleon deep inelastic scattering~\cite{Nucleon,DIS}.

We start by discussing the valence quark component of the model.
Next, we consider the contributions associated with the meson cloud dressing.
At the end, we explain how the formalism
is naturally extended to the nuclear medium.

\subsection{Model for the Valence Quark Contributions \label{secBare}}

The covariant spectator quark model is a constituent
quark model derived from the covariant spectator theory~\cite{Nucleon,Gross69,Gross97}.
In the model, the baryons are regarded as three-constituent quark systems,
where a quark is free to interact with electroweak fields in relativistic
impulse approximation while the two quarks
are spectators~\cite{Spectator-Review,Nucleon,Nucleon2,FixedAxis,Omega,AxialFF,GA-Medium1}.
{Integrating over the degrees of freedom of the two spectator quarks, we reduce the baryon system to a quark--diquark configuration, where the diquark can be represented as an on-shell particle with effective mass $m_D$.}
The effective quark--diquark wave function
simulates the effect of confinement and is consistent with the chiral symmetry
in the limit when the mass of
the quarks vanishes~\cite{Nucleon,Nucleon2,Savkli01a,Gross06a}.

The interaction of the probe with the quarks is described in terms of quark
electromagnetic and axial form factors that simulate the structure
associated with the gluon and quark--antiquark dressing of the quarks.
These constituent quark form factors are parametrized using
a vector meson dominance form 
calibrated in the study of the electromagnetic structure of the
nucleon and baryon decuplet~\cite{Nucleon,Omega,AxialFF}.

The baryon wave functions are derived from the spin-flavor-orbital
SU(6)$\otimes$O(3) symmetry associated with the
quark--diquark spin-flavor and radial configurations.
The radial wave functions are determined phenomenologically
from experimental data or lattice QCD data
associated with some ground state systems.
In the model, the effect of the SU(3) quark flavor symmetry-breaking 
is reflected at the level of the
baryon radial wave functions, considering different range
parameters for the systems, according to the quark \mbox{flavor content.}

We discuss now the structure of the wave functions of the octet baryons.
Following Refs.~\cite{Octet2,GA-Medium1},
we consider the octet baryon wave functions composed of  
a mixture of quark--diquark configurations in   
an $S$-state and a $P$-state
\ba
\Psi_B(P,k) = n_S \Psi_S (P,k) + n_P \Psi_P(P,k),
\label{eqPsi-total}
\ea
where $P$ is the momentum of the baryon, $k$ is the diquark momentum,
$n_P$ is the $P$-state admixture, and $n_S= \sqrt{1- n_P^2}$.
Based on the studies of the electromagnetic structure
of the nucleon and the baryon octet, we expect
the $S$-state to yield the dominant contribution~\mbox{\cite{Octet4,Octet,Nucleon}.}

{
The $S$- and $P$-components of the wave functions are defined as~\cite{GA-Medium1,AxialFF}}
\ba
 & &
\Psi_S (P,k) = \frac{1}{\sqrt{2}} \left[
\phi_S^0 \left| M_A \right> +  \phi_S^1 \left| M_S \right>
\right] \psi_S(P,k), \label{eq-PsiS}\\     
& &
\Psi_P (P,k) = {\tilde {\not k}} \frac{1}{\sqrt{2}} \left[
\phi_S^0 \left| M_A \right> +  \phi_S^1 \left| M_S \right>
\right] \psi_P(P,k),  \label{eq-PsiP} 
\ea
where $\phi_S^{0,1}$ are the spin wave functions labeled
by the diquark spin ($S=0,1$), and $\left|M_{S,A} \right>$ are the
mixed symmetric ($S$) and mixed anti-symmetric ($A$) 
flavor states in the exchange of quarks 1 and 2~\cite{Nucleon,Omega,Octet}.
The functions $\psi_S$ and $\psi_P$ are the radial wave functions,
and $\tilde k = k - \frac{P \cdot k}{M_B^2} P$
(recall that $P^2=M_B^2$ for on-shell baryon states). 
In the notation for $\Psi_B(P,k)$ and  $\left|M_{S,A} \right>$,
we omit for the simplicity
the labels associated with the spin of  baryon $B$.
The explicit expressions for the flavor are presented
in Appendix~\ref{appendixBare}.

The operator ${\tilde {\not k}}$ is included
to generate a $P$-state wave function. In the numerical calculations, 
we correlate the  $P$-state radial wave function
with the $S$-state radial wave function
using $\psi_P (P,k) = \psi_S(P,k)/\sqrt{- \tilde k^2}$
for simplicity
(notice that $\tilde k^2 \le 0$ by construction)~\cite{Nucleon2,AxialFF,GA-Medium1}.
In these conditions, compared with the $S$-state model ($n_P=0$),
no extra momentum range parameters 
are included in the model.

\subsubsection{Radial Wave Functions \label{secRWF}}

In the covariant spectator quark model,
the radial wave functions of a baryon $B$
are represented in terms of the dimensionless variable
\ba
\chi_{\ms B} = \frac{(M_B - m_D)^2 - (P-k)^2}{M_B m_D}.
\label{eqChi}
\ea

The $S$-state radial wave functions take the Hulthen form~\cite{Nucleon,Gross06a,NDeltaSL2}
\ba
\psi_B(P,k) = \frac{N_B}{m_D (\beta_1 + \chi_{\ms B})(\beta_l + \chi_{\ms B})},
\label{eqRWF}
\ea
where $N_B$ is a normalization constant, and 
$l=2$ ($N$), $l=3$ ($\Lambda$ and $\Sigma$), or $l=4$ ($\Xi$)
are momentum range parameters determined
in the study of the electromagnetic structure of
the octet baryon in free space
from the analysis of the lattice QCD data~\cite{Octet2,Lin09}. 
We consider here the parametrization from Ref.~\cite{Octet2}:
$\beta_1 = 0.0532$, $\beta_2= 0.809$, $\beta_3 = 0.603$,
and $\beta_4 = 0.381$.
We obtain then the order $\beta_2 > \beta_3 > \beta_4$
expected from the natural size of the baryons.
Systems with more strange quarks are more compact than
systems with more light quarks.
The normalization constant $N_B$ is determined by the
condition $\int_k |\psi_B(P,k)|^2=1$ when $P=(M_B,{\bf 0})$.
The integral on $k$ is defined below.

The form used for the radial wave functions (\ref{eqRWF}) is motivated by simplicity, 
by the asymptotic falloff induced in the electromagnetic form factors,
and by the possibility of breaking the SU(3) symmetry
in the simple form, as mentioned already~\cite{Nucleon,Octet,Omega,GA-Medium1}.
The presence of the factor $1/m_D$ in the definition of the radial wave functions
implies that the form factors are independent of the value of diquark mass~\cite{Nucleon,NDeltaSL2}.
This approximation is justified by the
effective parametrization of the radial wave functions, 
and by phenomenological determination of the values
of the momentum range parameters $\beta_l$  ($l=1,\ldots,4$).

\subsubsection{Electromagnetic Interaction with Quarks \label{secQEMcurr}}

The electromagnetic transition current associated
with the interaction with a baryon $B$
can be calculated in impulse approximation
using~\cite{Nucleon,Nucleon2,Octet,Octet2}
\ba
J_{0 B}^\mu = 3 \sum_{\Gamma} \int_k {\overline \Psi_B}(P',k) j_q^\mu \Psi_B(P,k),
\label{eqEM-bare}
\ea
where $\Psi_B(p,k)$ are the quark--diquark wave function described above,
$j_q^\mu$ is the quark current operator,
and  factor 3 takes into account the contributions
of all diquark pairs.
Hereafter, we use the subscript $0$ to label functions related
to the bare contributions.
In the calculation, we consider the sum into the intermediate
diquark polarizations $\Gamma = { s, \lambda}$,
including the scalar diquark ($s$) and the vectorial diquark polarization
$\lambda =0, \pm$, and integrate on the on-shell diquark momentum using 
\ba
\int_k \equiv \int \frac{d^3 {\bf k}}{2 E_D (2\pi)^3},
\hspace{.2cm} \mbox{where} \hspace{.1cm} E_D = \sqrt{m_D^2 + {\bf k}^2},
\ea
which is the diquark energy.

The relativistic impulse approximation current (\ref{eqEM-bare})
is based on a single quark operator (i.e.,~no exchange or interaction currents)
and is valid when we consider an effective quark--diquark wave function
with phenomenological radial wave functions~\cite{Omega,Nucleon,GA-Medium1}.

The quark current operator $j_q^\mu$ in Equation~(\ref{eqEM-bare}) has the general structure~\cite{Nucleon,Octet2,Octet}
\ba
j_q^\mu (q) = j_1 (Q^2) \gamma^\mu + j_2(Q^2) \frac{i\sigma^{\mu \nu} q_\nu}{2 M_N},
\label{eq-jq}
\ea
where $j_i$ are SU(3) flavor operators acting on the
third quark of the $\left|M_S  \right> $ and  $\left|M_A  \right> $ states.
The quark current (\ref{eq-jq}) includes an implicit connection
with the quark mass expressed by the dependence on
the nucleon mass $M_N$ on the Pauli term.

The operators $j_i$ ($i=1,2$) can be decomposed  into a sum
of operators acting on  quark 3 in the SU(3) space~\cite{Omega,Octet}
\ba
j_i (Q^2) = \frac{1}{6} f_{i+} (Q^2) \lambda_0 +  \frac{1}{2} f_{i-} (Q^2) \lambda_3 +
\frac{1}{2} f_{i0} (Q^2) \lambda_s,
\label{eq-jq2}
\ea
where $\lambda_0 = {\rm diag}(1,1,0)$, $\lambda_3 = {\rm diag}(1,-1,0)$, and
$\lambda_s = {\rm diag}(0,0,-2)$ are the flavor space operators
[see Appendix~\ref{appGM-matrices}].
These operators act on the wave functions in the flavor space $(u\, d\,   s)^T$.

The functions $f_{i\pm} (Q^2)$ ($i=1,2$)
are normalized by $f_{1 \pm} (0)=1$, 
$f_{2\pm} (0)= \kappa_\pm$, and $f_{1 0} (0)=1$,  $f_{2 0} (0)=\kappa_s$.
In this notation, $\kappa_+$ and  $\kappa_-$
are the isoscalar and  isovector quark anomalous magnetic moments
and $\kappa_s$ the strange quark anomalous magnetic moment.
The quark form factors are parametrized in terms of
a vector meson dominance form that includes contributions
from the $\rho$, $\omega$, and $\phi$ meson poles
plus an effective vector meson pole that simulates the short range structure.

The explicit expressions and parameters
related to the quark current and the radial wave functions 
are included in Appendix~\ref{appendixBare}.
The isoscalar and  isovector magnetic moments can be expressed
in terms of quarks $u$ and $d$ regarding anomalous magnetic moments
($\kappa_u$ and $\kappa_d$) [see Appendix \ref{appBareEMFF}].

The evaluation of the transition current (\ref{eqEM-bare}) 
with the  octet baryon wave functions (\ref{eq-PsiS})
leads to the following expressions for the
electromagnetic form factors
as defined by Equation~(\ref{eqJem})~\cite{Octet2,Octet}
\ba
&  & 
F_{1B}(Q^2) =
\left(\frac{3}{2} j_1^A + \frac{1}{2} \frac{3- \tau}{1 + \tau} j_1^S
- 2 \frac{\tau}{1 + \tau} \frac{M_B}{M_N} j_2^S
\right)  B(Q^2), \label{eqF10}\\
&  & 
F_{2B}(Q^2) =
\left[ \left( \frac{3}{2} j_2^A - \frac{1}{2} \frac{1- 3\tau}{1 + \tau} j_2^S \right)
  \frac{M_B}{M_N}
- 2 \frac{1}{1 + \tau} j_1^S
\right]  B(Q^2),
\label{eqF20}
\ea
where
\ba
\tau= \frac{Q^2}{4 M_B^2}, \hspace{.8cm}
B(Q^2) = \int_k \psi_B(P',k) \psi_B(P,k), 
\ea
and
\ba
j_i^A = \left<M_A \right| j_i \left| M_A\right>, \hspace{1.cm}
j_i^S = \left< M_S\right| j_i \left| M_S\right>,
\label{eqjiAS}
\ea
are the projections of the quark current operators
on the mixed anti-symmetric ($A$) and mixed symmetric ($S$) states.

The flavor structure of the octet baryons
is encoded in the coefficients $j_i^A$ and $j_i^S$.
These coefficients are presented in Appendix~\ref{appBareEMFF}.
The radial structure of the baryons
is included in the overlap integral
of the radial wave functions $B(Q^2)$,
normalized as $B(0)=1$.
The present expressions are based on a model
that includes only the $S$-state contribution
to the octet baryons~\cite{Octet,Octet2}.

The electric and magnetic form factors are calculated
using Equations~(\ref{eqF10}) and (\ref{eqF20}).
The previous relations take into account only the valence
quark contribution to the electromagnetic form factors.
For the discussion of dressed form factors
that take into account the meson cloud dressing,
discussed above, we consider the compact notation
\ba
\tilde e_{0B} \equiv F_{1B}(Q^2), \hspace{1.2cm}
\tilde \kappa_{0B} \equiv F_{2B}(Q^2).
\label{eqek-tilde}
\ea
Using this notation, we can calculate the valence quark 
contribution to the electric and magnetic form factors
\ba
G_{E0B}(Q^2) = \tilde e_{0B} - \tau  \tilde \kappa_{0B}, \hspace{1.cm}
G_{M0B}(Q^2) = \tilde e_{0B} +\tilde \kappa_{0B}.
\label{eqGEGMbare}
\ea

The combination of bare and meson cloud effects to the electromagnetic form factors is discussed in Section~\ref{secEM-pioncloud}.

\subsubsection{Weak Interaction with Quarks \label{sec-Axial}}

In the covariant spectator quark model, the weak transition
current associated with the $B \to B'$ transitions 
is determined in relativistic impulse approximation using~\cite{Nucleon,Omega,Nucleon2,AxialFF}
\ba
\hspace{-1.cm}
(J_5^\mu)_{B'B} =
3 \sum_{\Gamma}
\int_k \overline \Psi_{B'}(P',k) (j_{Aq}^\mu) \Psi_B(P,k), 
\label{eqJ5}
\ea 
where $j_{Aq}^\mu$ is the quark axial current operator.
We follow here the notation used for the electromagnetic transition current (\ref{eqEM-bare}).
The quark axial current operator includes the axial flavor
operators associated with the hadronic transition $B \to B'$,
as discussed next.    
The current (\ref{eqJ5}) can be used to
calculate the axial form factors based
on the general form (\ref{eqJaxial}).

The possible transitions
(allowed kinematically) between the octet baryon members are
presented in Table~\ref{tab-Axial-Transitions}.
We can divide the transitions into 3 kinds, 
depending on the flavor transition operator $X$:
transitions associated with the variation in isospin $\Delta I$
($u \leftrightarrow d$, operator $I_\pm$), transitions associated
with the variation in strangeness $\Delta S$
($d \leftrightarrow s$, operator $V_\pm$), and neutral current transitions ($I_0 = \lambda_3$).

{
In the weak axial interaction with the baryons, we consider the quark axial current operator
\ba
j_{Aq}^\mu = 
\left(  g_A^q (Q^2) \gamma^\mu +
g_P^q(Q^2) \frac{q^\mu}{2 M_N}
\right) \gamma_5 \frac{\lambda_a}{2},
\label{eqJAq}
\ea
where $\lambda_a$  ($a=1,\ldots,8$) are the Gell--Mann matrices.
In the previous equation, $g_A^q$ and $g_P^q$ are the quark 
axial-vector and quark-induced pseudoscalar form factors, respectively.
The flavor operators $\lambda_a$
act on the quark flavor states ($\left|M_{A,S} \right>$), and the Lorentz
operators act on the baryon spin states.}
The explicit expressions for $g_A^q$ and $g_P^q$
are presented in Appendix~\ref{appBareAxial}.

As for the quark electromagnetic current $j_q^\mu$,
the current $j_{Aq}^\mu$ is defined in terms of the nucleon mass for convenience.
For $g_A^q$, we assume that the function is equivalent
to the Dirac isovector form factor $f_{1-}$ due
to its isovector character~\cite{AxialFF}.

The induced pseudoscalar form factor $G_P$ can be decomposed 
in the bare and pole contributions~\cite{AxialFF}
\ba
G_P (Q^2)= G_P^{\rm B} (Q^2) + G_P^{\rm pole} (Q^2),
\label{eqGP-BP}
\ea
where the pole contribution takes the form
\ba
G_P^{\rm pole} (Q^2) = \frac{4 M_{BB'}^2}{\mu^2 + Q^2} G_A^{\rm B} (Q^2),
\label{eqGP-pole}
\ea
where $G_A^{\rm B}$ is the bare contribution to $G_A$,
and $\mu$ is the mass of the meson related
 to the weak transition (the pion for $|\Delta I|=1$ transitions and
 neutral current transitions, and the kaon for the $|\Delta S|=1$ transitions).

The bare contributions to the axial transitions
can be expressed in the form~\cite{AxialFF}
\ba
G_A^{\rm B} (Q^2)&=& g_A^q {\cal F}\left\{
  \frac{3}{2} n_S^2 B_0 - 3 n_{SP} \frac{\tau}{1 + \tau} B_1 + \frac{6}{5} n_P^2
  \left[ \tau B_2 - (1 + \tau) B_4 \right]
  \right\}, \label{eqGA-bare}\\
  G_P^{\rm B} (Q^2) &= & g_A^q {\cal F}\left\{
  - 3 n_{SP} \frac{1}{1 + \tau} B_1
  + \frac{3}{2} n_P^2 \left[
    B_5 + 2(B_2 -B_4)
    \right]
  \right\} + \label{eqGP-bare} \\
  & & g_P^q {\cal F} \frac{M_{BB'}}{M_N}\left\{
  \frac{3}{2} n_S^2 B_0 - 3 n_{SP} B_1 +
   \frac{3}{2} n_P^2 \left[ \tau B_2 + B_3  - (2 + \tau)B_4 \right]
    \right\},
\nonumber
\ea
where ${\cal F}$ is a coefficient dependent on the baryon flavor, determined by
\ba
{\cal F} = f_X^A - \frac{1}{3} f_X^A , \hspace{.5cm}
f_X^A = \left<M_A \right| X \left| M_A\right>, \hspace{.5cm}
f_X^S = \left< M_S\right| X \left| M_S\right>.
\label{eq-calF}
\ea
The expressions for $f_X^{A,S}$ and ${\cal F}$ are presented in Appendix~\ref{appBareAxial}.

In Equations~(\ref{eqGA-bare}) and (\ref{eqGP-bare}), 
$n_{SP} = n_S n_P$ and the functions $B_i$ ($i=0,\ldots,5$) are
overlap integrals of the radial functions $\psi_S$ and $\psi_P$.    
The normalization of the radial wave function leads to $B_0(0)=1$.
The integrals $B_i$ ($i=0,..,5$) are defined in Appendix~\ref{appBareAxial}.

In the calculation of the octet baryon axial form factors,
we use the radial wave functions and the Dirac isovector form
factor $f_{1-}$ determined by the study of the octet baryon electromagnetic form factors~\cite{Octet2}.
The free parameters of our model of the axial form factors:
the admixture parameter $n_P$ and the parameters of $g_P^q$
[see Appendix~\ref{appBareAxial}] are determined by the
fit to lattice QCD results for the nucleon axial form factors
from Ref.~\cite{Alexandrou11a}.

The meson cloud contributions to the the axial form factors are discussed in \mbox{Section~\ref{sec-Axial-MC}.}

\subsubsection{Extension of the Model to the Lattice QCD Regime \label{secLattice}}

The quark model, discussed in the previous sections
for baryons in the free space, can be extended to
lattice QCD regimes associated with a given pion mass.
The extension of the formalism to the lattice QCD regime
is based on the properties of the radial wave
functions and the vector meson dominance form of the electromagnetic quark currents~\cite{Octet,Octet2,AxialFF,GA-Medium1}.

The determination of the parameters of the model by with the lattice QCD data
with large pion masses provides a clear estimate
of the pure valence quark degrees of freedom since
meson cloud effects are suppressed for large pion masses~\cite{Lattice,LatticeD,Omega,Octet,Octet2}.

The radial wave functions presented in Section~\ref{secRWF},
in terms of the mass of the baryon ($M_B$),
are determined in the lattice QCD  regime replacing the
baryon mass by the baryon mass in lattice.
As for the quark electromagnetic $j_q^\mu$
and axial $j_{Aq}^\mu$ currents defined 
in terms of vector meson poles, they are redefined
in lattice QCD in terms of the vector mass poles associated
with the lattice QCD regime (labeled by the pion mass of the simulation).
In the second term of the currents, we also
replace  the nucleon mass $M_N$ by the nucleon mass in lattice.
The coefficients associated with the vector meson dominance parametrizations are kept unchanged
in the lattice QCD regime.
The quark electromagnetic and axial currents
are discussed in Sections~\ref{secQEMcurr} and \ref{sec-Axial}, and in 
Appendices~\ref{appBareEMFF} and \ref{appBareAxial}.

With this procedure, we have a method that can
be used to calculate elastic and electromagnetic transition form factors,
which can be compared with numerical results form lattice QCD simulations.
We expect the comparison to be accurate for
simulations associated with large pion masses
since the effects of the meson cloud excitations are small.
The formalism has been tested successfully in the comparison
with nucleons, nucleon to $\Delta(1232)$ transitions,
and nucleon to $N(1440)$ transitions~\cite{Lattice,Roper1}.

More recently, the extension of the model to lattice QCD
has been used to determine the parameters associated
with the valence quark properties of the systems
using fits of the 
radial wave functions of the $\Delta(1232)$~\cite{LatticeD,Siegert1,Siegert2},  
of the octet baryons, and of the
decuplet baryons~\cite{Omega,Octet,Octet2,GA-Medium1}.

Our study of the octet baryon electromagnetic structure
was based on the lattice QCD results from Lin et~al.~\cite{Lin09}.
Reference~\cite{Lin09} includes lattice QCD data 
for $p$, $n$, $\Sigma^+$, $\Sigma^-$, $\Xi^0$, and $\Xi^-$
from the range $m_\pi=$ 350--700 MeV,
a range appropriated for the calibration of the model.

Our study of the octet baryon axial form factors
was based on the results from Ref.~\cite{Alexandrou11a}.
Reference~\cite{Alexandrou11a} includes a systematic study of the $G_A$
and $G_P$ nucleon form factors with several sets of pion masses
in the range of $m_\pi=370$--470 MeV
that can be used for an accurate calibration
of the unknown parameters in range $Q^2=0$--2 GeV$^2$.
There has been significant progress
in lattice QCD simulations for the octet baryon axial form factors
and octet baryons axial coupling constants using different methods and pion mass ranges~\cite{Alexandrou11a,Alexandrou13a,ARehim15a,Jang20a,Djukanovic22a,Bali23a}.
{Lattice QCD simulations near the physical limit are in agreement with the experimental value of the nucleon axial-vector coupling $G_A(0)$.
For finite $Q^2$, however, there are still discrepancies between lattice QCD simulations and experimental data~\cite{Meyer22a,Chang18a}.}
{
Combined studies of the $G_A$ experimental data and lattice simulations suggest that the octet baryon weak axial-vector form factors can be described by a dominant contribution associated with the valence quark and a component associated with the meson cloud dressing of the baryon cores~\cite{AxialFF,GAholography}.}

Once the parameters of the model are calibrated
by the lattice QCD data, one can calculate the valence
quark contributions to the electromagnetic and form factors in the physical limit
(using the physical masses) and combine
the bare contributions with the meson cloud
contributions to obtain the final result for the
electromagnetic and axial form factors~\cite{Lattice,LatticeD,Omega,Octet,Octet2,AxialFF}.

\subsection{Meson Cloud Contributions}

The physical  baryon $\left|B  \right>$ state
can, in general, be represented as a combination of a bare three-quark state and term associated with
the meson cloud excitations~\cite{Octet,Octet2,Octet4,AxialFF,GA-Medium1}
\ba
\left|B  \right> = \sqrt{Z_B} \left[
\left| qqq  \right> + c_B \left| {\rm MC}  \right>
\right] .
\label{eqDressedWF}
\ea
In this representation, $\left| qqq  \right>$ takes into account
the pure valence quark contributions discussed in
the previous sections, 
and $c_B \left| {\rm MC}\right>$ represents the baryon--meson
state associated with the meson cloud dressing.
The coefficient $c_B$ is determined by the normalization 
$Z_B (1+ c_B^2)=1$,
assuming that  $\left| {\rm MC}\right>$ is  normalized 
to unity.

The factor $Z_B= \sqrt{Z_B} \sqrt{Z_B}$ provides the probability of 
finding the $qqq$ state in the physical baryon state.
The probability of being associated with the meson cloud
component is then $1- Z_B$, in general a small
fraction of the probability associated with the three-quark state.
The meson cloud terms $\left| {\rm MC}  \right>$
are associated with baryon--meson states
like $\left| \pi N \right>$,  $\left| \sigma N \right>$,
$\left| \rho N \right>$, $\left| \pi \Delta \right>$, etc.
The corrections associated with baryon--meson--meson
states are usually very small~\cite{Octet2,GA-Medium1}.

We recall that  the separation between
valence quark and meson cloud effects
is intrinsically model-dependent.
In our applications, we try to reduce the model
dependence using the comparison with lattice QCD simulations
with large pion masses (reduce meson cloud effects)
to determine parameters of the model
associated with the valence quark physics~\cite{Octet,Octet2,AxialFF}.

\subsubsection{Electromagnetic Transitions \label{secEM-pioncloud}}

In the first studies of the octet baryon electromagnetic structure~\cite{Octet,Octet2,Octet4}, we look for the more relevant meson cloud contributions to the nucleon system.
According to the chiral perturbation theory, the pion, the lightest meson,
has the largest  contribution to the meson cloud~\cite{Jenkins93,Meissner97,Kubis99,Lu01b,Miller02,Franklin02,Cheedket04,Leinweber05,Boinepalli06,Wang09}.
It is possible, however,  in some transitions between baryon states,
that the pion cloud contributions are small, and that the
next meson contribution, the kaon cloud,  became important.
It has been shown that the kaon cloud contributions
are relevant, in particular, for the octet baryon to decuplet baryon
transitions~\cite{OctetDecuplet2,OctetDecupletD1,OctetDecupletD2}.
In  the first approximation, we consider for simplicity
that the pion has the more relevant contribution to
the meson cloud and to the nucleon and to the octet baryon
and restrict the calculation to the pion cloud.

The electromagnetic interaction with the valence quarks
was discussed in Section~\ref{secBare}.
The processes associated with the pion cloud contributions
can be decomposed into the two processes
displayed in Figure~\ref{fig-Meson-Cloud}:
the direct photon  interaction with the pion
and interaction with the baryon when
the pion is ``on the air''.
The transition current can then be expressed in the form 
\ba
J_B^\mu = Z_B \left[ J_{0B}^\mu + J_\pi^\mu + J_{\gamma B}^\mu \right],
\label{eqJtotal}
\ea
where $J_{0B}^\mu$ stands for the direct interaction
with the quark core (\ref{eqEM-bare}) and can be written as 
\ba
J_{0 B}^\mu = \tilde e_{0B} \gamma^\mu +  \tilde \kappa_{0B} \frac{i \sigma^{\mu \nu} q_\nu }{2 M_B},
\label{eqEM-bare2}
\ea
$J_\pi^\mu$ describes the coupling with the pion, and $J_{\gamma B}^\mu$
the coupling with the intermediate baryon states.
The factor $Z_B$ is a normalization constant defined by Equation~(\ref{eqDressedWF}).

For the photon interaction with the pion--baryon states, we consider~\cite{Octet4,Octet2}
\ba
& &
J_\pi^\mu = \left(b_1 \gamma^\mu + b_2 \frac{i \sigma^{\mu \nu} q_\nu }{2 M_B}   \right) G_{\pi B},
\label{eqJpi} \\
& &
J_{\gamma B}^\mu = \left(c_1 \gamma^\mu + c_2 \frac{i \sigma^{\mu \nu} q_\nu }{2 M_B}   \right) G_{e B}+
\left(d_1 \gamma^\mu + d_2 \frac{i \sigma^{\mu \nu} q_\nu }{2 M_B}   \right) G_{\kappa B},
\label{eqJgammaB}
\ea 
where $b_1$, $b_2$, $c_1$, $c_2$, $d_1$, and $d_2$
are phenomenological functions of $Q^2$,
and $G_{\pi B}$, $G_{e B}$, and $G_{\kappa B}$
are the operators acting on the SU(3) baryon--meson space.
$G_{\pi B}$ take into account the photon--pion interaction,
while $G_{e B}$ and $G_{\kappa B}$ describe the Dirac and Pauli
couplings, respectively, with the intermediate baryon.

The currents (\ref{eqEM-bare2})--(\ref{eqJgammaB})
are constrained by the octet baryon charges.
As a consequence, $b_1(0)= c_1(0)$ and $d_1(0)= 0$.
See Refs.~\cite{Octet2,Octet,Octet4} for a discussion on the subject.

The operators $G_{\pi B}$, $G_{e B}$, and $G_{\kappa B}$
are defined by SU(3) operators~\cite{Octet4}.
In an SU(3) model, the product of the functions $b_i$, $c_i$, and $d_i$
with the operators $G_{\pi B}$, $G_{e B}$, and $G_{\kappa B}$
can be represented as a product of the ratios
$\beta_B = g_{\pi B B}^2/g_{\pi NN}^2$ and  
$Q^2$-dependent functions associated
with the baryon--meson loop integrals.
In the calculations, we use factors $\beta_N$, 
$\beta_\Lambda$, $\beta_\Sigma$, and $\beta_\Xi$ ($\beta_N=1$ by construction)~\cite{Octet,Octet2}.

The falloffs of the functions $b_1$, $b_2$, $c_1$, $c_2$, $d_1$, and $d_2$
are adjusted to a form consistent with the result expected from pQCD
and valence quark sum rules, leading to suppression of $1/Q^4$ in comparison
with the leading order contribution~\cite{Octet2}.
In addition, the $Q^2$ dependence of the functions $b_1$ and $b_2$
takes into account chiral constraints on the Dirac and Pauli
square radii near the chiral limit~\cite{Bernard95a,Perdrisat07a}.

The numerical values of  couplings  $\beta_N$, 
$\beta_\Lambda$, $\beta_\Sigma$, and $\beta_\Xi$
are calculated using an SU(3) baryon--meson symmetry model.
Combining the different terms of the current (\ref{eqJtotal}), we \mbox{can write}
\ba
& &
F_{1B}  (Q^2)=
Z_B \left[ \tilde e_{0B} + a_{1} b_1(Q^2)
  + a_{2} c_1 (Q^2) + a_{3} d_1  (Q^2)\right],
\label{eqF1B}
\\
& &
F_{2B} (Q^2)=
Z_B \left[ \tilde \kappa_{0B} + a_{1} b_2 (Q^2)
  + a_{2} c_2 (Q^2) + a_{3} d_2 (Q^2) \right],
\label{eqF2B}
\ea
where the bare contributions are defined by Equations~(\ref{eqF10}), (\ref{eqF20}), and (\ref{eqek-tilde}),
and the coefficients $a_j$ ($j=1,2,3$)
are presented in Appendix~\ref{appMesonCloud1}.
The argument $Q^2$ is omitted in the coefficients $a_j$ for simplicity.

Notice that the functions $a_j$ include combinations
of the bare electromagnetic form factors $\tilde e_{0B} $, $\tilde \kappa_{0B}$.
Notice also that the coefficients $a_j$ are the same
for $F_{1B} $ and  $F_{2B}$.

The values of the constants $\beta_B$ and the explicit parametrizations of the functions $b_i$, $c_i$ and $d_i$ ($i=1,2$) are presented in Appendix~\ref{appMesonCloud1}.

The normalization constants $Z_B$ are determined by the constants $\beta_B$
and the by value of $b_1(0)$,
associated with the octet baryon self-energies~\cite{Octet4}:
\ba
& &
Z_N= \frac{1}{1+ 3 \beta_N b_1(0)}, \hspace{1cm}
Z_\Lambda= \frac{1}{1+ 3 \beta_\Lambda b_1(0)}, \nonumber \\
& &
Z_\Sigma= \frac{1}{1+ (\beta_\Lambda + 2 \beta_\Sigma) b_1(0)}, \hspace{1cm}
Z_\Xi= \frac{1}{1+ 3 \beta_\Xi b_1(0)}.
\label{eqZBs}
\ea

The parameters associated with the bare contributions
(radial wave function momentum range parameters $\beta_l$, $l=1,\ldots,4$)
are determined by the fit to the lattice QCD
form factor data from Ref.~\cite{Lin09}.
The remaining parameters of the model associated with the pion/meson cloud 
for the free space are then adjusted using
\begin{itemize}
\item
  the proton electromagnetic form factor data~\cite{JLab00a,JLab02a,JLab10a,Zhan11,Arrington07a} and 
  the neutron  electromagnetic form factor data~\cite{Ostrick99,Herberg99,Glazier05,Passchier99,Eden94,Zhu01,Warren03,Madey03,Riordan10,Schiavilla01,Bosted95,Kubon02,Anklin98,Anklin94,Lachniet09},
  including the proton and neutron magnetic moments;   
\item
  the octet baryon magnetic moments of the $\Lambda$, $\Sigma^+$, $\Sigma^-$, $\Xi^-$,
  and $\Xi^0$~\cite{PDG2010};
\item
\textls[-15]{the nucleon square electric and magnetic radii, and 
the $\Sigma^+$  square electric radius~\mbox{\cite{PDG2010,Eschrich01}.}}
\end{itemize}

The numerical results for the octet baryon electromagnetic form factors
are presented in Section~\ref{secResults1EM}.

\subsubsection{Axial Transitions \label{sec-Axial-MC}}

The meson cloud contributions are also relevant for the axial transitions.
In the first studies of the axial transitions, we considered
a simplified form for the meson cloud contributions,
associated with the axial coupling with the intermediate baryons
(in the elastic electromagnetic
  transitions, we need to take into account the two diagrams from
  Figure~\ref{fig-Meson-Cloud} in order to reproduce
  the charge of the baryon within a baryon--meson system.
  For the charged current axial transitions, we need to take into account that the diagrams include two baryon--meson couplings proportional to the factor $g_{\pi NN}^2$)
[diagram (b) from Figure~\ref{fig-Meson-Cloud}].

Similarly to the electromagnetic case, we represent 
the combination of the valence and meson cloud contributions
to the $B \to B'$ axial form factor in the form
\ba
G_A(Q^2) = \sqrt{Z_B Z_{B'}}
\left[ G_A^{\rm B} (Q^2) + G_A^{\rm MC} (Q^2)
  \right],
\ea
where $G_A^{\rm MC}$ is determined by
\ba
G_A^{\rm MC} (Q^2) = \eta_{BB'} G_{AN}^{\rm MC} (Q^2), \hspace{1.cm}
G_{AN}^{\rm MC} (Q^2) =  G_{AN}^{\rm MC0} \left( \frac{\Lambda^2}{\Lambda^2 + Q^2} \right)^4,
\label{eqGAN-MC}
\ea
where $G_{AN}^{\rm MC}$ is the parametrization
of the meson cloud contribution to the nucleon axial form factor
($n \to p$ transition) and $\eta_{BB'}$ is an SU(3) coefficient
associated with the transition.
The function $G_{AN}^{\rm MC}$ is defined by the relative
contribution to the axial form factor
($G_A(0) = Z_N \left[G_A^{\rm B} (0)  + G_{AN}^{\rm MC0}  \right]$),
and the cutoff is $\Lambda = 1.05$ GeV.
The falloff of $G_{AN}^{\rm MC}$ is compatible
with the expected falloff of a baryon system
with 5 constituents~\cite{AxialFF}.
A similar value for $\Lambda$ can be used to
parametrize the nucleon axial form factor
$G_{AN}^{\rm exp} (Q^2)= 1.2723 \left( 1 + \frac{Q^2}{\Lambda^2} \right)^{-2}$.
The details of the parametrization are presented in
Appendix~\ref{appMesonCloud2}.

For large $Q^2$, we can write~\cite{AxialFF}
\ba
G_A (Q^2) \simeq \sqrt{ Z_B Z_{B'}} G_A^{\rm B}(Q^2).
\label{eqGAasymp}
\ea
This relation can be used to estimate the constants $\sqrt{Z_B}$
and the impact of the meson cloud dressing in the axial form factors.

Also, the induced pseudoscalar form factor has
contributions associated with the meson cloud
\ba
G_P (Q^2) =  \sqrt{Z_B Z_{B'}}
\left[ G_P^{\rm B} (Q^2) + G_P^{\rm pole} (Q^2) + 
   G_P^{\rm MC} (Q^2)
  \right],
\ea
where 
\ba
G_P^{\rm MC} (Q^2) = \frac{4 M_{BB'}^2}{\mu^2 + Q^2} G_A^{\rm MC} (Q^2).
\ea
Notice that meson cloud contribution for $G_P$ has no adjustable parameters
because it is determined directly from $G_A^{\rm MC}$.

The parametrizations for $G_A^{\rm MC}$ and $G_P^{\rm MC}$
can be regarded as the global contribution of the meson cloud,
not just the pion since they are estimated by empirical data
(all meson cloud effects included).

The parametrizations of the meson cloud $G_{AN}^{\rm MC}$ are determined 
from the study of the lattice QCD data for the nucleon,
as discussed in Section~\ref{sec-Axial}, 
and from the empirical data for the
nucleon axial form factors and the axial-vector coupling constant.
The parameters involved are $Z_N$ [related to $b_1(0)$] and two coefficients included in $\eta_{BB'}$
[see Appendix~\ref{appMesonCloud2}].
The value of $Z_N$ or $b_1(0)$ is determined using the relation (\ref{eqGAasymp})
applied to the parametrization $G_{A N}^{\rm exp} (Q^2)$,
while the coefficients in $\eta_{BB'}$
are adjusted to the $\Lambda \to p$, 
$\Sigma^- \to n$, $\Xi^- \to \Lambda$, and $\Xi^- \to \Sigma^+$
experimental axial-vector coupling constants~\cite{PDG2010}.

The model calculations for $G_A$ are in agreement
with the experimental data for the $n \to p$
axial form factor~\cite{AxialFF,Bernard02,Park12,PDG2022}. 
Also, the calculations for $G_P$ are in agreement with
the data obtained at very low $Q^2$ by muon capture and
from pion electroproduction~\cite{Bernard02,Choi93}.
Notice that, apart from the valence quark contributions
for $G_P$, estimated from lattice QCD data (small relative contributions),
the model calculations of $G_P$ are predictions.

\subsection{Extension of the Model to the Nuclear Medium  \label{secMedium1}}

The formalism discussed in the previous sections
can be extended from the free space to nuclear matter.
We consider here the simplest case, the symmetric nuclear matter
associated with a medium with equal density of protons and neutrons.

In symmetric nuclear matter, the hadrons behave like free particles
with effective masses 
($m_h^\ast$ for mesons and $M_B^\ast$ for baryons) 
modified by the interaction with the nuclear medium and modified baryon--meson couplings.
These medium modifications are calculated using the QMC model,
where the interaction with the medium is described
using the self-consistent interaction
with Lorentz--scalar--isoscalar $\sigma$ and
Lorentz--vector--isovector $\omega$ fields.
The calculations of the baryon--meson coupling constants, discussed below, use the MIT bag model/QMC model bare axial couplings~\cite{Lu01a}, meaning that they take into account only the valence quark contributions.
The effective masses of the mesons and baryons are calculated
using the QMC model~\cite{QMCReview,Tsushima22a,CMartinez17a}.
The values for the densities $\rho=0$,  $0.5\rho_0$, and $\rho_0$
are presented in Table~\ref{table-Masses}.


\begin{table}[H]
  \caption{Meson and baryon masses are in GeV~\cite{Tsushima22a,CMartinez17a}.
  \label{table-Masses}}
\begin{tabularx}{\textwidth}{LCCCCCCCCC}
\toprule
\boldmath{$\rho/\rho_0$}  & \boldmath{$m_\pi^\ast$} &  \boldmath{$m_\rho^\ast$} & \boldmath{$m_\phi^\ast$} & \boldmath{$m_K^\ast$} & \boldmath{$M_N^\ast$} &  \boldmath{$M_\Lambda^\ast$} & \boldmath{$M_\Sigma^\ast$} & \boldmath{$M_\Xi^\ast$}\\
\midrule
0.0  &   0.138	& 0.7753 & 1.0195 & 0.4937 & 0.9390 &        1.1157 &	1.1931 &	1.3181 \\
0.5  &   0.138	& 0.7031 & 1.0080 &  0.4573 & 0.8311 &	1.0438	& 1.1213 &	1.2822	\\
1.0  &   0.138	& 0.6516  & 1.0010 & 0.4305 & 0.7544 &	0.9929	& 1.0707 &	1.2568 \\
\bottomrule
\end{tabularx}
\end{table}

The calculations of the coupling constants are based
on the Goldberger--Treiman relation~\cite{GTrelation,Octet2}:
\ba
\frac{g_{\pi BB}^\ast}{g_{\pi BB}} & = &
\left(\frac{f_\pi}{f_\pi^\ast}\right)
\left(\frac{g_A^{B \ast}}{g_A^B} \right)
\left(\frac{M_B^\ast}{M_B} \right),  \nonumber \\
& = &
\left(\frac{f_\pi}{f_\pi^\ast}\right)
\left(\frac{g_A^{N \ast}}{g_N^N} \right)
\left(\frac{M_B^\ast}{M_B} \right),
\label{eqGT-relation}
\ea
where $g_A^{B \ast}$ represent the baryon $B$
bare axial coupling [bare contribution to $G_A^{B}(0)$
  in medium],
and $f_\pi^\ast$ is the in-medium pion decay constant
(units of energy).
In the second equation, we approximate
$g_A^{B \ast}/g_A^B \simeq g_A^{N \ast}/g_A^N$.
In Equation~(\ref{eqGT-relation}), we consider
the diagonal case ($g_{\pi BB}^\ast$), but
the case $g_{\pi \Lambda \Sigma}^\ast$ can be calculated with minor
modifications~\cite{Octet2}.

The constant $f_\pi^\ast$ depends on $\rho$, and it is calculated using
a chiral perturbation theory expression derived
in Ref.~\cite{ChPT97}, assuming also that the pion mass
is almost unchanged in medium ($m_\pi^\ast \simeq m_\pi$),
where $m_\pi$ is the mass of the pion in free space.


The extension of the covariant spectator quark model to nuclear matter takes into account the medium modifications determined by the QMC model.
In the valence quark component of the model, we consider the medium modifications in the hadron masses in the quark current and the baryon mass modification ($M_B^\ast$) in the radial wave function $\psi_B$.
As for the meson cloud component, we consider the modifications on the baryon--meson coupling constants $g_{\pi BB'}^\ast$ as presented in Table~\ref{table-gpiBB1}.
In the table, we use $g_A \equiv g_A^N$ and $g_A^\ast \equiv g_A^{N \ast}$ for simplicity.

In the electromagnetic transitions,
we take into account the modification in $\beta_B^\ast$~\cite{Octet2}
\ba
& &
\beta_N^\ast = \left(\frac{g_{\pi NN}^\ast}{g_{\pi NN}} \right)^2 \beta_N,
\hspace{.5cm}
\beta_\Lambda^\ast = \left(\frac{g_{\pi \Lambda \Sigma}^\ast}{g_{\pi \Lambda \Sigma}} \right)^2 \beta_\Lambda, \nonumber \\
& &
\beta_\Sigma^\ast = \left(\frac{g_{\pi \Sigma \Sigma}^\ast}{g_{\pi \Sigma \Sigma}} \right)^2 \beta_\Sigma,
\hspace{.65cm}
\beta_\Xi^\ast = \left(\frac{g_{\pi \Xi \Xi}^\ast}{g_{\pi \Xi \Xi}} \right)^2 \beta_\Xi,
\ea
where the power 2 takes into account the double pion--baryon coupling of the processes from Figure~\ref{fig-Meson-Cloud}.

In the axial transitions, we take advantage of the fact that the different baryon--meson couplings can be represented as linear combinations of $g_{\pi NN}$ to write~\cite{GA-Medium1}
\ba
G_A^{\rm MC \ast} (Q^2)=   \left(\frac{g_{\pi NN}^\ast}{g_{\pi NN}} \right)^2 G_A^{\rm MC} (Q^2),
\label{eqGAMC2}
\ea
where the medium modifications on cutoff $\Lambda$ are neglected in the first approximation.
A consequence of the relation (\ref{eqGAMC2}) is that the meson cloud function $b_1$
is also modified in medium, and 
\ba
b_1^\ast (0) = \left(\frac{g_{\pi NN}^\ast}{g_{\pi NN}} \right)^2 b_1(0).
\ea

The medium variation in $b_1(0)$ implies that  the normalization factors of the wave functions due to the meson cloud are also modified in the form
\ba
Z_B^\ast = \frac{1}{1 + 3 a_B b_1^\ast(0)},
\ea
where the coefficients $a_B$ are determined by Equation~(\ref{eqZBs}).

\begin{table}[H]
\caption{
  Ratios $g_A^{N \ast}/g_A^N$,  $f_\pi^\ast/f_\pi$ and coupling constant ratios (no dimensions). We use $g_A \equiv g_A^N$ and $g_A^\ast \equiv g_A^{N \ast}$.
  The values for $g_A^\ast/g_A$ are from Ref.~\cite{Lu01a},
  and the values for $f_\pi^\ast/f_\pi$ are from Ref.~\cite{ChPT97}.
  \label{table-gpiBB1}}
  
\begin{tabularx}{\textwidth}{lCCcccc}
\toprule
\boldmath{$\rho/\rho_0$}  & \boldmath{$g_A^\ast/g_A$} &   \boldmath{$f_\pi^\ast/f_\pi$} & \boldmath{$g_{\pi NN}^\ast/g_{\pi NN}$}
& \boldmath{$g_{\pi \Lambda \Sigma}^\ast/g_{\pi \Lambda \Sigma}$}
& \boldmath{$g_{\pi \Sigma \Sigma}^\ast/g_{\pi \Sigma \Sigma}$}
& \boldmath{$g_{\pi \Xi \Xi}^\ast/g_{\pi \Xi \Xi}$} \\
\midrule
0.0 &  1.0000 &	\sp 1.0000 & 1.0000     & 1.0000 & 1.0000 & 1.0000\\
0.5 & 0.9404 &	\sp 0.9180 & 0.9067	& 0.9584 & 0.9628 & 0.9965 \\
1.0 & 0.8920 &	\sp 0.8279 & 0.8656	& 0.9588 & 0.9669 & 1.0273\\
\bottomrule
\end{tabularx}
\end{table}

The values of $n_P$ are calculated in Refs.~\cite{AxialFF,GA-Medium1}
\ba
& & \hspace{2.35cm}
n_P= -0.507 \; \mbox{for} \; \rho=0, \nonumber  \\
& &
n_P= -0.536 \; \mbox{for} \; \rho=0.5 \rho_0, \hspace{.5cm} 
n_P= -0.560 \; \mbox{for} \; \rho=\rho_0.
\ea
For each density, the value of $n_P$ is determined
imposing that $G_A^{{\rm B} \ast}(0) = g_A^{N \ast}$ for consistency,
with the QMC model used in the calculation of in-medium
masses and coupling constants~\cite{GA-Medium1}.


\subsection{Summary of the Formalism}

{
In the previous sections, we reviewed the formalism
associated with the calculation of the electromagnetic and axial form factors
of octet baryons in free space and in nuclear matter.
The contributions to the form factors are decomposed into valence quarks and meson cloud contributions.}
For the calculations in nuclear matter, we combine the covariant quark model and the QMC model to take into account the medium modifications associated with the hadron masses and baryon--meson coupling constants and calculate the electromagnetic and axial form factors in terms of the medium density.

{
The formalism discussed is mainly based on Refs.~\cite{GA-Medium1,Octet2,Octet3,AxialFF,Octet}.
The electromagnetic and axial form factors of the octet baryons are first studied in free space in Refs.~\cite{AxialFF,Octet}.
The study of the electromagnetic form factors is then extended to nuclear matter in Refs.~\cite{Octet2,Octet3}.
The in-medium axial form factors are discussed in Ref.~\cite{GA-Medium1}.
The present work is the first time that the electromagnetic and axial form factors of the octet baryons are discussed in a combined form.}

{
The covariant spectator quark model was originally developed for the study of the
nucleon electromagnetic form factors and nucleon parton distribution functions
in deep inelastic scattering~\cite{Nucleon,Nucleon2,DIS} and has since been extended
to the study of the electromagnetic structure of baryons
based on the SU(3) flavor-symmetry~\cite{Octet4,Omega}, 
as well as a significant number of nucleon resonances~\cite{Spectator-Review,Ramalho-Pena23}
in different spacelike and timelike regions.
Details of the formalism and references to different applications of models can be found in Sections~\ref{secIntro} and \ref{secModel}.}

{
The motivation to the extension of the covariant spectator quark model 
formalism to the nuclear medium has been the development of tools that can be used in the study
of electromagnetic and weak interactions of baryons
in dense nuclear matter based on the observable degrees of freedom in free space.}

{
In the next section, we present and discuss numerical calculations of the electromagnetic and axial form factors in free space and in nuclear medium for the octet baryons and transitions between octet baryon states.}

In Section~\ref{secResults2}, we discuss the extension of the formalism for densities above the normal nuclear matter and explain how the baryon form factors can be used to calculate neutrino/antineutrino--baryon cross-sections in terms of the neutrino energies and the square transfer momentum.

{
The discussion about the advantages and limitations of the formalism,
the contribution of the present work to the field, 
and future developments will be presented in Section~\ref{secDiscussion}.}

\section{Electroweak Form Factors in Nuclear Medium \label{secResults1}}

The formalism discussed in the previous section
has been used in the calculation of the octet baryon electromagnetic form
factors in the nuclear medium~\cite{Octet2}
and the weak axial transition form factors between octet baryon members in the nuclear medium~\cite{GA-Medium1}.
In the calculations, we used the densities $\rho= 0.5 \rho_0$ and $\rho=\rho_0$ in comparison with calculations in free space ($\rho=0$).

In the study of the electromagnetic form factors, we consider the simplest approximation, when the octet baryon structure is represented by a quark--diquark $S$-state~\cite{Octet2,Octet} [based on Equation~(\ref{eq-PsiS})].
In the study of the axial form factors, we consider a combination of $S$- and $P$-states since it was demonstrated that the inclusion of a quark--diquark $P$-state is necessary for an accurate description of the nucleon axial transition form factors~\cite{AxialFF,GA-Medium1}.

In this section, we omit the discussion of the results
for the neutral baryons, except for the neutron,
because in these cases we cannot expect very accurate estimates
for the electromagnetic and axial form factors.
Recall that the electric form factors of neutral baryons are proportional to $Q^2$ near $Q^2=0$.
The limitations in the electromagnetic form factors are a consequence of the lack of lattice data for $\Lambda$ and $\Sigma^0$.
As for axial form factors, the $\Lambda$ and $\Sigma^0$
axial-vector form factors are dominated by the meson cloud contribution
(poorly constrained by the data) since the valence quark
contributions vanish~\cite{AxialFF,GA-Medium1}.

{
For the comparison with the in-medium experimental data, it is important to know the uncertainties associated with the model calculations in nuclear matter.
The precision of our calculations is determined by the precision of the QMC parametrizations of the hadron masses and baryon--meson couplings.
Based on the parametrizations of the hadron masses~\cite{Tsushima22a,CMartinez17a}, we can estimate the uncertainties of the model calculations as better than 2\%, assuming a similar precision for the axial-vector couplings.
The same relative precision should be expected for calculations of form factors at finite $Q^2$ due to the dominance of the linear terms for small variations in the input parameters.}


\subsection{Electromagnetic Form Factors \label{secResults1EM}}

We now review  the calculations of the electromagnetic form factors of the octet baryons from Ref.~\cite{Octet2}.
The electromagnetic form factors include the valence quark and the pion cloud contributions.
In general, the valence quarks provide the dominant contribution, with more than 80\% of the total, while the pion cloud contributions are at most 10\%.
In the nuclear medium, the valence quark contributions
are more modified than the pion cloud contributions~\cite{GA-Medium1}.
The exceptions to these conclusions are the results for
the electric form factors of neutral baryons
(neutron, $\Lambda$, $\Sigma^0$, and $\Xi^0$) that have
in general smaller magnitudes than the charged baryons.

We start by discussing the medium modifications on the form factors $G_E$ and $G_M$.
Since the in-medium form factors $G_E$ and $G_M$ cannot be directly measured at the moment~\cite{Octet3}, in a second stage, we discuss the medium effects on the ratio $G_E/G_M$.

\subsubsection{Electromagnetic Ratios $G_E^\ast/G_E$ and $G_M^\ast/G_M$ \label{secEM-ratios}}

As mentioned already, a simple way to study the impact of the medium effects on the form factors is the calculation of the ratios between the functions ($G_E$ and $G_M$) in medium and in vacuum for the same value of $Q^2$.

\begin{figure}[H]
  \hspace*{-2mm}
\begin{tabular}{cc}
\includegraphics[width=2.4in]{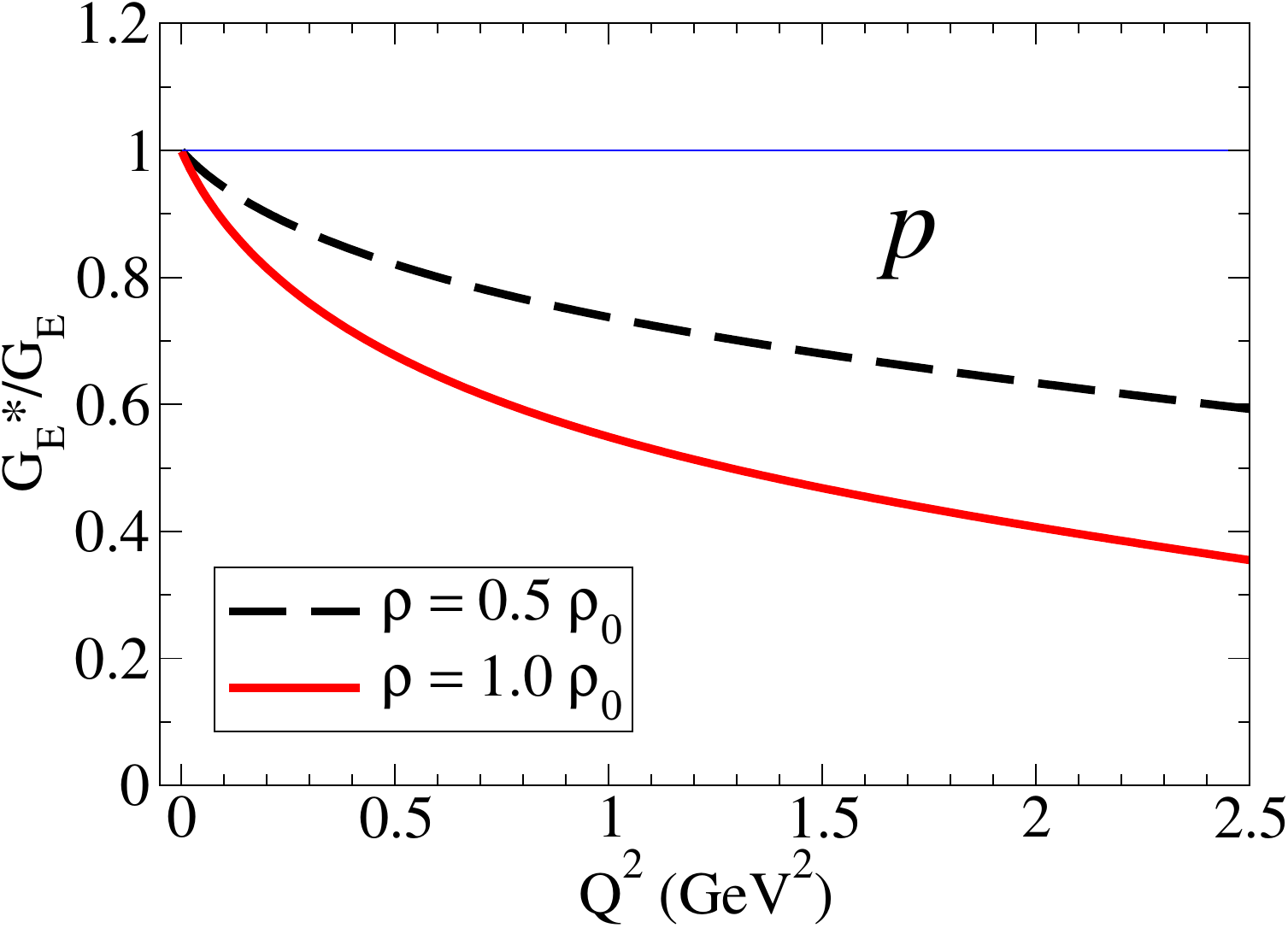} &
\includegraphics[width=2.4in]{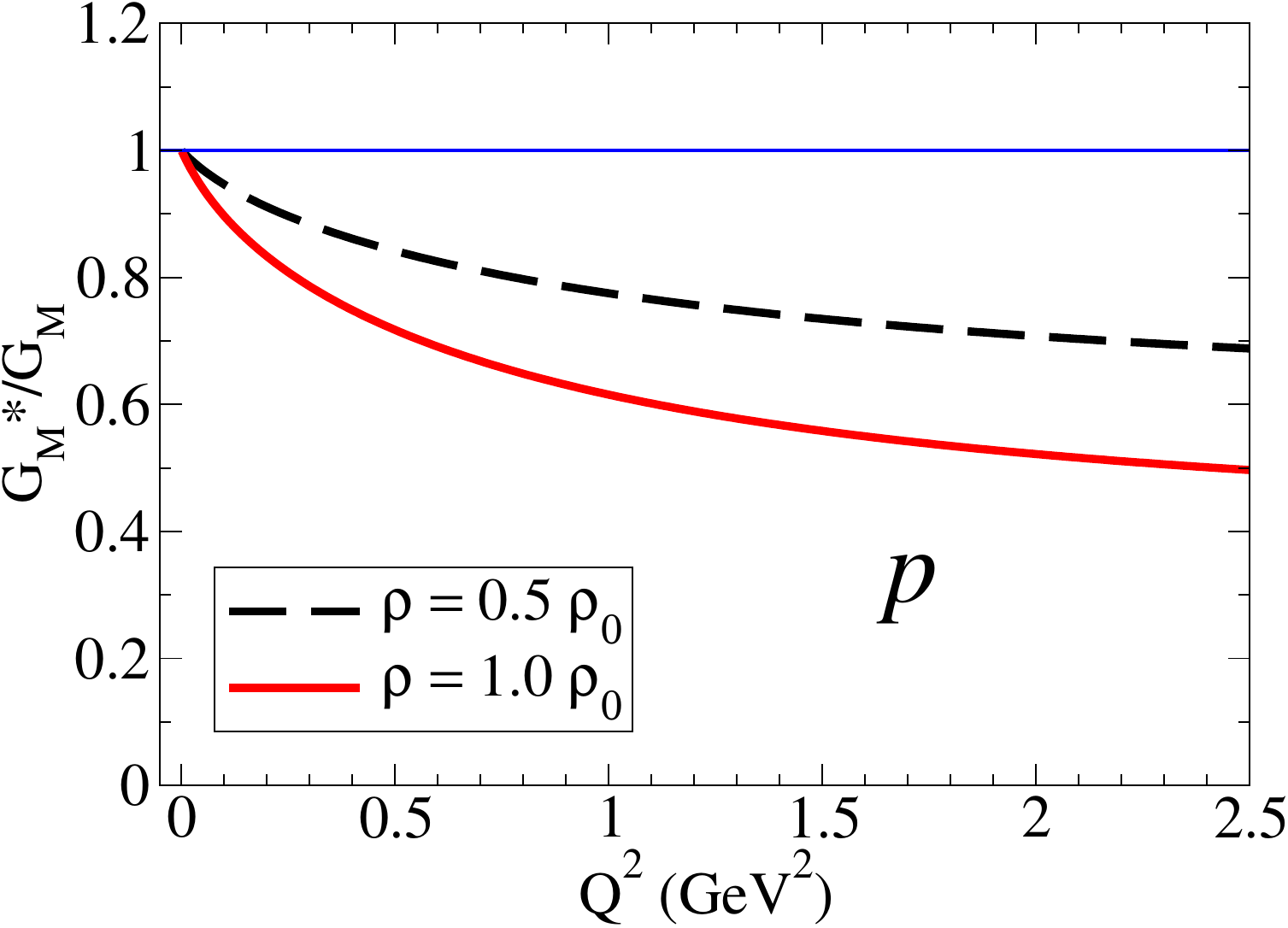}\\
\includegraphics[width=2.4in]{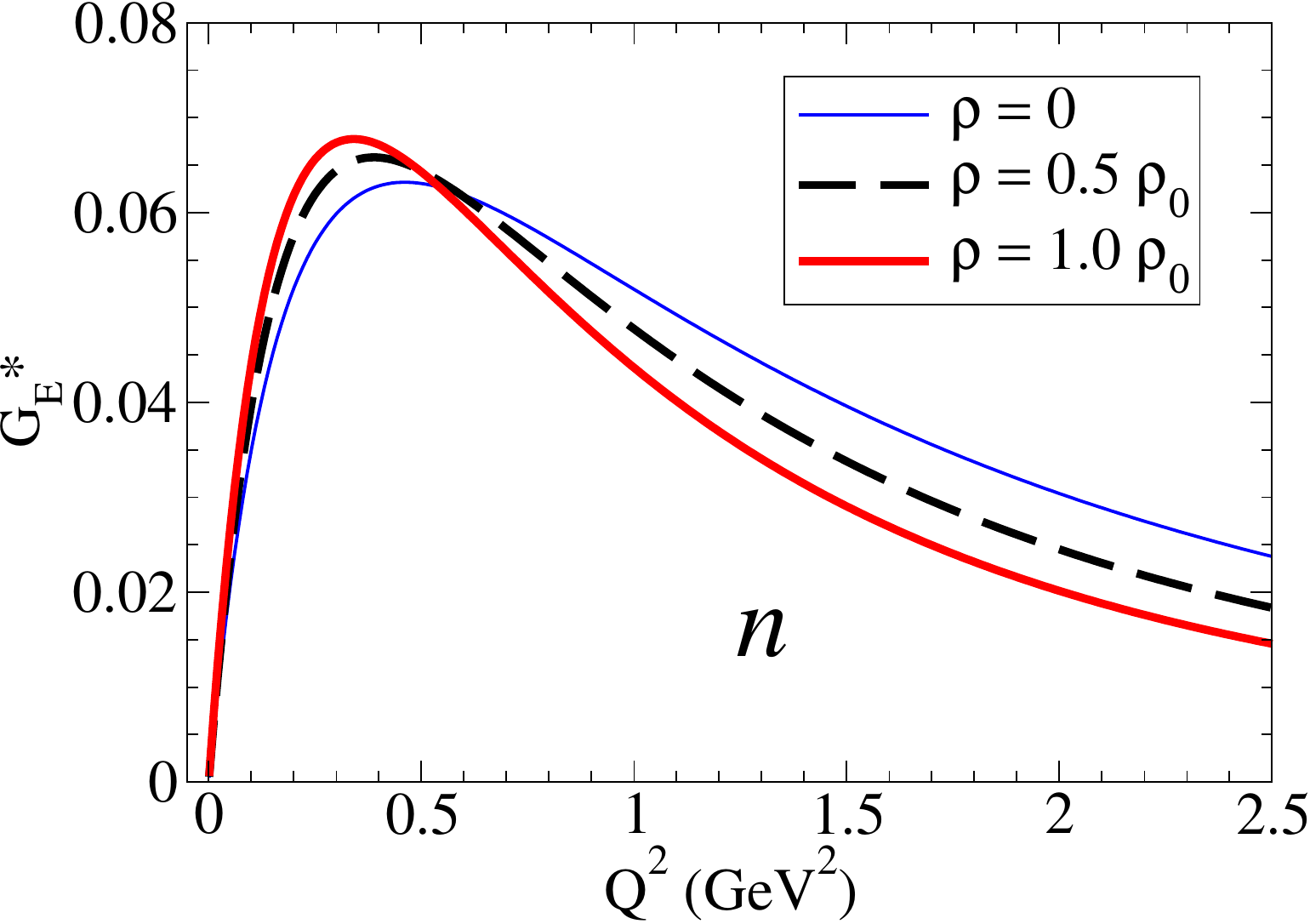} &
\includegraphics[width=2.4in]{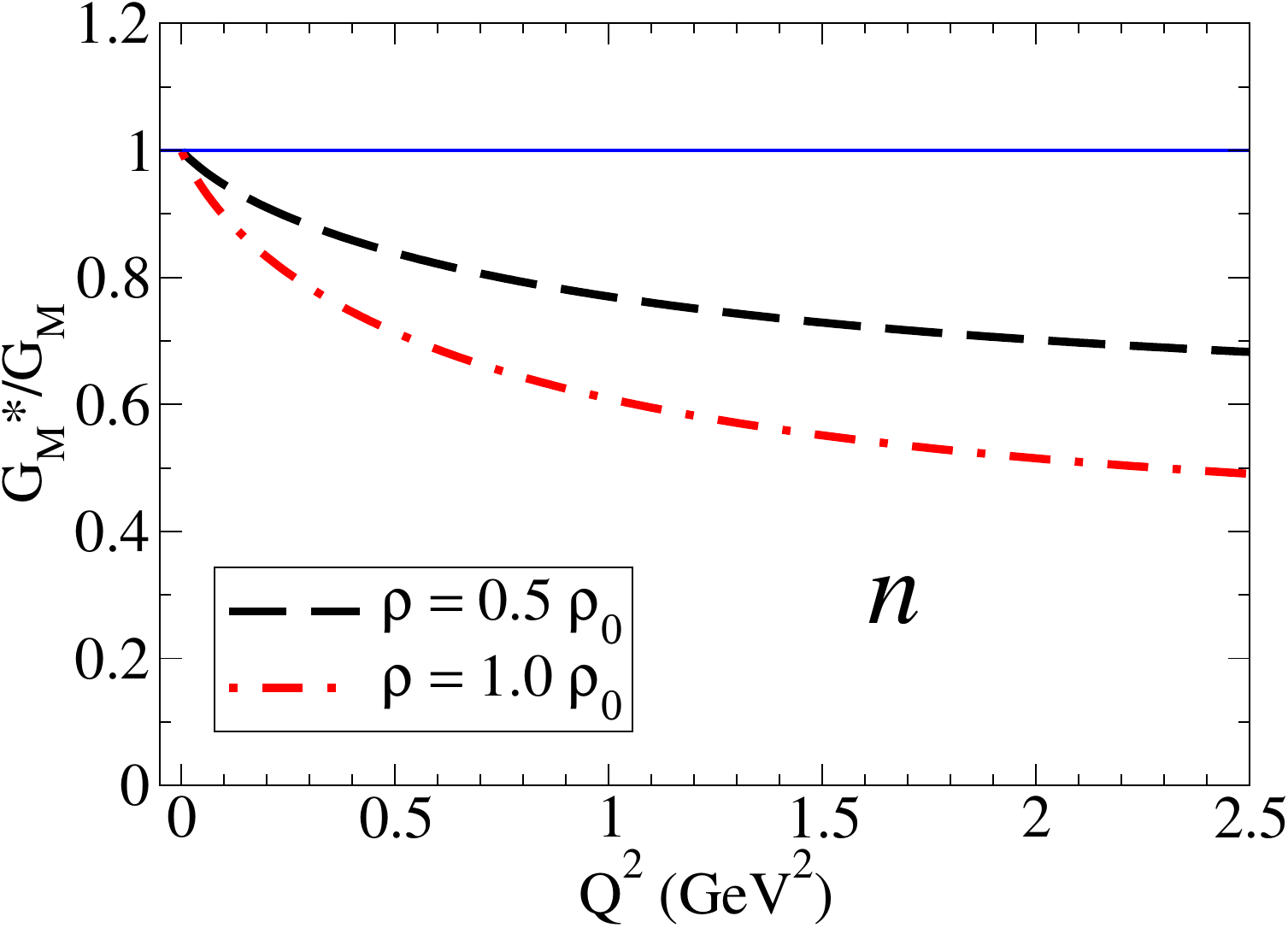}\\
\includegraphics[width=2.4in]{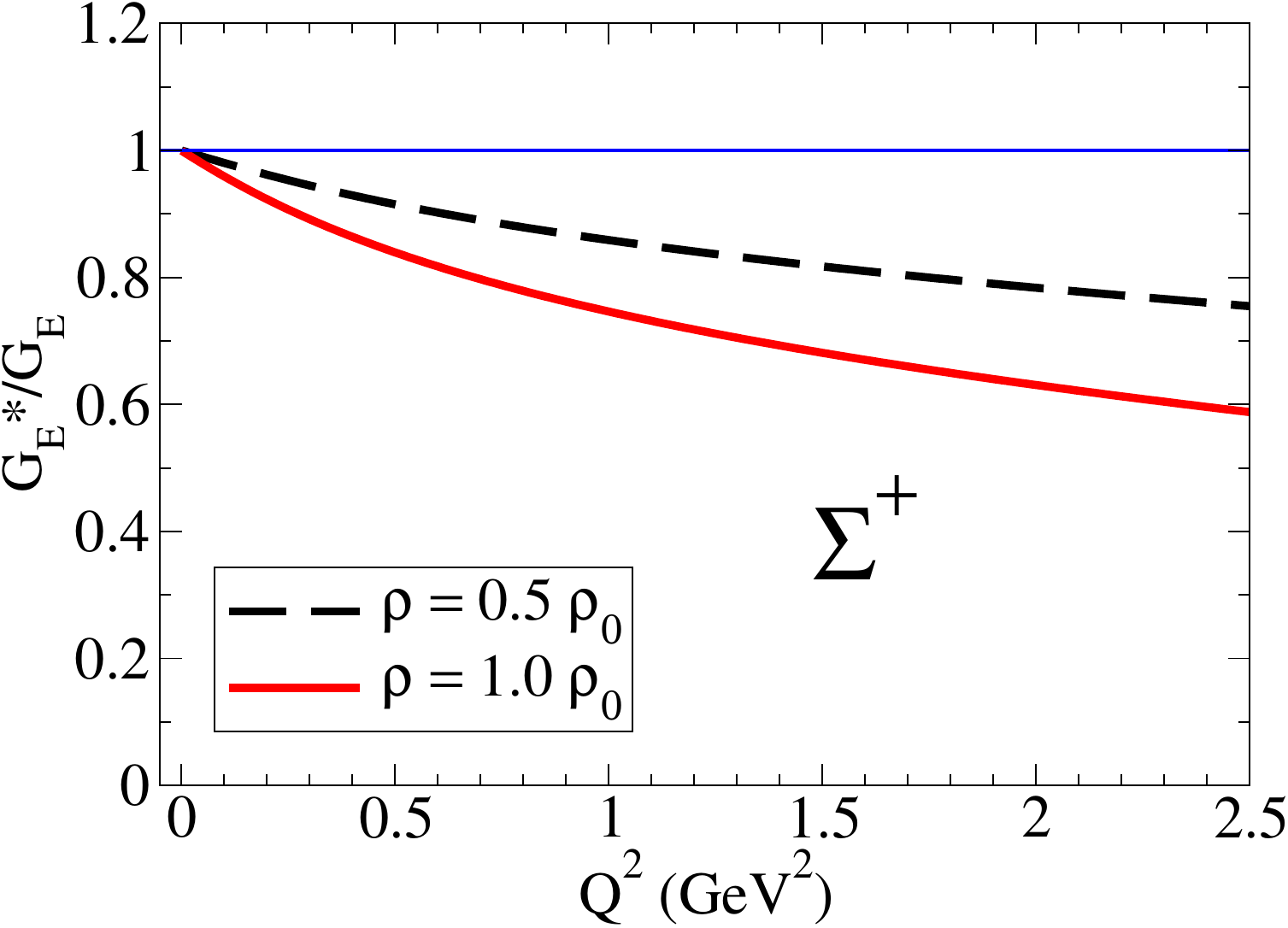}&
\includegraphics[width=2.4in]{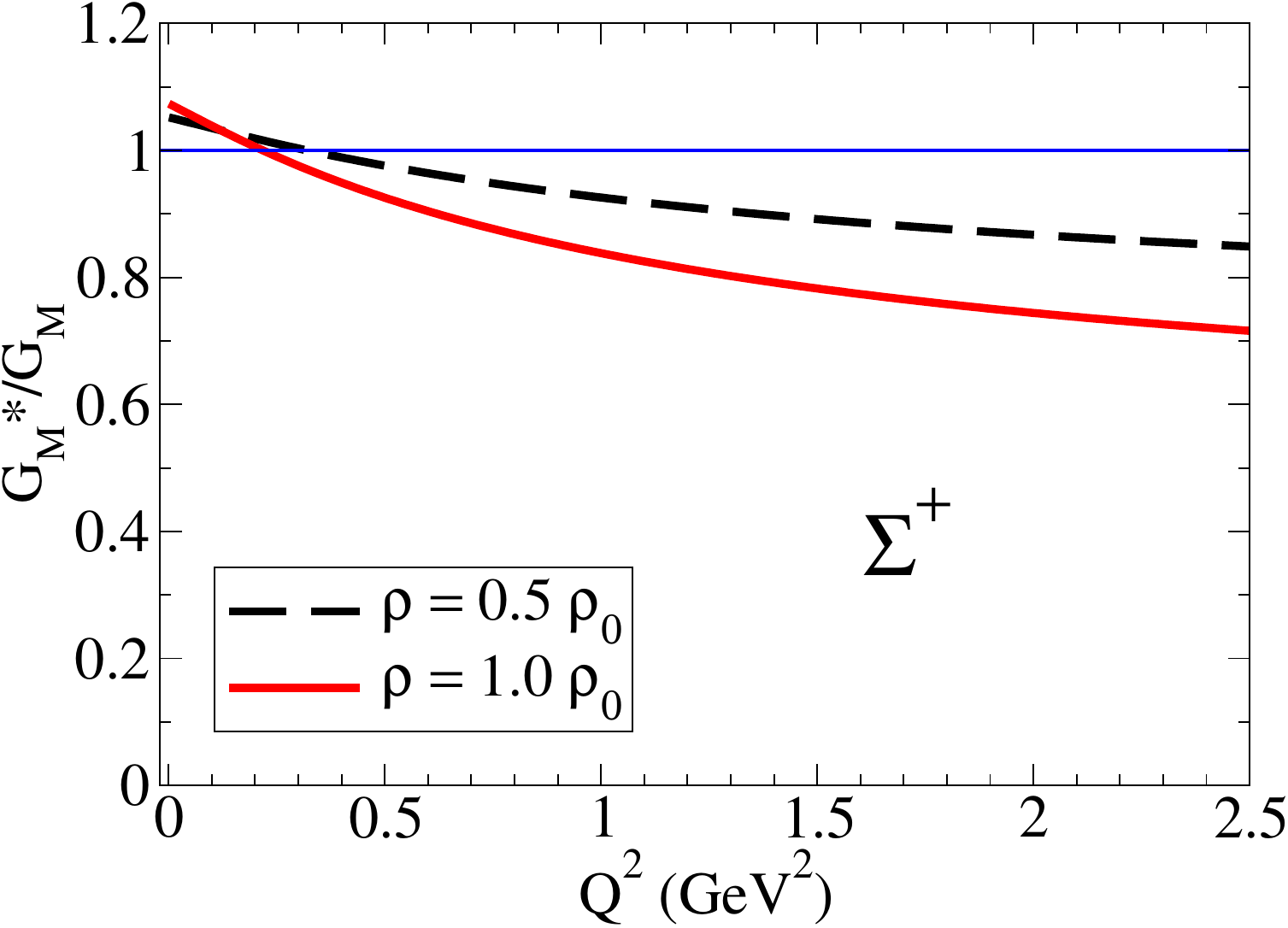}\\
\includegraphics[width=2.4in]{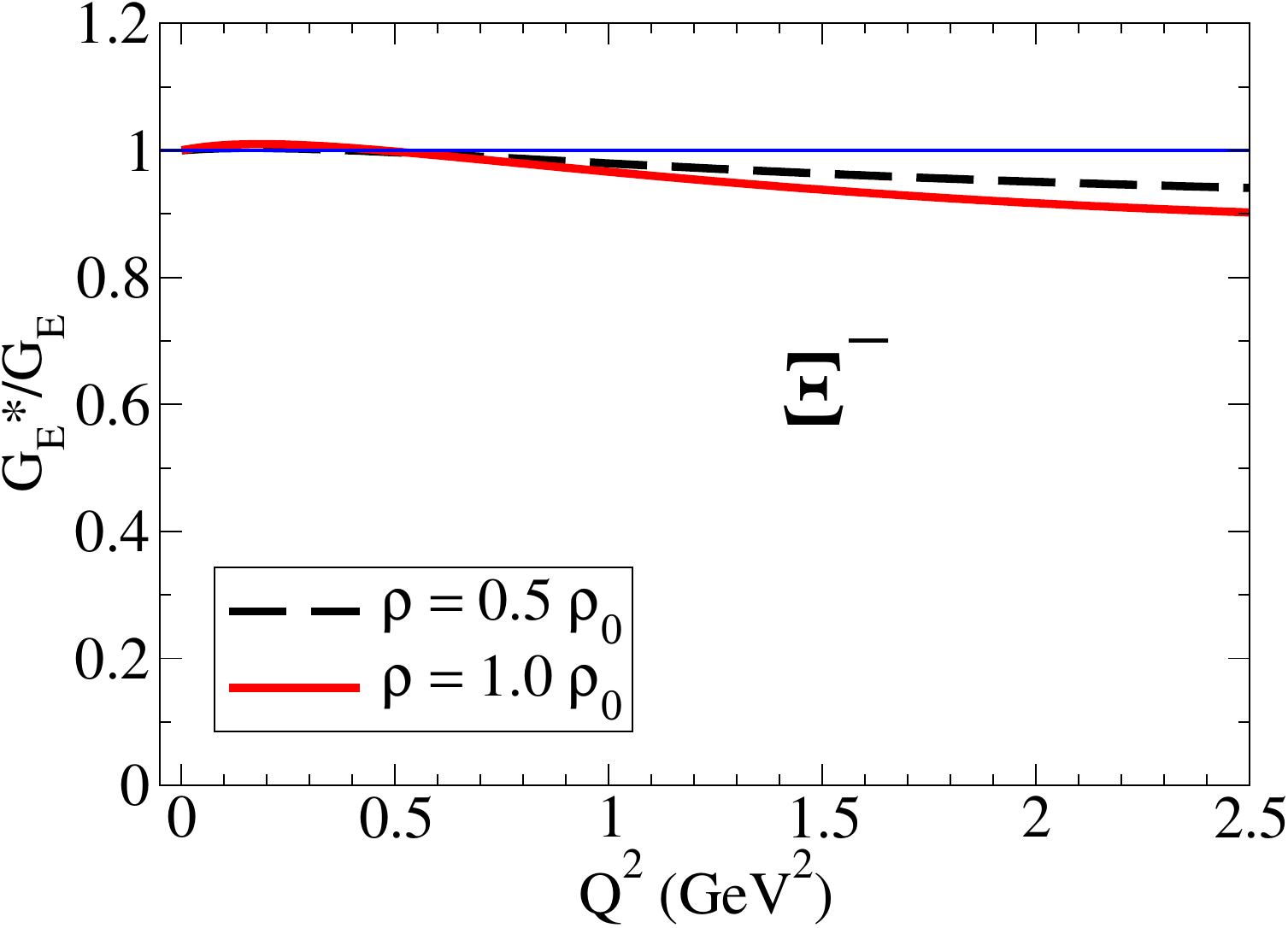} &
\includegraphics[width=2.4in]{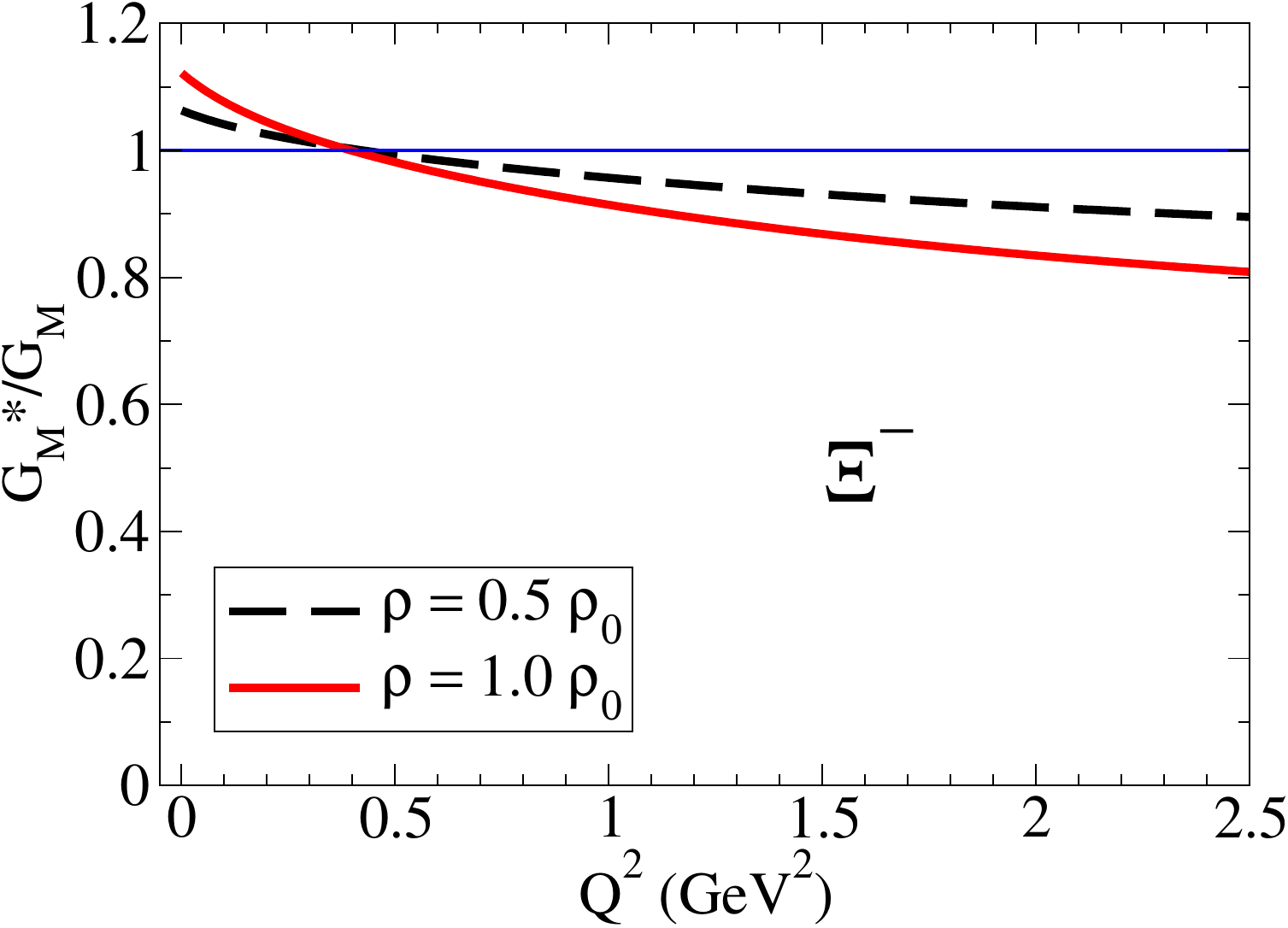}\\
\end{tabular}
  \caption{Electric and magnetic form factor ratios
      for the nucleon: proton, neutron,  $\Sigma^+$, and $\Xi^-$.
      The magnetic form factors are in natural units
      ($\frac{e}{2 M_B}$  and $\frac{e}{2 M_B^\ast}$).
      The horizontal line is included to represent the ratio in free space.
      In the case of the neutron, we display $G_E^\ast$
      instead of the ratio $G_E^*/G_E$ for a cleaner comparison.
\label{figNucleonR}}
\end{figure}

We then calculate  the ratios $G_E^\ast/G_E$ and $G_M^\ast/G_M$ for the baryon $B$: proton, neutron, $\Sigma^+$, and $\Xi^-$ for the densities $\rho=0.5 \rho_0$ and $\rho=\rho_0$.
Later on, we discuss the similarities with the remaining cases.
The numerical results are presented in Figure~\ref{figNucleonR} on the left side for $G_E$ and on the right side for $G_M$.
In the case of the neutron, due to the small magnitude and the particular dependence on $Q^2$,
it is more appropriate to compare the magnitudes of the form factors directly.
In the case of the $\Lambda$ and $\Sigma^0$ (neutral particles), we may observe the divergence of $G_E^\ast/G_E$ due to zero of the electric form factors in free space.
See Ref.~\cite{Octet3} for a more detailed discussion.

From the results for the ratio $G_E^\ast/G_E$, we can conclude that the form factors for  $p$, $\Sigma^+$, and $\Xi^-$ are suppressed in the nuclear medium, and also that the suppression increases with the density.
We can also conclude that suppression with $Q^2$ is more effective for systems with light quarks than for systems with one or two strange quarks.
In the case of the $\Xi^-$, the ratio is almost the unity.

From the results for $G_E^\ast$ for the neutron,
the main conclusions are that the magnitudes of $G_E$ are small
compared to the charged cases ($|G_{E}(0)|=1$), and the variations
due to the medium are small in magnitude.
One can also conclude that $G_E^\ast/G_E$ is reduced in medium below $Q^2=0.25$ GeV$^2$.
We can understand this result noticing that, 
at low $Q^2$, one has $G_{E} \simeq - \frac{1}{6} r_{En}^2 Q^2$.
As a consequence, $G_E^\ast/G_E = r_{En}^{\ast 2}/ r_{En}^2$, where $r_{En}^{\ast 2}$ and $r_{En}^2$ are the neutron square charge radius in medium and in vacuum, respectively.

The results for the ratio $G_M^\ast/G_M$, displayed on the right side of Figure~\ref{figNucleonR} show a trend similar to that observed for $G_E^\ast/G_E$.
Differently from Ref.~\cite{Octet2}, we represent here the magnetic form factors in natural units [see Equation~(\ref{eqGMdef})].
The ratios $G_M^\ast/G_M$ for $\Sigma^0$ and $\Sigma^-$ are similar to the ratio for $\Sigma^+$.
There are also similarities between the cases $\Lambda$ and $\Sigma^0$ and $\Xi^-$ and $\Xi^0$.
These similarities are a consequence of the approximated SU(3) symmetry in the expressions for the bare and meson cloud contributions.

The results from Figure~\ref{figNucleonR} for the $G_E$ for the proton, $\Sigma^+$, and $\Xi^+$ are interpreted as the enhancement of the absolute value of the charge radius in medium ($\sqrt{r_{EB}^2}$).
The results for the proton and neutron magnetic moments can also be interpreted as an enhancement of the magnetic radius in medium ($\sqrt{r_{MB}^2}$).
We follow here the interpretation of Ref.~\cite{Cloet09a}.
A more detailed discussion of the octet baryon electric and magnetic ratios can be found in Ref.~\cite{Octet2}.

The ratios $G_E^\ast/G_E$ and $G_M^\ast/G_M$ for the nucleon have been calculated
using Skyrme and quark--soliton models~\cite{Meissner87a,Meissner89a,Smith04,Christov95a,Yakhshiev03},
QMC models~\cite{Lu99a,QMCEMFFMedium2,QMCEMFFMedium3,QMCEMFFMedium4,Smith04}, 
Nambu--Jona--Lasinio, and light front quark models~\cite{Frank96,Cloet09a,deAraujo18,Christov96a}.
In general, one observes a reduction in both $G_E$ and $G_M$ in the nuclear medium at low $Q^2$.
As in our model calculations, the suppression is more significant for $G_E$ than for $G_M$, increases with the density, and with $Q^2$ up to $Q^2=1$ GeV$^2$.
See for instance Refs.~\cite{Lu99a,QMCEMFFMedium2,Yakhshiev03,Smith04,Frank96,Meissner89a}.
In the case of the neutron,  an enhancement of the electric form factor~\cite{Cloet09a,deAraujo18} is expected.
The magnitude and shape of the functions $G_E^\ast/G_E$ and $G_M^\ast/G_M$ depend on the specific nucleus and on the considered orbital states~\cite{Lu99a}.
In the case of bag and QMC models, the trend of the ratios changes after a certain value of $Q^2$ (0.5 or 1 GeV$^2$)~\cite{Lu99a,QMCEMFFMedium2}.    

It is worth mentioning that, when we  convert $G_M^\ast/G_M$ for the nucleon to units of nuclear magneton, the numerical results are modified by the factor $\frac{M_N}{M_N^\ast}$ (see Section~\ref{secEMFFdef}).
In that case, we expect to obtain, at $Q^2=0$,
a ratio larger than unity (enhancement of magnetic moment in medium).
This enhancement of the magnetic moment in medium [in units $e/(2M_N)$] has been reported using different frameworks dominated by the valence quark degrees of freedom~\cite{Cloet09a}.



\subsubsection{Electromagnetic Double Ratios of Octet Baryons \label{secEMFF-2ratios}}

Since the measurements at JLab of the ratio between the electric and magnetic form factors $G_E/G_M$ using the polarization transfer method for the proton~\cite{Perdrisat07a,JLab00a,JLab02a,JLab04a,JLab05a,JLab10a,JLab12a,JLab17a,Zhan11} and for the neutron~\cite{Madey03}, there is the perspective that the method may be extended to other baryons and to baryons bound to a nucleus.
The interaction with baryons bound to a nucleus can be regarded as the interaction with a baryon immersed in a nuclear medium (quasi-elastic reaction)~\cite{Octet3}.
At the moment, there are only complete measurements of the ratio $G_E^\ast/G_M^\ast$ for protons on $^4$He targets~\cite{Dieterich01,Strauch03a,Paolone10,Malace11a,JLab-data}.
There is the possibility that the experiments can be extended for $^2$H, $^8$O, and $^{12}$C targets~\cite{Malov00,MAMI2021,Kolar23a}.
Under study are extension of the experiments that measure the ratio $G_E^\ast/G_M^\ast$ for bound neutrons~\cite{PAC35}.

Experiments at JLab and MAMI measured the ratios  $G_E/G_M$ for bound protons ($G_E^\ast/G_M^\ast$) and for free protons ($G_E/G_M$).
The ratio between the medium and vacuum
\ba
{\cal R} (Q^2) = \frac{G_E^\ast/G_M^\ast}{G_E/G_M}
\label{eqDRbaryons}
\ea
can be used to measure the effect of the nuclear medium
on the ratio $G_E/G_M$.
If ${\cal R} \approx 1$, there are no medium effects.
If ${\cal R}< 1$, the ratio $G_E/G_M$ is reduced in  nuclear medium (quenching effect).
If ${\cal R} > 1$, the ratio $G_E/G_M$ is enhanced in nuclear medium.

The medium effects on the magnetic moments are usually discussed and measured in units of nuclear magneton, as discussed in Section~\ref{secEMFFdef}.
For that reason, in the following, we convert the double ratios ${\cal R}$ from Equation~(\ref{eqDRbaryons}) into units of nuclear magneton in vacuum $\frac{e}{2 M_N}$.
In the conversion of magnetic form factors from natural units to nuclear magneton, we multiply by the factor $M_N/M_B$ in free space and by $M_N/M_B^\ast$ in the medium.
The overall conversion for the double ratio is then
$(M_B^\ast/M_N)/(M_B/M_N)= M_B^\ast/M_B$.
We  then use the factor $C_B= M_B^\ast/M_B$ in the conversion of  ${\cal R}$ to nuclear magneton.

\begin{figure}[H]
  \hspace*{-3mm}
\begin{tabular}{cc}
\includegraphics[width=2.4in]{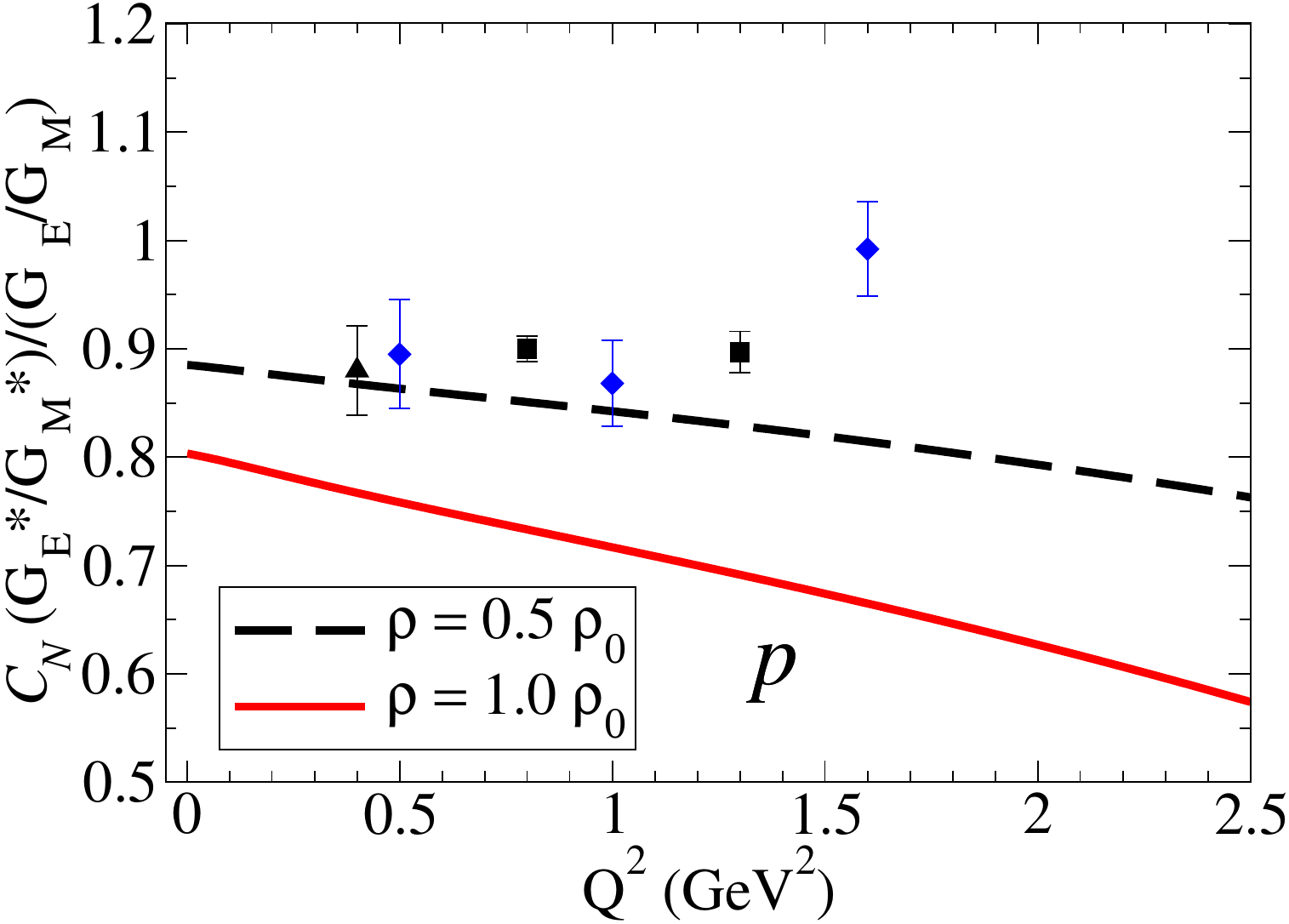} &
\includegraphics[width=2.4in]{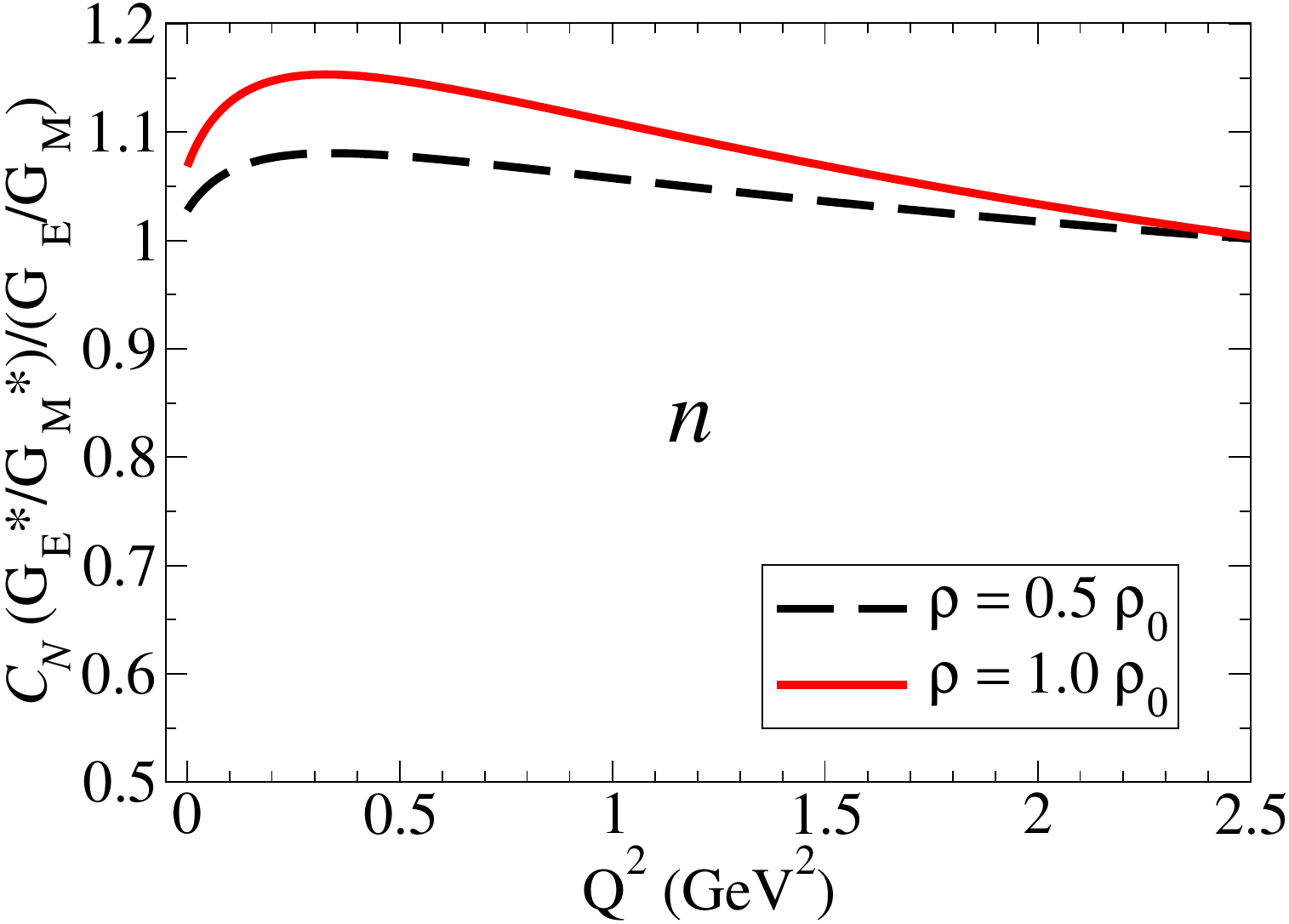}\\
\end{tabular}
\caption{Proton and neutron double ratios in units of nuclear magneton. $C_N= \frac{M_N^\ast}{M_N}$.
  The proton data are from MAMI~\cite{Dieterich01} (black) and JLab~\cite{Strauch03a} (blue).
\label{figNucleonDR}}
\end{figure}

\begin{figure}[H]
  \hspace*{-3mm}
\begin{tabular}{cc}
\includegraphics[width=2.4in]{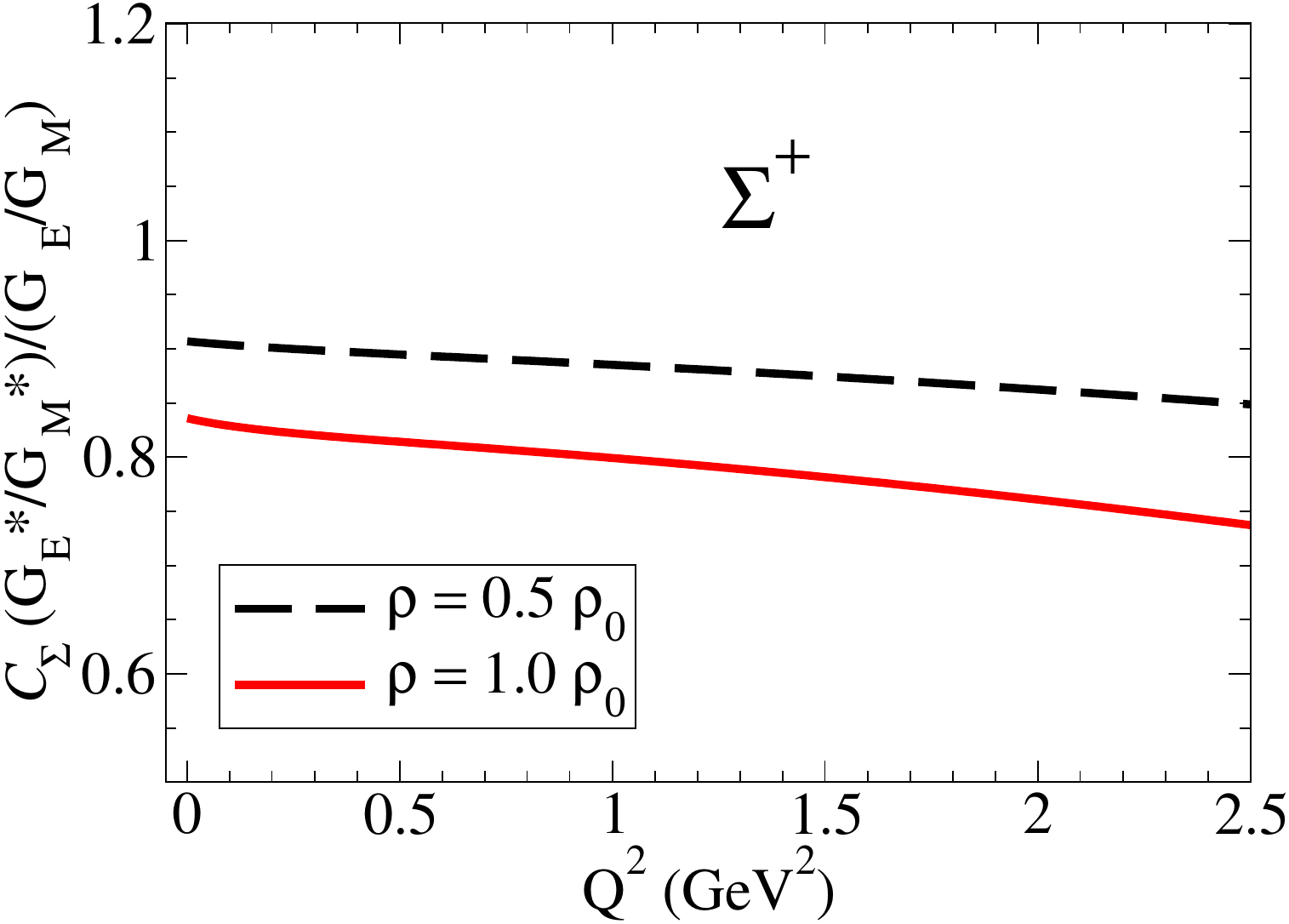}&
\includegraphics[width=2.4in]{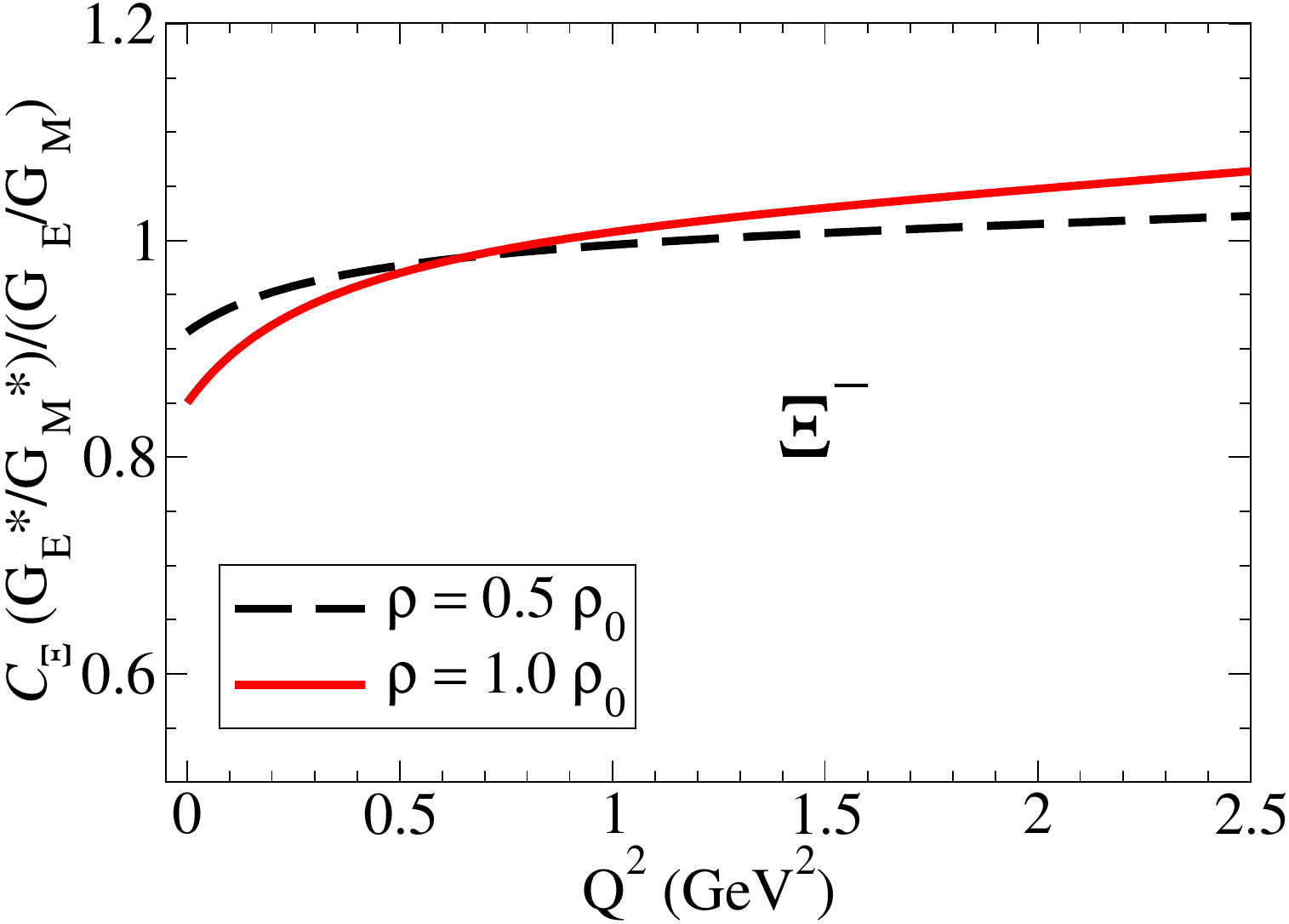}\\
\end{tabular}
\caption{$\Sigma^+$ and $\Xi^-$ double ratios in units of nuclear magneton. 
    $C_\Sigma= \frac{M_\Sigma^\ast}{M_\Sigma}$ and
    $C_\Xi= \frac{M_\Xi^\ast}{M_\Xi}$.
\label{figHyperonDR}}
\end{figure}

Our calculations of the double ratios for the nucleon and $\Sigma^+$ and $\Xi^-$, converted to nuclear magneton, are presented Figures~\ref{figNucleonDR} and \ref{figHyperonDR}, respectively.
As before, we consider the densities $\rho=0.5 \rho_0$ and $\rho= \rho_0$.

On the right side of Figure~\ref{figNucleonDR}, the model calculations for the proton are compared with the experimental results from MAMI and JLab~\cite{Dieterich01,Strauch03a}.
The calculations of the proton double ratio, presented on the right side of Figure~\ref{figNucleonDR}, suggest that $G_E/G_M$ is suppressed in the nuclear medium.
Noticing that the double ratio can also be written as $(G_E^\ast/G_E)/(G_M^\ast/G_M)$, we can then conclude that the suppression of $G_E/G_M$ is a consequence of the stronger suppression of $G_E^\ast$ comparatively to $G_M^\ast$.
A more detailed analysis~\cite{Octet3} shows that $G_E/G_M$
is suppressed in the nuclear medium because both the electric square radius $r_{Ep}^2$ and the magnetic square radius  $r_{Mp}^2$ are enhanced in the nuclear medium, but $r_{Ep}^2$ dominate over $r_{Mp}^2$
 (in the low-$Q^2$ region, 
  we can write  $G_E/G_M \propto 1 - \frac{1}{6} (r_{Ep}^2 -r_{Mp}^2)Q^2$~\cite{Cloet09a}). 
From the comparison of the model calculations with the data, we conclude that the $^4$He data are better described by a calculation with a smaller density ($\rho = 0.5 \rho_0$) than the calculation with dense nuclear matter ($\rho = \rho_0$).
{
The uncertainty of the data (about 5\%) is in the present case larger than the model uncertainty (about 2\%).
The difference in the double ratio predictions for the two densities (about 10\%) suggests that future experiments may distinguish the double ratios for low dense nucleus ($\rho \simeq 0.5 \rho_0$) from more dense nucleus  ($\rho \gtrsim 0.75 \rho_0$).}

\vspace{-6pt}

The predictions for the neutron 
contrast with the model calculations for the proton (quenched effect).
According to the results from left side of Figure~\ref{figNucleonDR},
we expect now an enhancement of $G_E/G_M$ for the neutron for $Q^2 < 1$ GeV$^2$.
We can understand this trend by recalling that $G_E^\ast/G_E = r_{En}^{\ast 2}/ r_{En}^2$.
The enhancement of $G_E/G_M$ in medium is then a consequence of the enhancement of $|r_{En}^2|$ in medium, an effect that dominates over the multiplicative factor $C_N = \frac{M_N^\ast}{M_N} < 1$.
When $Q^2$ increases, however, the enhancement effect is reduced, and one may expect a suppression of the double ratio for $Q^2 > 2.5$ GeV$^2$.

{From the analysis of the model calculations of the neutron double ratio, we can conclude that a maximum is expected for the region $Q^2=0.2$--0.5 GeV$^2$.
We can also conclude that the difference between the two calculations (about 7\%) is maximal in the same region.
The maximum on the double ratios is a consequence of the enhancement of $G_{E}^\ast$ in the nuclear medium in the region, as shown in Figure~\ref{figNucleonR} for the neutron combined with the suppression of $G_{M}^\ast/M_N^\ast$.
Above $Q^2=0.5$ GeV$^2$, $G_{E}^\ast$ and $G_{M}^\ast/M_N^\ast$ are enhanced with similar rates with dominance of $G_{M}^\ast$, leading to a smooth convergence to the unity (almost no medium effect) near $Q^2 = 2.5$ GeV$^2$.
In principle, measurements of double ratios for nuclei with intermediate density ($\rho \simeq 0.5 \rho_0$) can be distinguished from double ratios of nuclei with higher densities ($\rho \simeq \rho_0$) when the precision of the measurements is closer to the model accuracy (2\%).
In general, we can expect to measure an enhancement of about 10\% near $Q^2=0.5$ GeV$^2$ for a nucleus of intermediate density ($\rho \simeq 0.75 \rho_0$).}

To summarize the results for nucleon double ratio:
In the case of the proton, $G_E^\ast/G_E$ and $G_M^\ast/G_M$ are both reduced in the nuclear medium, but there is a dominance of $G_E$ (stronger suppression).
In the case of the neutron, $G_E$ is enhanced, $M_N/G_M$ is enhanced, and, as a consequence, the double ratio is positive ($G_E/G_M$ is enhanced in units of \mbox{nuclear magneton).}

The double ratio for the nucleon has been calculated
using the QMC model~\cite{Lu99a}, quark--soliton model~\cite{Smith04},
and light front quark models~\cite{Cloet09a,deAraujo18}.
Model calculations predict in general the quenching of the ratio $G_E/G_M$ for the proton in the range of 5\% to 10\% at low $Q^2$ depending on the density~\cite{Cloet09a,deAraujo18,Smith04}.
As for the neutron, Refs.~\cite{Cloet09a,deAraujo18} predict, as our model, an enhancement of $G_E/G_M$ for $Q^2 < 1$ GeV$^2$.
Calculations based on light front quark model and Nambu--Jonas--Lasinio model, with manifest dominance of the valence quark effects, lead to the conclusion that, near $Q^2=0$, the neutron double ratio can be estimated by ${\cal R} (0) \approx \left(\frac{M_N}{M_N^\ast} \right)^2$, corresponding to an enhancement of about 20\%~\cite{Cloet09a}.

The model calculations for $\Sigma^+$ and $\Xi^-$ double ratios are presented in Figure~\ref{figHyperonDR}.
The results for the $\Sigma^-$ (right side) are similar to the results for the proton, except for the slower falloff with $Q^2$.
We can understand these results noticing that we should expect slower falloffs comparatively to the case of the proton for both $G_E$ and $G_M$ as a consequence of the reductions in $r_{E}^2$ and $r_M^2$ compared to the proton, as expected from a system with a strange quark (more compact system).

The analyses of the results for $\Xi^-$, on left side of Figure~\ref{figHyperonDR}, require some care.
At low $Q^2$, the quenched effect is a consequence of the near independence of $G_E$ in medium (notice in Figure~\ref{figNucleonR} that $G_E^\ast/G_E \simeq 1$) and the enhancement of $G_M^\ast/G_M$.
At large $Q^2$, one can notice a very slow increment regarding the double ratio, very close to unity, the signature of the smooth medium variations, as expected in a system dominated by strange quarks.

A more detailed discussion of the double ratios, including the neutral baryons ($\Lambda$, $\Sigma^0$, and $\Xi^0$) is presented in Ref.~\cite{Octet3}.


\subsection{Axial Form Factors \label{secResults1Axial}}

We discuss now the calculations of the axial form factors $G_A$ and $G_P$
in nuclear medium based on the formalism from Section~\ref{secModel}.
These calculations use the model developed in Ref.~\cite{AxialFF} for the octet baryon axial form factors in free space.
The valence quark contributions are calculated using the expressions presented in Section~\ref{sec-Axial}, where the quark axial form factors are determined by the analyses of the lattice QCD data for the nucleon from Ref.~\cite{Alexandrou11a}, while meson cloud contributions are estimated using the nucleon form factor data for $G_A$ and the octet baryon axial couplings
(see Section~\ref{sec-Axial-MC}).

As mentioned already, there are 12 possible transitions associated with charged currents between octet baryon members: 6 associated with $|\Delta I|=1$ transitions, and 6 associated with $|\Delta S|=1$ transitions
(see Table~\ref{tab-Axial-Transitions}).
There are also 8 elastic transitions associated with neutral currents.
We focus here on the inelastic transitions (charged currents) since the elastic form factors can  also be related to the $|\Delta I|=1$ transition form factors.

It is also worth noticing that, in our formalism, the form factors associated with the $\Sigma^+ \to \Lambda$ and $\Sigma^- \to \Lambda$ transitions differ by a sign, and the form factors associated with the $\Sigma^- \to \Sigma^0$ and  $\Sigma^0 \to \Sigma^+$ are identical.
There are then only 4 independent  $|\Delta I|=1$ transitions to discuss.

The discussion about the magnitude and shape of the 10 independent axial transition form factors is simplified when we consider a few typical cases associated with each channel ($|\Delta I|=1$ or $|\Delta S|=1$), depending on the baryon masses.
Notice that the mass increases with the number of strange quarks of the baryon.

The extension of the calculations of the octet baryon axial form factors from the free space to the nuclear medium was motivated by the quality of the description of the nucleon lattice QCD and the nucleon physical data for $G_A$ and $G_P$, combined with a fair description of the $\Lambda \to p$, $\Sigma^- \to n$ $\Xi^- \to \Lambda$, and $\Xi^0 \to \Sigma^+$ axial couplings~\cite{AxialFF,GA-Medium1}.

\subsubsection{Axial-Vector Form Factor}

A detailed discussion of the axial-vector form factors in vacuum can be found in Refs.~\cite{AxialFF,GA-Medium1}.
Here, we mention only the general properties.
The available free space data are well described by a combination of valence quark and meson cloud contributions.
The magnitude of the axial-vector form factors varies with the transition under discussion.
At $Q^2=0$, the absolute values of $G_A$ are between 0.2 and 1.3.
The valence quark contributions are dominant, but the inclusion of the processes associated with meson cloud helps to improve the description of the physical data.

The calculations for the nucleon axial-vector
form factor ($n \to p$ transition) are in agreement with the known data for the $n \to p$ axial-vector
form factor in the regions $Q^2=0$--1 GeV$^2$ and $Q^2=2$--4 GeV$^2$~\cite{Bernard02,Schindler07a,Park12}.

\begin{figure}[H]  

  \hspace*{-3mm}
\begin{tabular}{cc}
\includegraphics[width=2.4in]{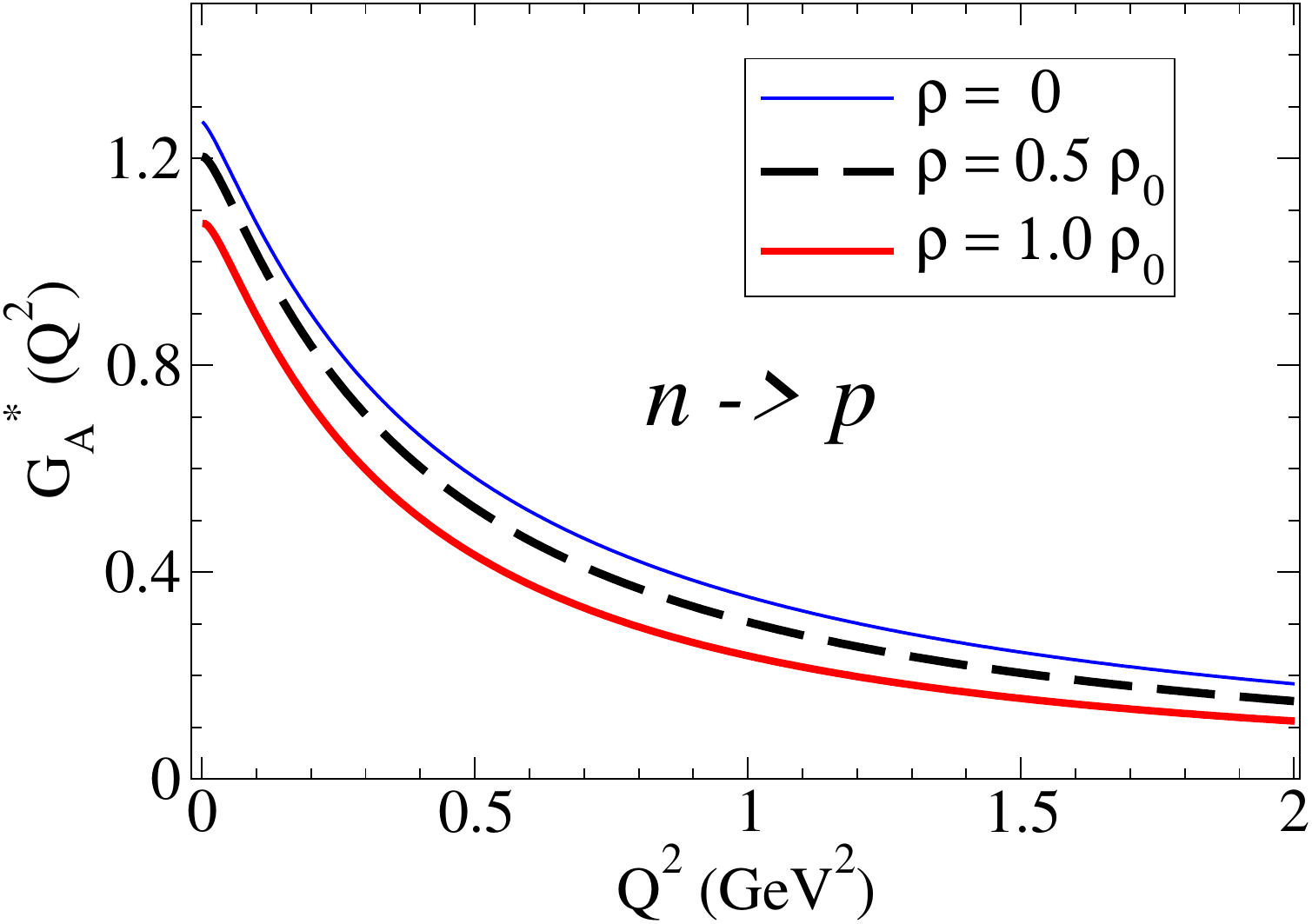} &
  \includegraphics[width=2.4in]{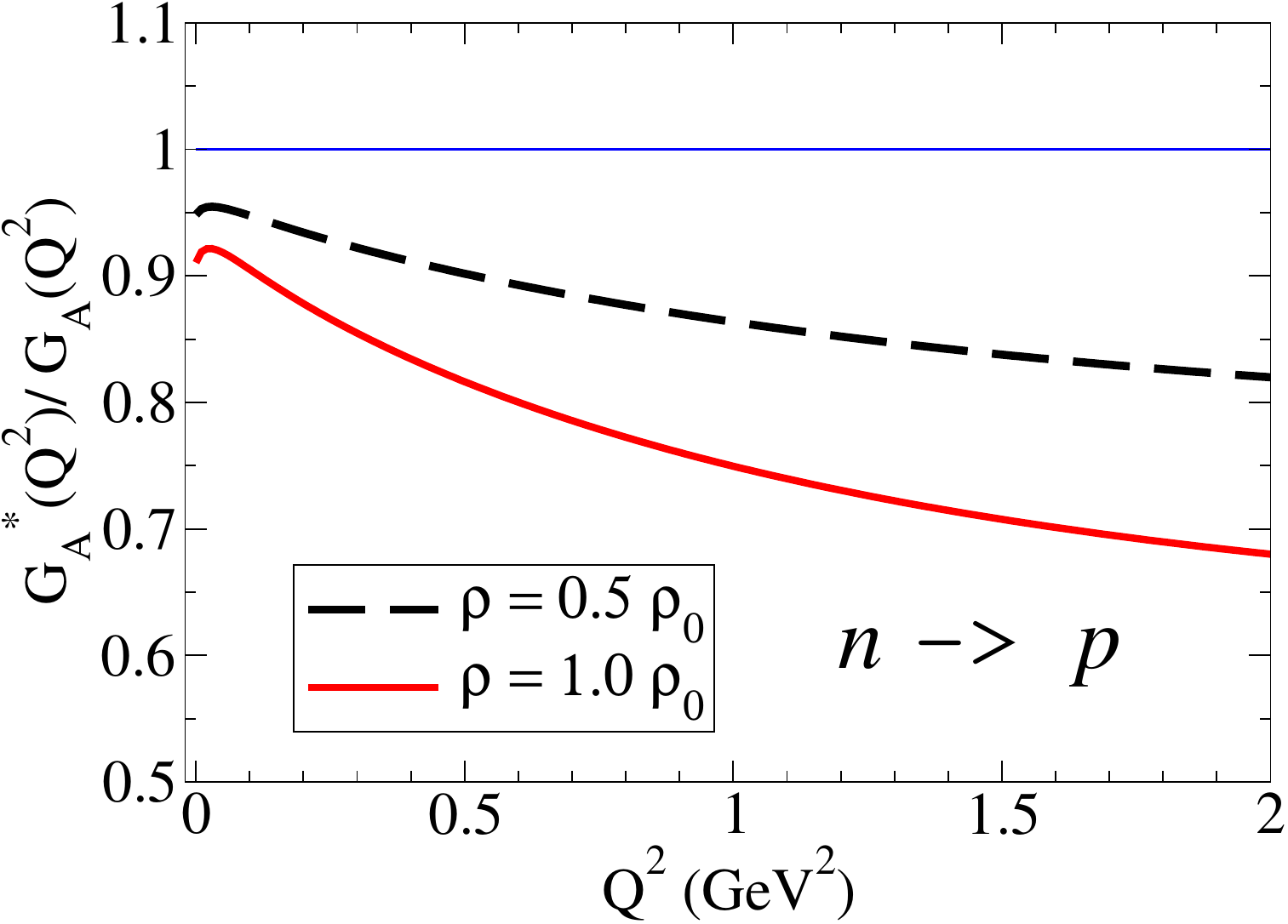}\\
\includegraphics[width=2.4in]{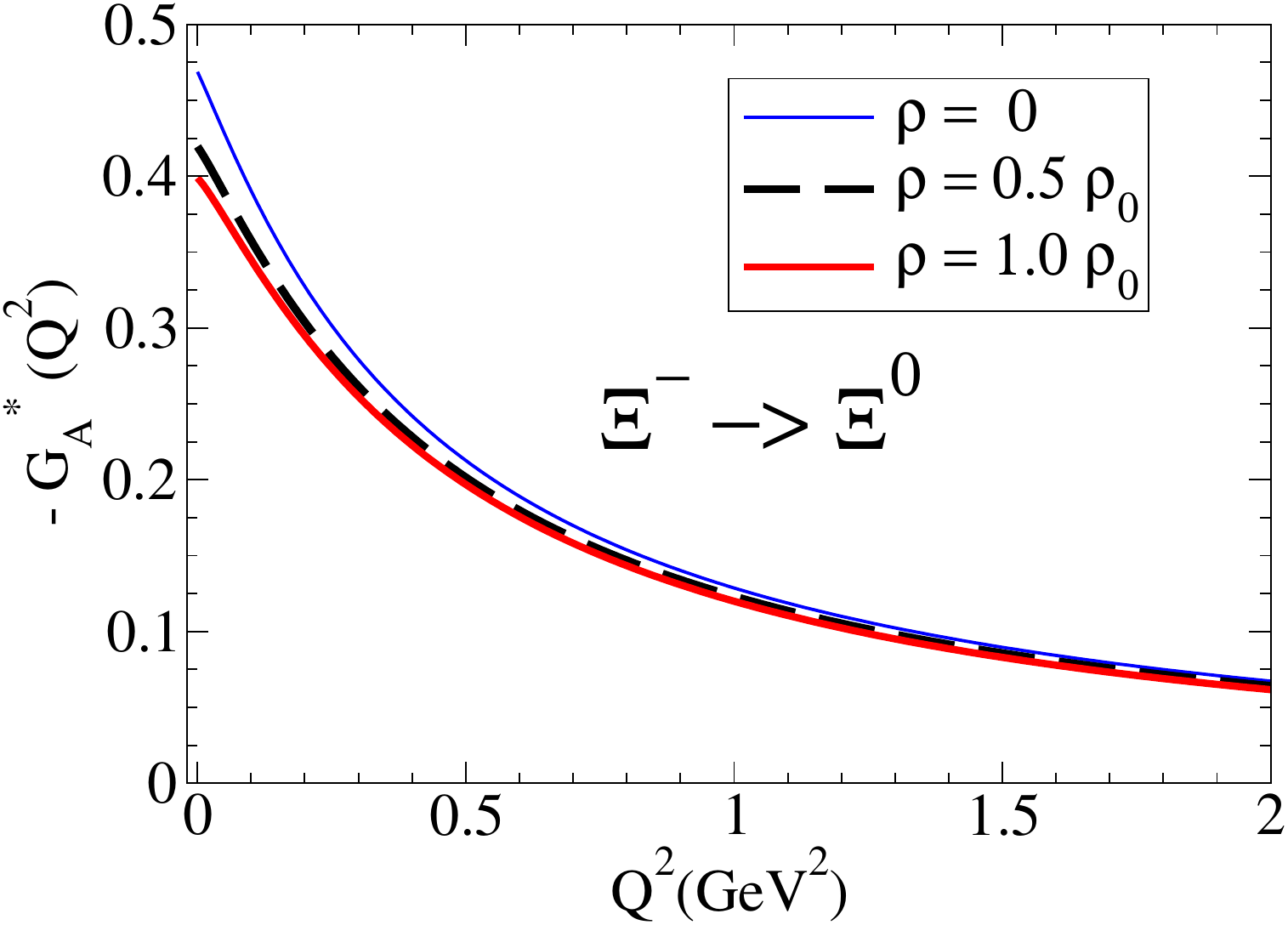} &
  \includegraphics[width=2.4in]{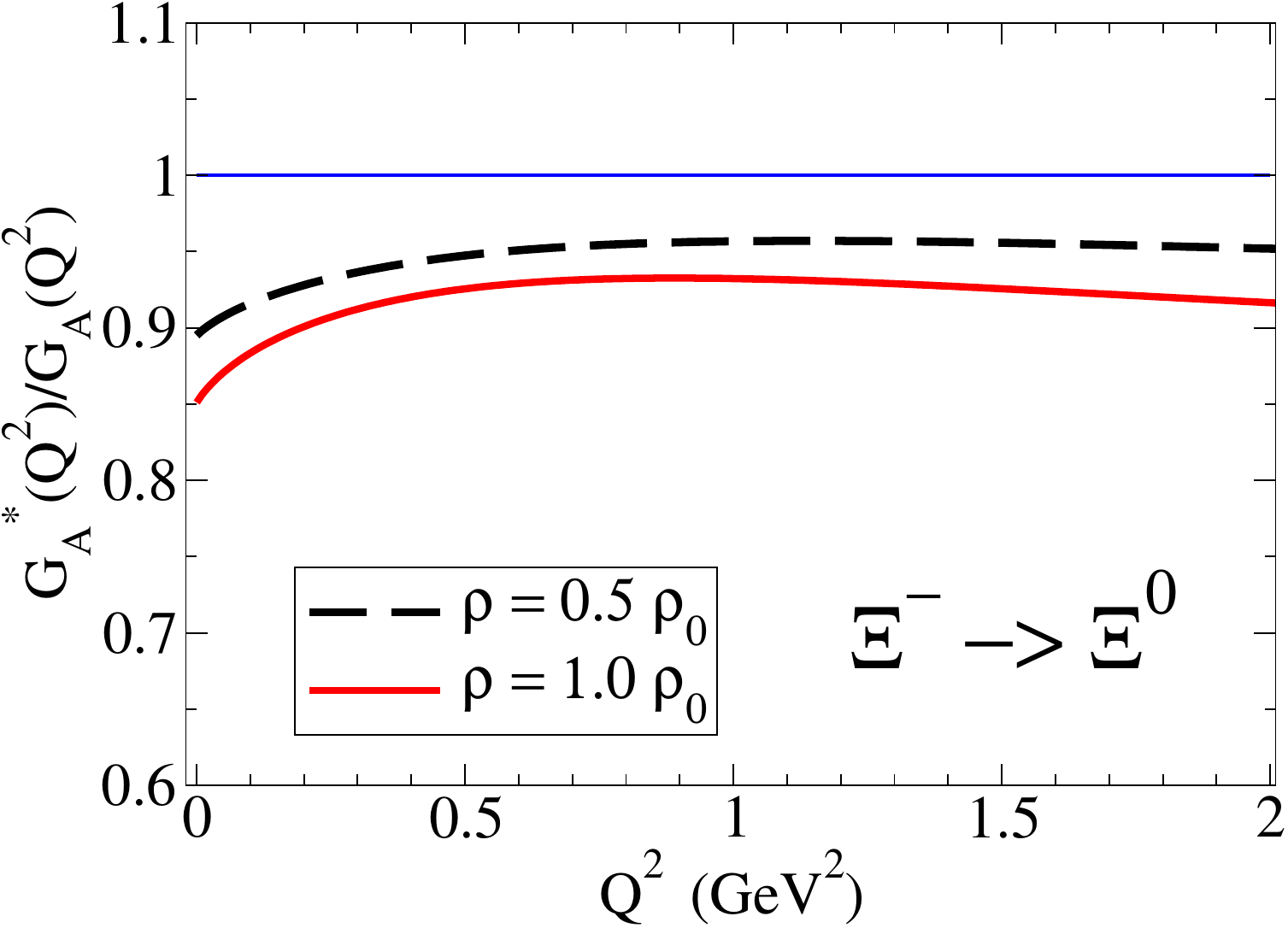}\\
  \end{tabular}
    \caption{$n \to p$ and $\Xi^- \to \Xi^0$ axial form factors in nuclear medium ($|\Delta I|=1$ transitions).
    We use  $-G_A$ for the negative functions for an easy comparison of magnitudes. 
    The horizontal line ($G_A^\ast /G_A \equiv 1$) is included to represent the ratio in free space.
\label{figGA-p1}}
\end{figure}

\begin{figure}[H]
  \hspace*{-3mm}
\begin{tabular}{cc}
  \includegraphics[width=2.4in]{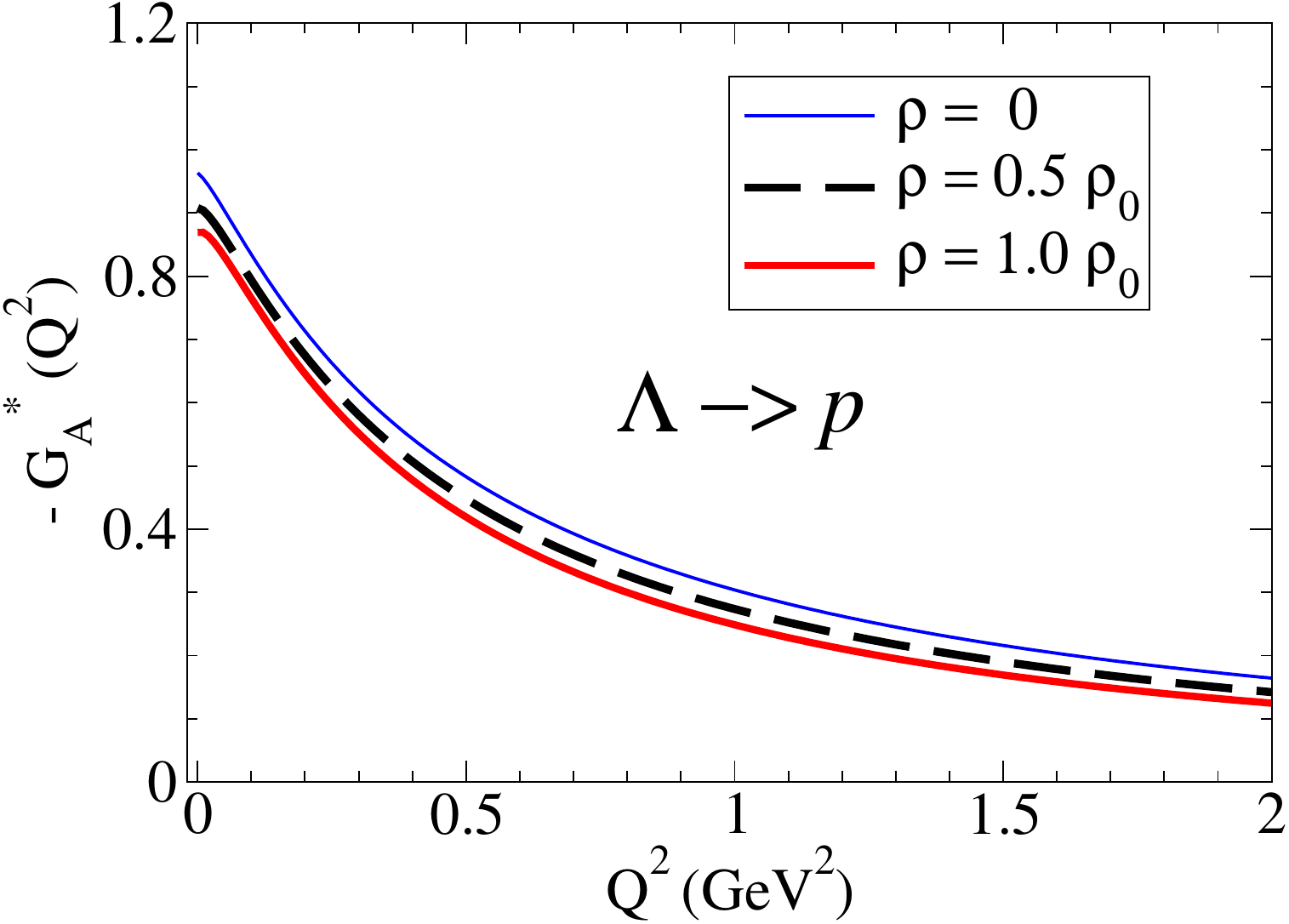}&
  \includegraphics[width=2.4in]{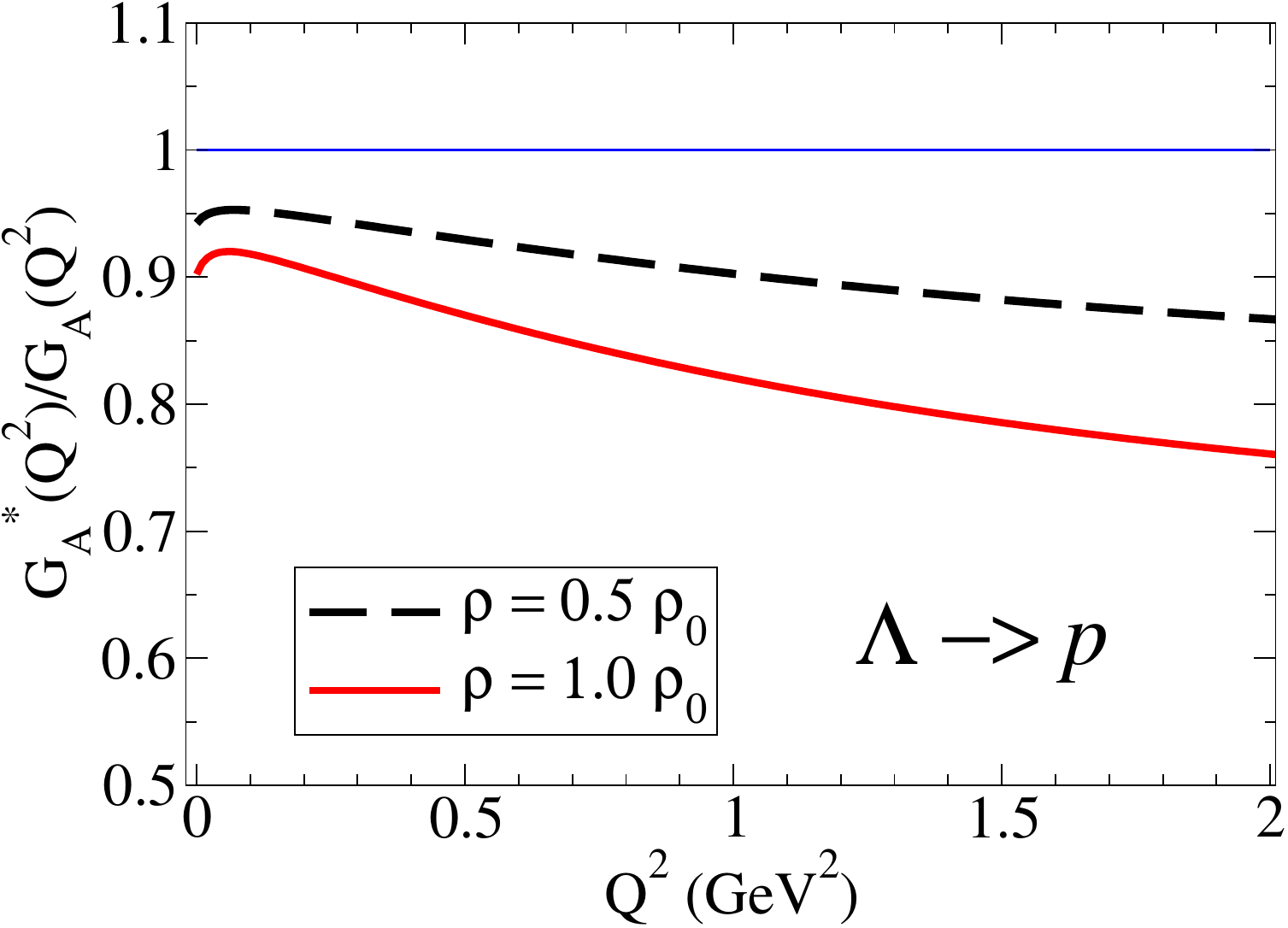}\\
\includegraphics[width=2.4in]{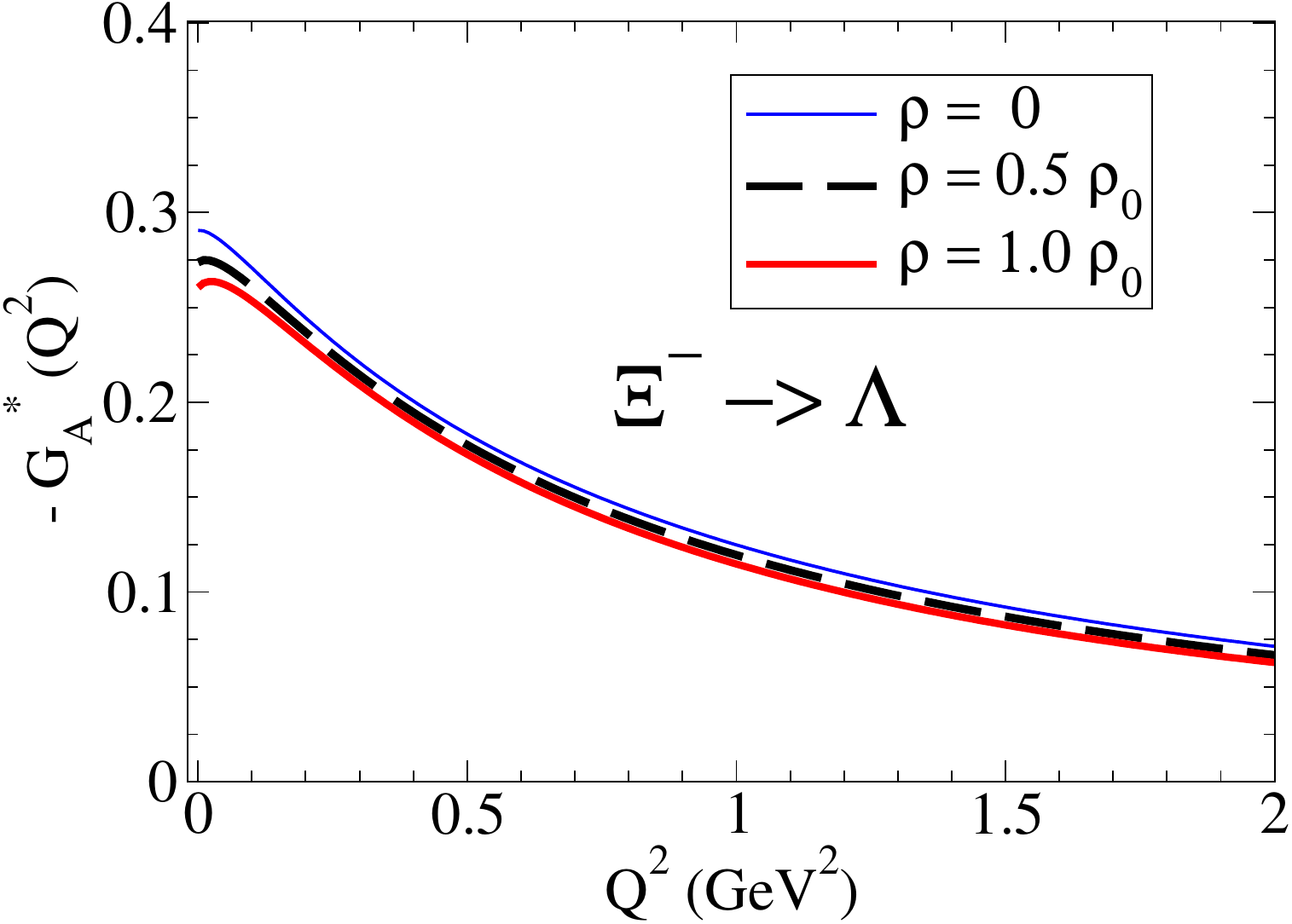} &
  \includegraphics[width=2.4in]{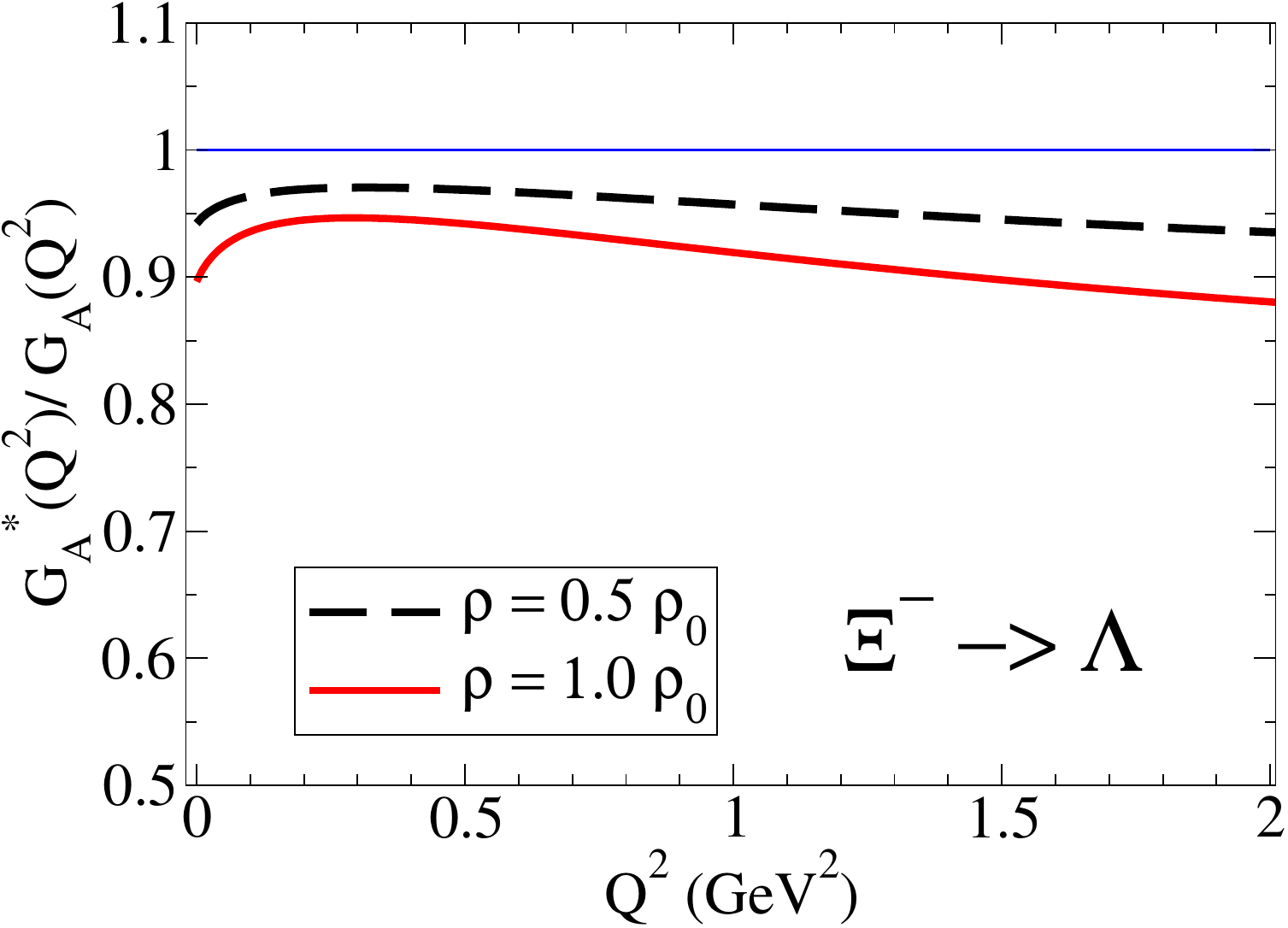}\\
  \end{tabular}
    \caption{$\Lambda \to p$ and $\Xi^- \to \Lambda$ axial-vector form factors in nuclear medium ($|\Delta S|=1$ transitions).
     We use  $-G_A$ for the negative functions for an easy comparison of magnitudes. 
     The horizontal line ($G_A^\ast /G_A \equiv 1$) is included to represent the ratio in free space.
\label{figGA-p2}}
\end{figure}

Calculations of $|\Delta I|=1$ transitions and $|\Delta S|=1$ transitions are presented in \mbox{Figures~\ref{figGA-p1} and \ref{figGA-p2}}, respectively, for the vacuum ($\rho=0$), $\rho= 0.5 \rho_0$, and  $\rho= \rho_0$.
On the left side, we present the explicit form factors; on the right side, we consider the ratio $G_A^\ast/G_A$.

From Figures~\ref{figGA-p1} and \ref{figGA-p2}, we can conclude that transitions associated with the lighter baryons ($n \to p$ and $\Lambda \to p$) are strongly suppressed in nuclear medium.
In contrast, in the transitions associated with heavier baryons ($\Xi^- \to \Xi^0$ and $\Xi^- \to \Lambda$),
the impact of the medium is milder, and the suppression is less significant.
Notice that the lines became flatter for larger values of $Q^2$.
The suppression increases with $Q^2$, particularly for transitions with lighter baryons.
As expected, the medium effects are more noticeable for more dense nuclear media, represented here by $\rho = \rho_0$.
For the normal nuclear matter ($\rho= \rho_0$), the suppression for $Q^2 \simeq 2$ GeV$^2$ can be about 30\% for lighter baryons ($n \to p$ transition) and about 15\% for heavier baryons ($\Xi^- \to \Lambda$ transition).

Overall, the valence quark and meson cloud contributions are suppressed
in medium but in different proportions.
The suppression of the meson cloud contribution is more noticeable at low $Q^2$.
The impact of the medium on the bare contributions is more visible at large $Q^2$, when the meson cloud contributions are almost irrelevant, and the falloffs of the form factors are regulated by quark power counting rules.

For the axial-vector form factors associated with the remaining transitions,
not presented here, we can observe an intermediate suppression in medium.

\vspace{-6pt}

The calculations presented here for the nuclear medium
are based on the QMC model and the bag model~\cite{Lu01a}.
We use the QMC estimates for the nucleon axial-vector coupling constant associated with the density $\rho$
(see Section~\ref{secMedium1} and Table~\ref{table-gpiBB1}).
The QMC model predicts, at $Q^2=0$, the quenching of 5\% for the density $\rho= 0.5 \rho_0$, and 10\% for the normal nuclear matter ($\rho= \rho_0$).
Similar proportions are obtained in our final results when we take into account the contributions from the meson cloud.

We discuss now the literature about the nucleon axial-vector form factor in nuclear medium.

Measurements of beta decay rates in heavy nuclei, more than 50 years ago, showed that the axial-vector of the nucleon is reduced by about 25\%~\cite{Brown85a,Gysbers19a}, providing evidence that the axial-vector form factors are quenched in nuclear medium.

Calculations based on the effective  theory that take into account two-body interactions~\cite{Gysbers19a} are consistent with the quenched effect of 0.75, observed on nucleon beta-decay rates of heavy nuclei~\cite{Brown85a}.

Skyrme, soliton, and quark--soliton models~\cite{Meissner87a,Meissner89a,Christov96a,Rakhimov98a}
are particularly useful to estimate the medium effects on
the axial-vector of baryons in general and the nucleon in particular
because the reduction in in-medium quark mass can be taken into account in a simple form.
In the leading order, the calculation underestimates the observed
nucleon axial-vector coupling, but more accurate results are obtained when higher-order
corrections are taken into account~\cite{Christov96a,Meissner87a,Meissner89a}.
Skyrme and soliton model calculations also predict the quenching
of the axial-vector coupling constant and the function $G_A$.
Explicit calculations can be found in Refs.~\cite{Christov96a,Rakhimov98a} for different densities.

To finish our discussion, we compare our estimates with the QMC model~\cite{Lu01a,Cheoun13b,Cheoun13a} used in the calibration of our model (ratio $g_A^{N \ast}/g_A^N)$.
The main difference between the calculations is that we include a contribution for the meson cloud.
Apart from small differences at low $Q^2$, the calculation differs in the shape of the ratio $G_A^\ast/G_A$.
In the bag model, the ratio decreases until $Q^2=0.5$--1 GeV$^2$ and starts to increase after that value, displaying an enhancement for larger values of $Q^2$, due to the relativistic Lorentz contraction effect.
In our calculations, the quenching effect increases with $Q^2$.

\subsubsection{Induced Pseudoscalar Form Factor}

The literature about the induced pseudoscalar form factor is scarce because it is hard to measure since the contributions to the neutrino/antineutrino cross-sections are suppressed by a factor $(m_\ell/M_{BB'})^2$, where $m_\ell$ is the lepton mass associated with the weak transition~\cite{Bernard02,Gorringe04,Cai23a}.
The available data for the nucleon $G_P$
came from pion electroproduction experiments
and  muon capture by nucleons~\cite{Bernard02,Choi93}.

The results of the induced pseudoscalar form factor $G_P$ differ significantly in magnitude for the channels $|\Delta I|=1$ and $|\Delta S|=1$.
This result is a consequence of the pole term (\ref{eqGP-pole}), which dominates the function 
$G_P$ near $Q^2=0$, leading to $G_P(0) \propto 1/\mu^2$, where $\mu= m_\pi$ for  $|\Delta I|=1$ and $\mu= m_K$ for  $|\Delta S|=1$.

Calculations of the induced pseudoscalar form factors $G_P$ of the octet baryons based on the covariant spectator formalism in vacuum are discussed in Refs.~\cite{AxialFF,GA-Medium1}.
The results for the  $|\Delta I|=1$ transitions are, in general, dominated by the pole term, although the meson cloud contribution (no adjustable parameters) is also relevant to the agreement of the model with the experimental data for the $n \to p$ transition.
As for the $|\Delta S|=1$ transitions, the magnitude of the form factors is significantly reduced, and the dominance of the pole term is less effective.
In that case, the bare term has a larger relative contribution and cancels parts of the pole and meson cloud terms~\cite{AxialFF}.
These general properties are also observed in the nuclear medium~\cite{GA-Medium1}.

  The calculations of $G_P$ for the $n \to p$ transition are in good agreement with the electroproduction data extracted with low-energy theorem~\cite{Choi93} and with the muon capture data~\cite{Bernard02} in the range $Q^2=0$--0.2 GeV$^2$.
  The model calculations are also in close agreement with the lattice QCD simulations for pion masses in the range $m_\pi  = 210$--265 MeV for $Q^2 > 0.2$ GeV$^2$~\cite{Alexandrou13a}.
These results are discussed in detail in Ref.~\cite{AxialFF}.

\begin{figure}[H]  
\hspace*{-3mm}
\begin{tabular}{cc}
  \includegraphics[width=2.45in]{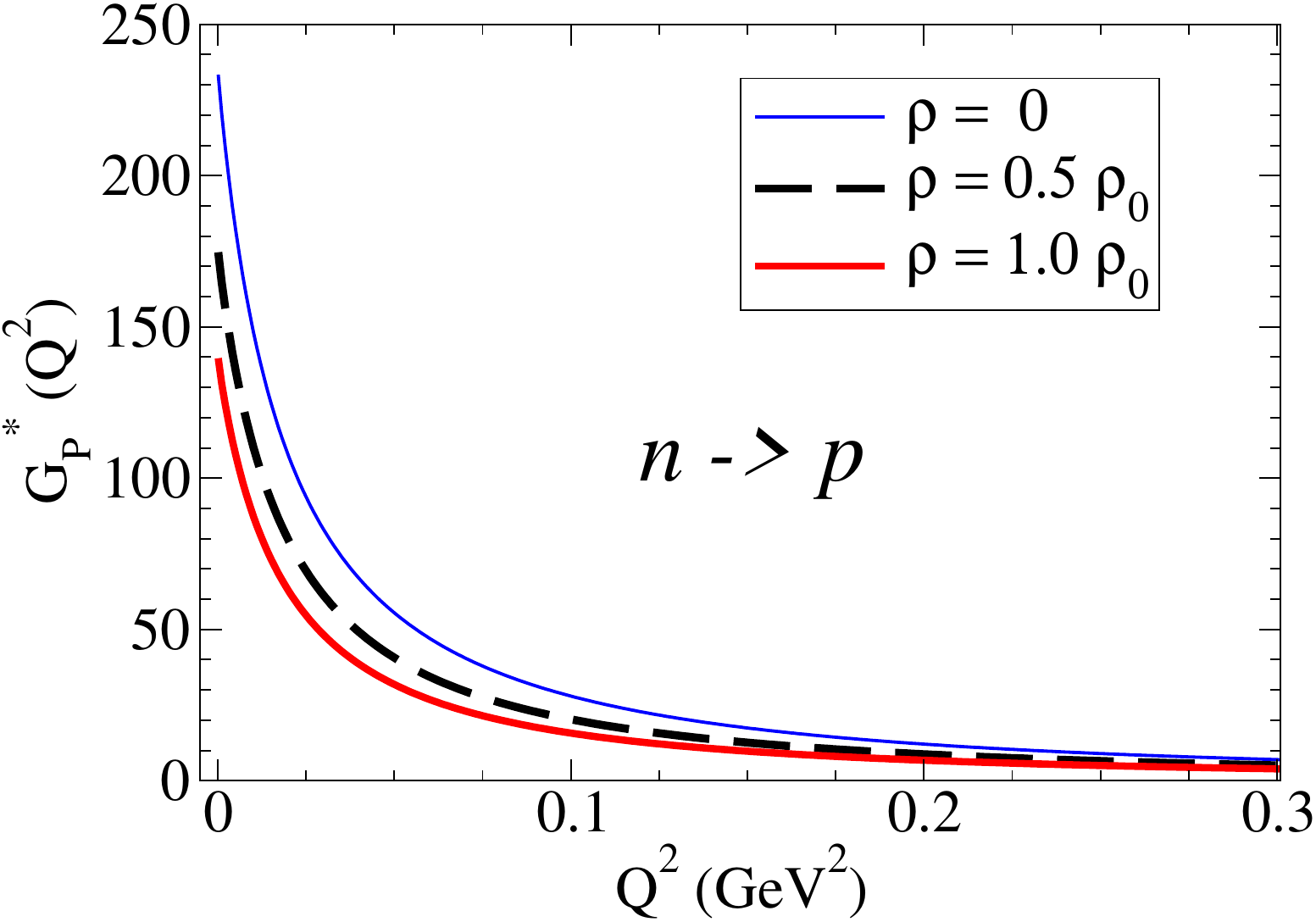} &
  \includegraphics[width=2.4in]{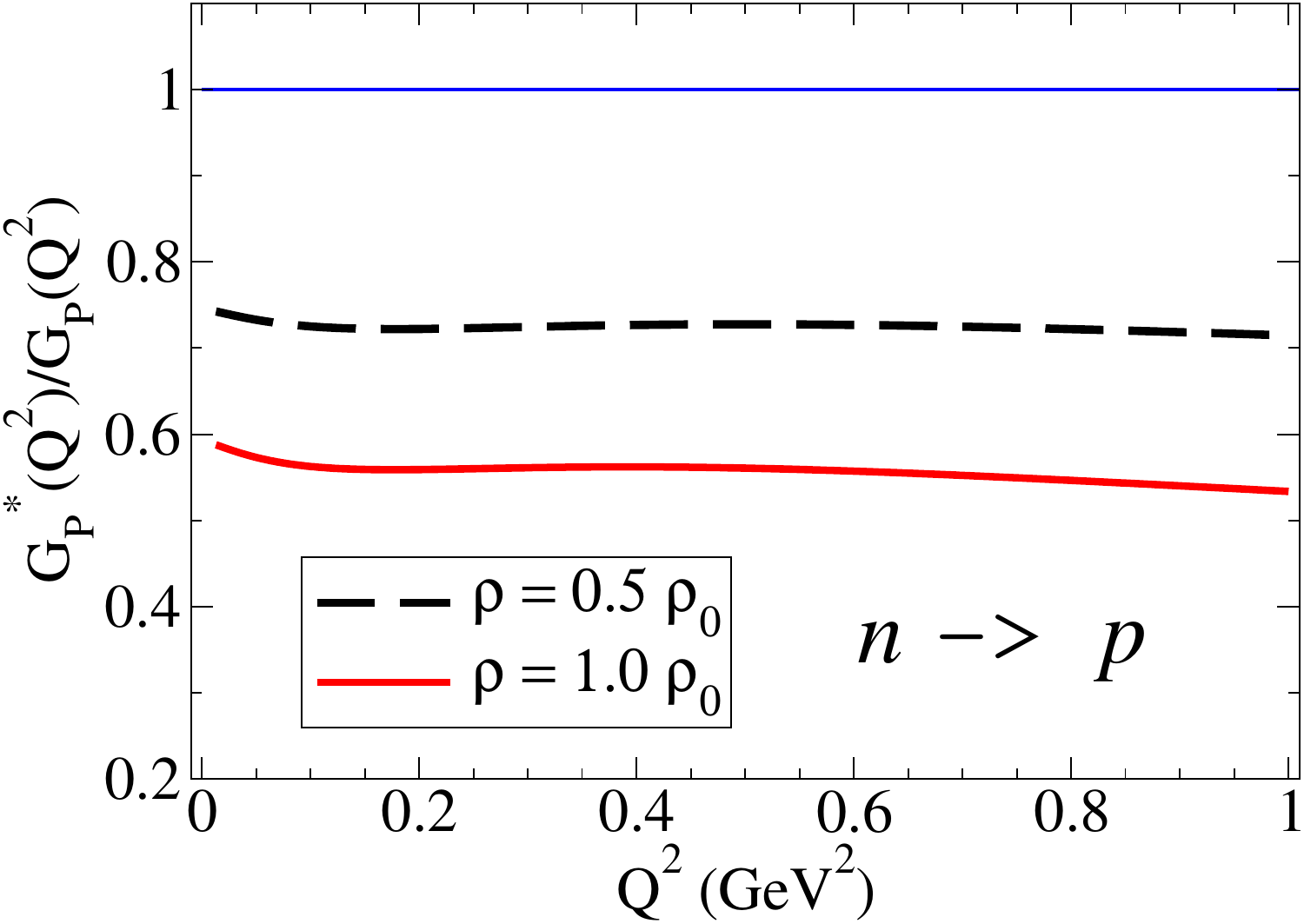}\\
\includegraphics[width=2.4in]{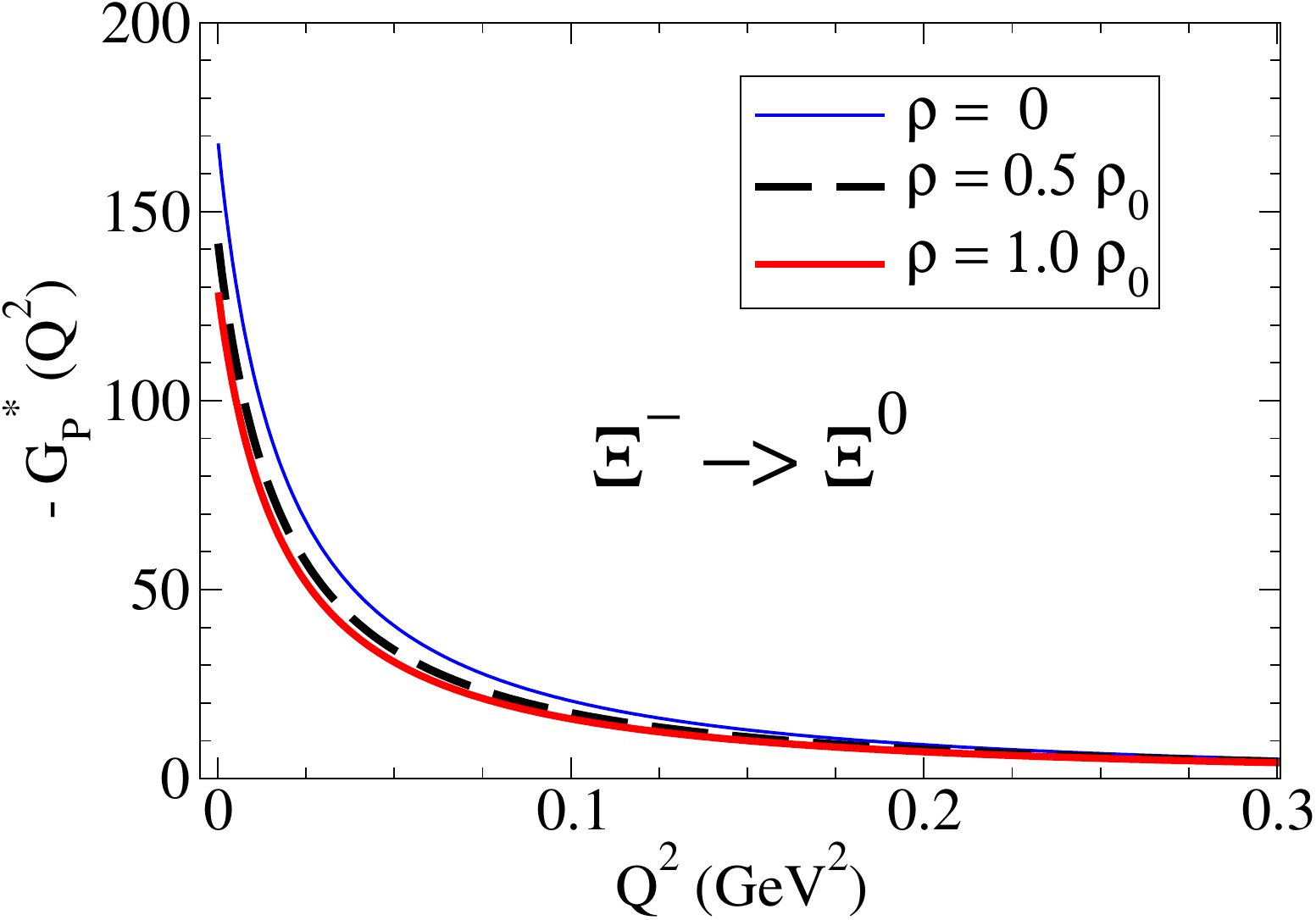} &
  \includegraphics[width=2.4in]{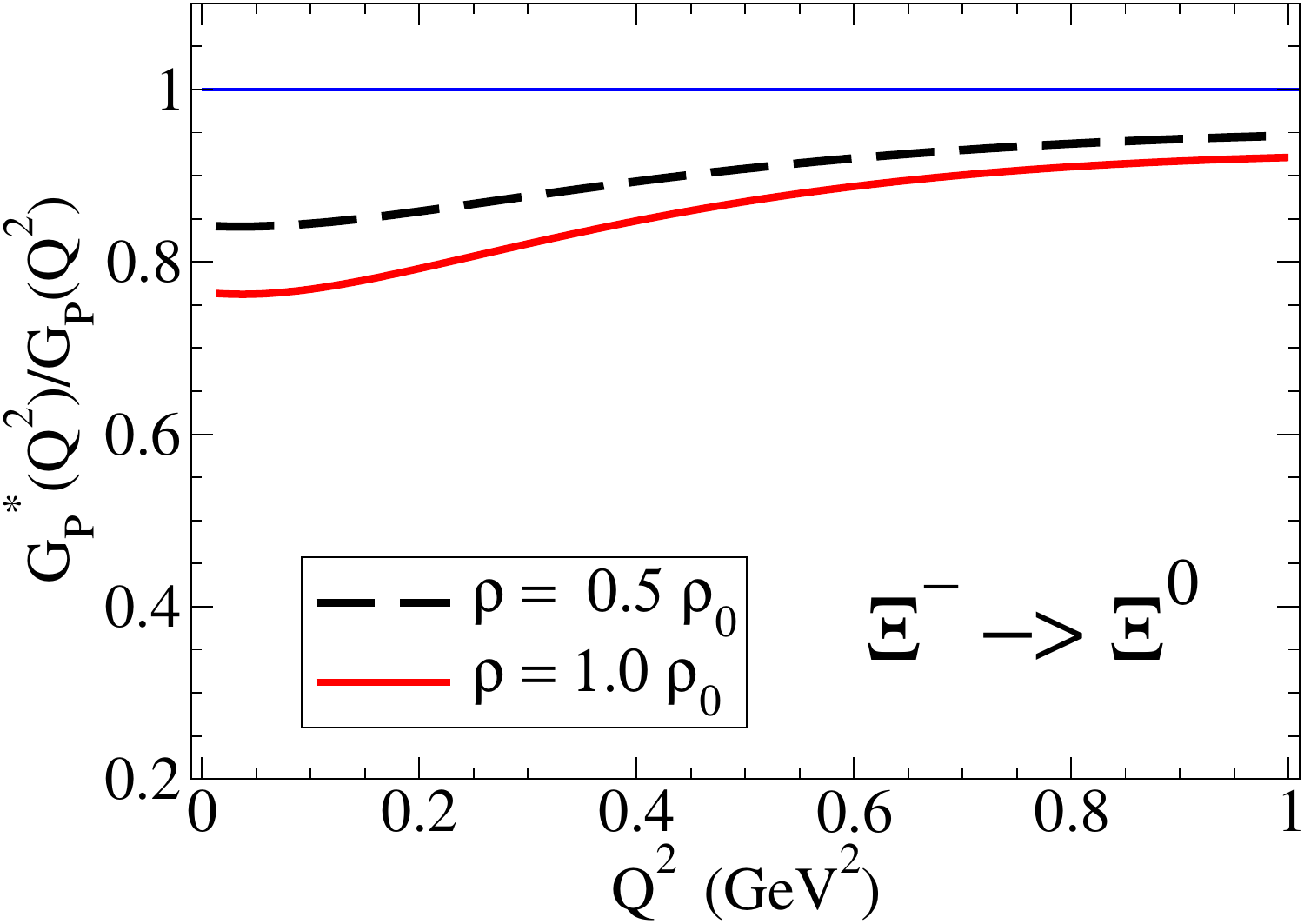}\\
\end{tabular}
    \caption{$G_P$ form factors for the $n \to p$ and $\Xi^- \to \Xi^0$ transitions ($|\Delta I|=1$) in nuclear medium. 
      For an easy comparison of magnitudes, we use $-G_P$ for the negative functions.
    The horizontal line ($G_P^\ast /G_P \equiv 1$) is included to represent the ratio in free space.
\label{figGP-p1}}
\end{figure}

\begin{figure}[H]  
\hspace*{-3mm}
\begin{tabular}{cc}
  \includegraphics[width=2.4in]{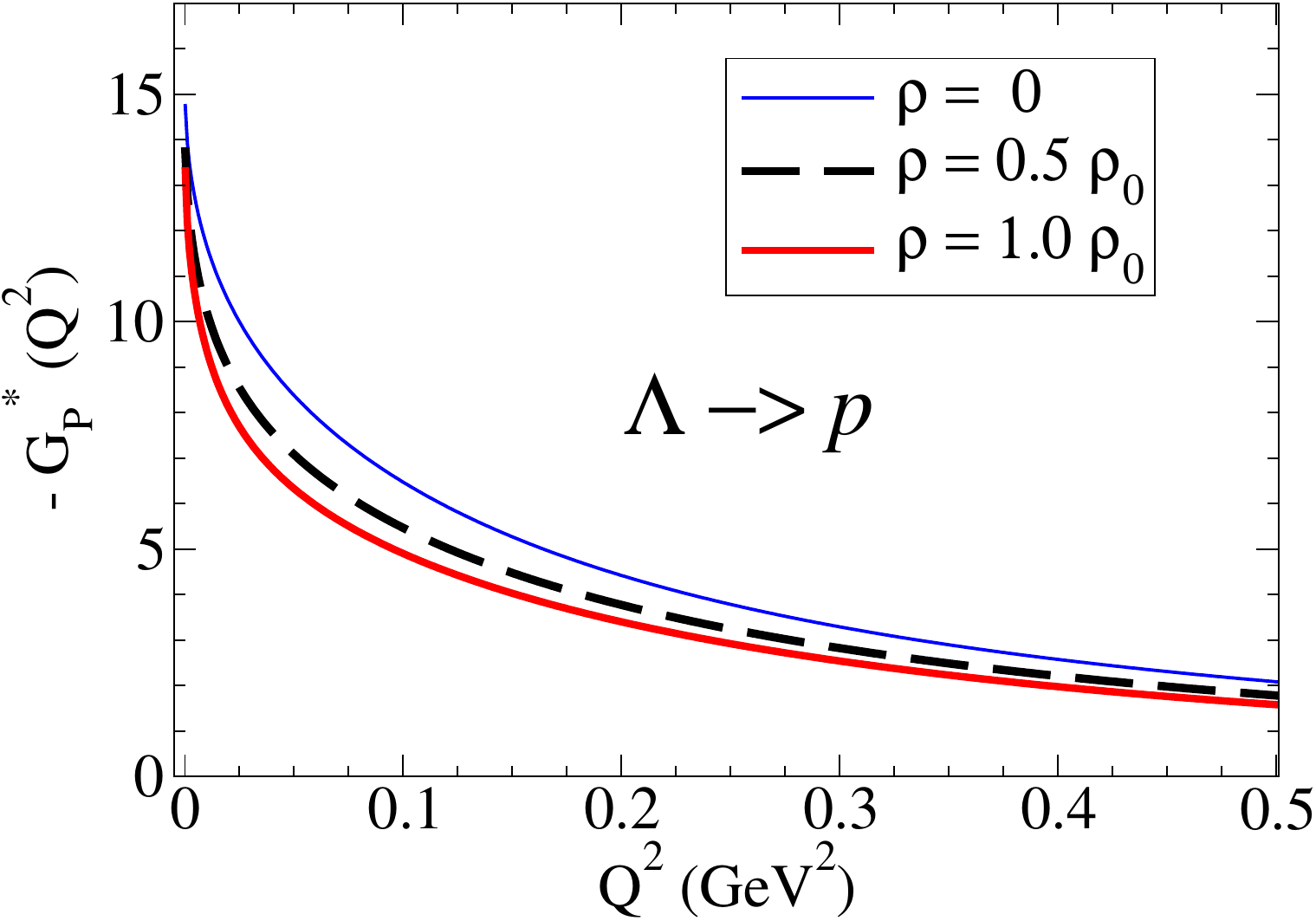}  &
  \includegraphics[width=2.4in]{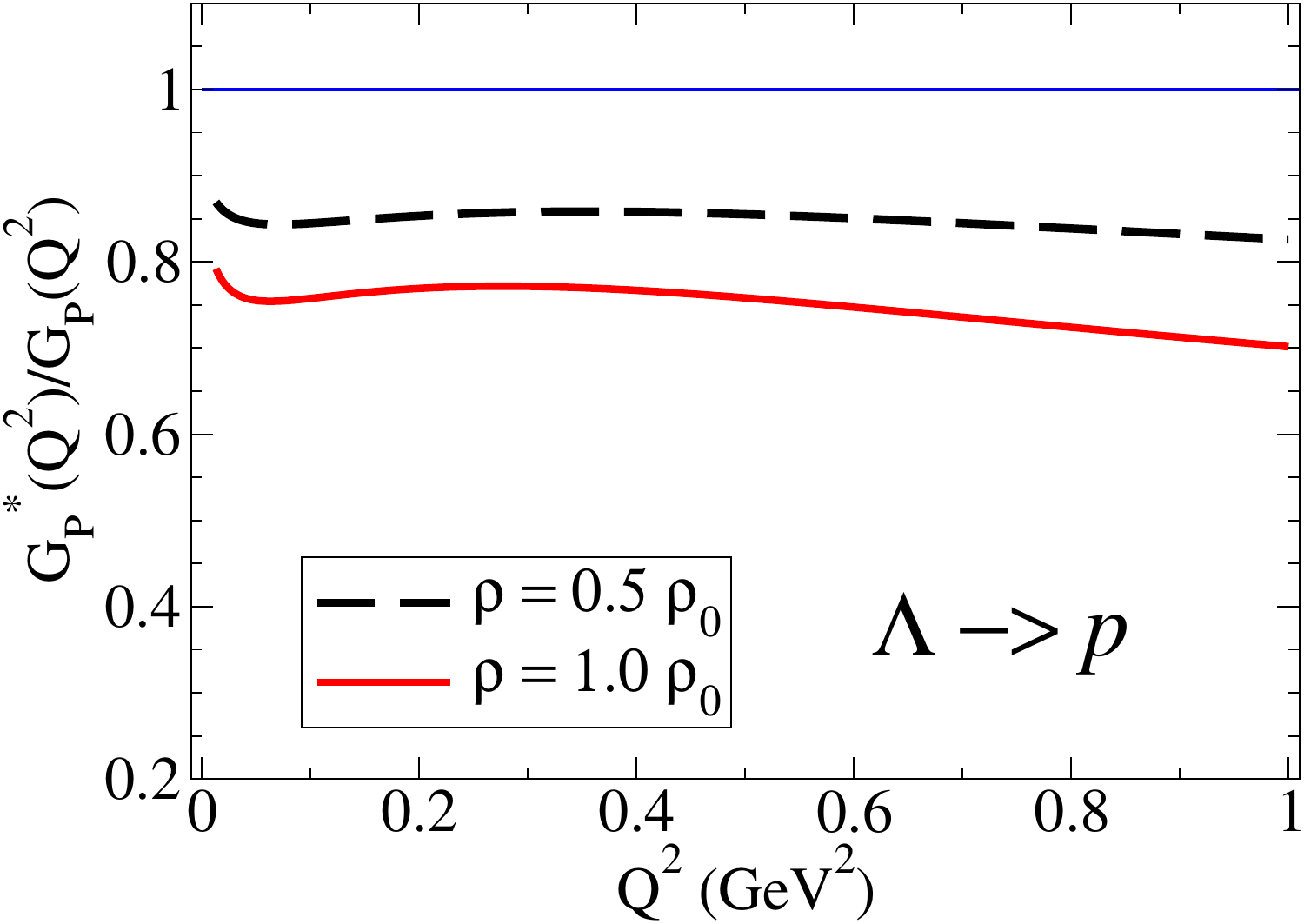}\\
  \includegraphics[width=2.4in]{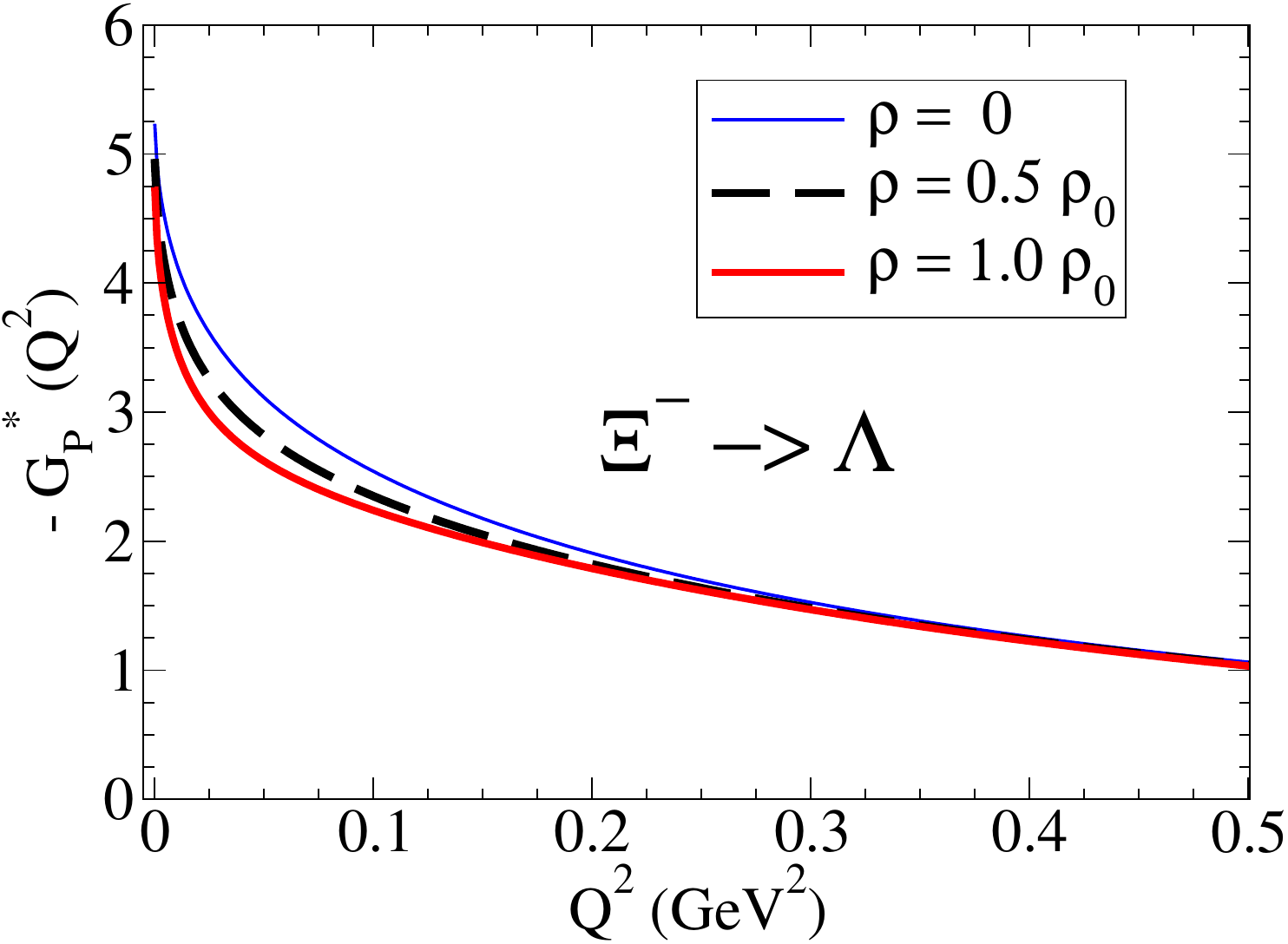}&
  \includegraphics[width=2.4in]{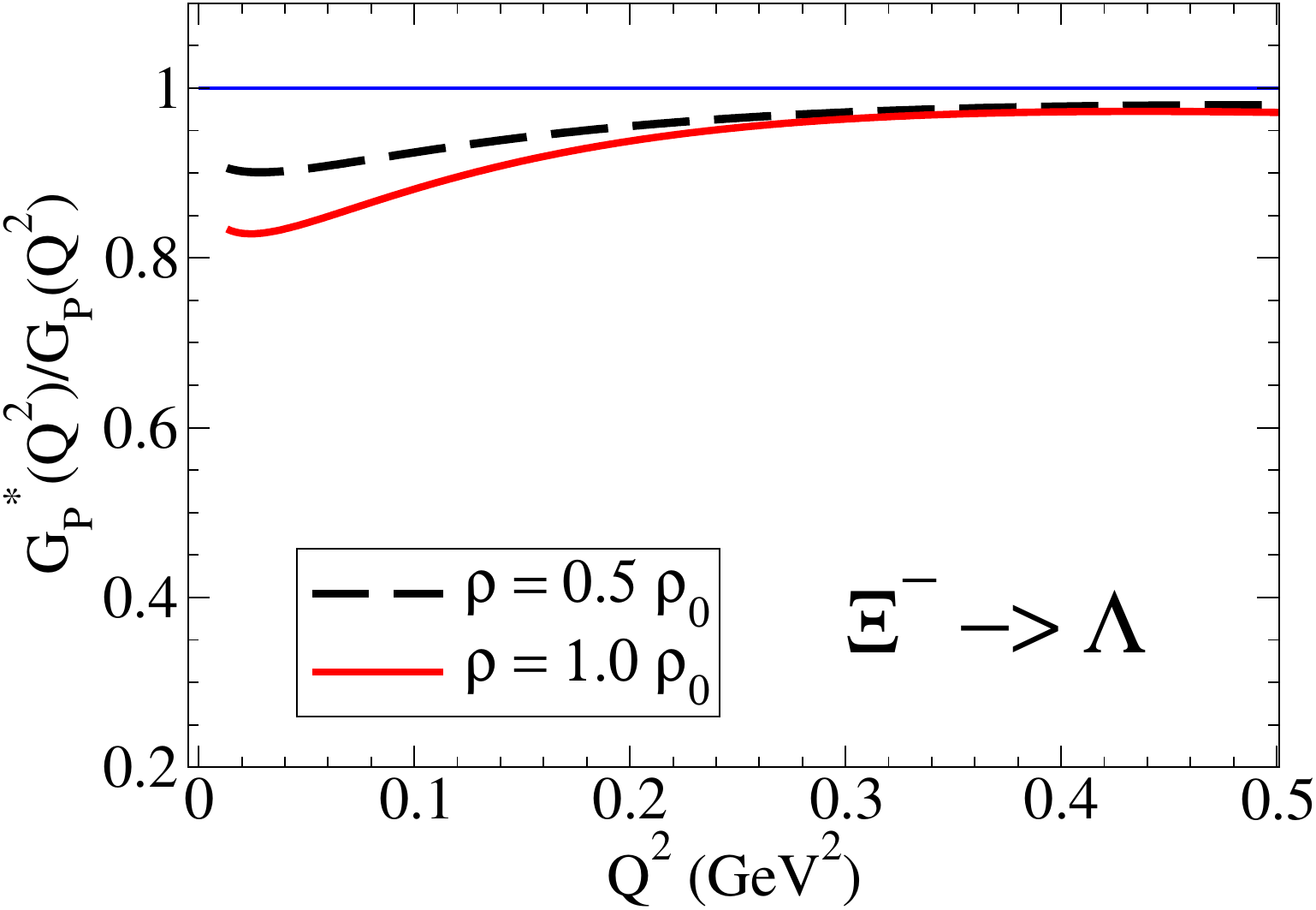}\\
\end{tabular}
  \caption{$G_P$
 form factors for the  $\Lambda \to p$ and $\Xi^- \to \Lambda$ transitions ($|\Delta S|=1$) in nuclear medium.
    For an easy comparison of magnitudes, we use $-G_P$ for the negative functions.
    The horizontal line ($G_P^\ast /G_P \equiv 1$) is included to represent the ratio in free space.
\label{figGP-p2}}
\end{figure}

Calculations of $G_P$ for the $|\Delta I|=1$ transitions: $n \to p$ and $\Xi^- \to \Lambda$ in vacuum and in medium are presented in Figure~\ref{figGP-p1}.
Calculations for the $|\Delta S|=1$ transitions: $\Lambda \to p$ and $\Xi^- \to \Lambda$ are presented in Figure~\ref{figGP-p2}.
The results for the form factors $G_P^\ast$ are on the left side, and the ratios to the vacuum ($G_P$) are on the right side.
Notice the difference of about an order of magnitude for $G_P(0)$ between the $|\Delta I|=1$ or $|\Delta S|=1$ channels.

In both cases, one can observe the reduction in the function $G_P$ in nuclear medium.
The difference in magnitudes of the two channels is clear in the figures.
The values for $G_P(0)$ are about an order of magnitude larger for $|\Delta I|=1$ transitions than for $|\Delta S|=1$ transitions.

The calculations for the $|\Delta I|=1$ transitions (Figure~\ref{figGP-p1}) show a significant suppression in the nuclear medium.
This result is mainly a consequence of the dominance of the pole contribution discussed above.
Using this property, we can write
$\frac{G_P^\ast}{G_P} \simeq\left(\frac{M_{B'}^\ast + M_B^\ast}{M_{B'} + M_B} \right)^2 \frac{G_A^\ast}{G_A}$.
The reduction in the masses in medium (power 2) and the suppression of $G_A$ in medium both contribute  to the significant suppression of the function $G_P$ at low $Q^2$.
Notice that the suppression is more significant for light baryons ($n \to p$) than for heavy baryons ($\Xi^- \to \Lambda$).
For heavy baryons, the more significant suppression at low $Q^2$ is a consequence
of the meson cloud effects on $G_A$.    


In the case of the $|\Delta S|=1$ transitions (Figure~\ref{figGP-p2}), one can notice, apart the difference in magnitudes mentioned above, the slower falloff of $G_P$ with $Q^2$, a consequence of the dependence of pole term on the mass of the kaon.

In the low-$Q^2$ region, the suppression of $G_P$ in nuclear medium is clear.
This result is a consequence of the suppression on the meson cloud contribution, included in $G_A$, and the dominance of the pole term $\frac{G_P^\ast}{G_P} \approx  \left(\frac{M_{B'}^\ast + M_B^\ast}{M_{B'} + M_B} \right)^2 \left(\frac{M_K}{M_K^\ast} \right)^2 \frac{G_A^\ast}{G_A}$, where the impact of the baryon mass reduction in medium is partially canceled by the reduction in the in-medium \mbox{kaon mass}.

In the large-$Q^2$ region, the effect of the kaon mass is irrelevant, and we recover the result
$\frac{G_P^\ast}{G_P} \simeq \left(\frac{M_{B'}^\ast + M_B^\ast}{M_{B'} + M_B} \right)^2 \frac{G_A^\ast}{G_A}$,
where $G_A^\ast$ and $G_A$ are well approximated by their bare results.
As a consequence, there is a significant reduction in $\frac{G_P^\ast}{G_P}$ compared with unity (reduction due to the in-medium baryon mass reduction and the in-medium reduction in $G_A$)
for transitions associated with light baryons, like for $\Lambda \to p$, or almost no medium modifications  $\frac{G_P^\ast}{G_P} \approx 1$ for transitions associated with heavy baryons, like for $\Xi^- \to \Lambda$.

For the remaining transitions, the magnitude of the suppression is between the
magnitudes discussed here~\cite{GA-Medium1}.

\section{Applications  \label{secResults2}}

We consider now two useful extensions of the formalism discussed in the previous sections.
First, we consider calculations of the nucleon form factors for intermediate densities, below the normal nuclear matter density ($0 < \rho < \rho_0$). 
After that, we look for the extension of the calculations of the octet baryon form factors for more dense nuclear matter ($\rho > \rho_0$).
The first application can be used in the study of protons and neutrons bound to a nucleus.
The second application is suitable for studies of heavy-ion collisions and  cores of compact stars.

\subsection{Nucleon Bound to a Nucleus}

The study of the electromagnetic and axial structure of nucleons bound to a nucleus can be conducted considering the parametrizations of the hadron masses (baryons and mesons) and the values of baryon--meson coupling constants for the nucleus average density $\rho$.
In practice, we obtain a value between free space ($\rho = 0$) and the normal nuclear matter ($\rho = \rho_0$).
We follow the method discussed in Section~\ref{secMedium1} and calculate the hadron masses and the coupling constants for the average densities $\rho$ of some nuclei.
The average densities calculated by the QMC model~\cite{QMCEMFFMedium5} are presented in Table~\ref{tab-nucleus}.

\begin{table}[H] 
  \caption{Average densities of nucleus calculated by the QMC model~\cite{QMCEMFFMedium5}.
     \label{tab-nucleus}} 
\begin{tabularx}{\textwidth}{LC}
\toprule
\textbf{Nucleus}     &  \boldmath{$\rho/\rho_0$}     \\
\midrule
12C    &     0.533 \\
16O    &    0.612 \\
40Ca   &    0.711  \\
90Zr   &   0.767  \\
97Au   &    0.783 \\
208Pb  &    0.786 \\
\bottomrule
\end{tabularx}
\end{table}

\begin{figure}[H]
   \hspace*{-3mm}
    \begin{tabular}{ccc}

\includegraphics[width=1.7in]{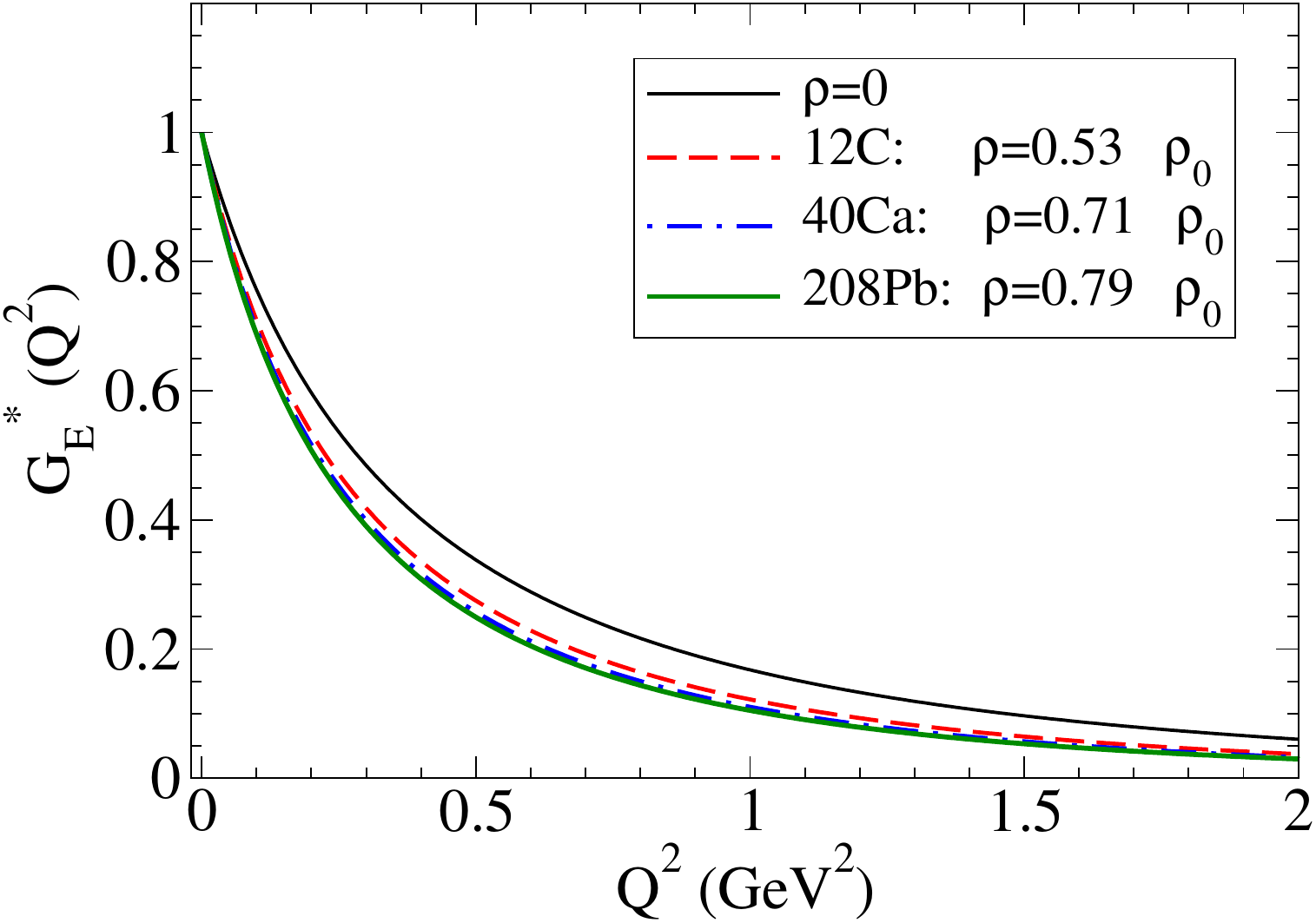} &
\includegraphics[width=1.7in]{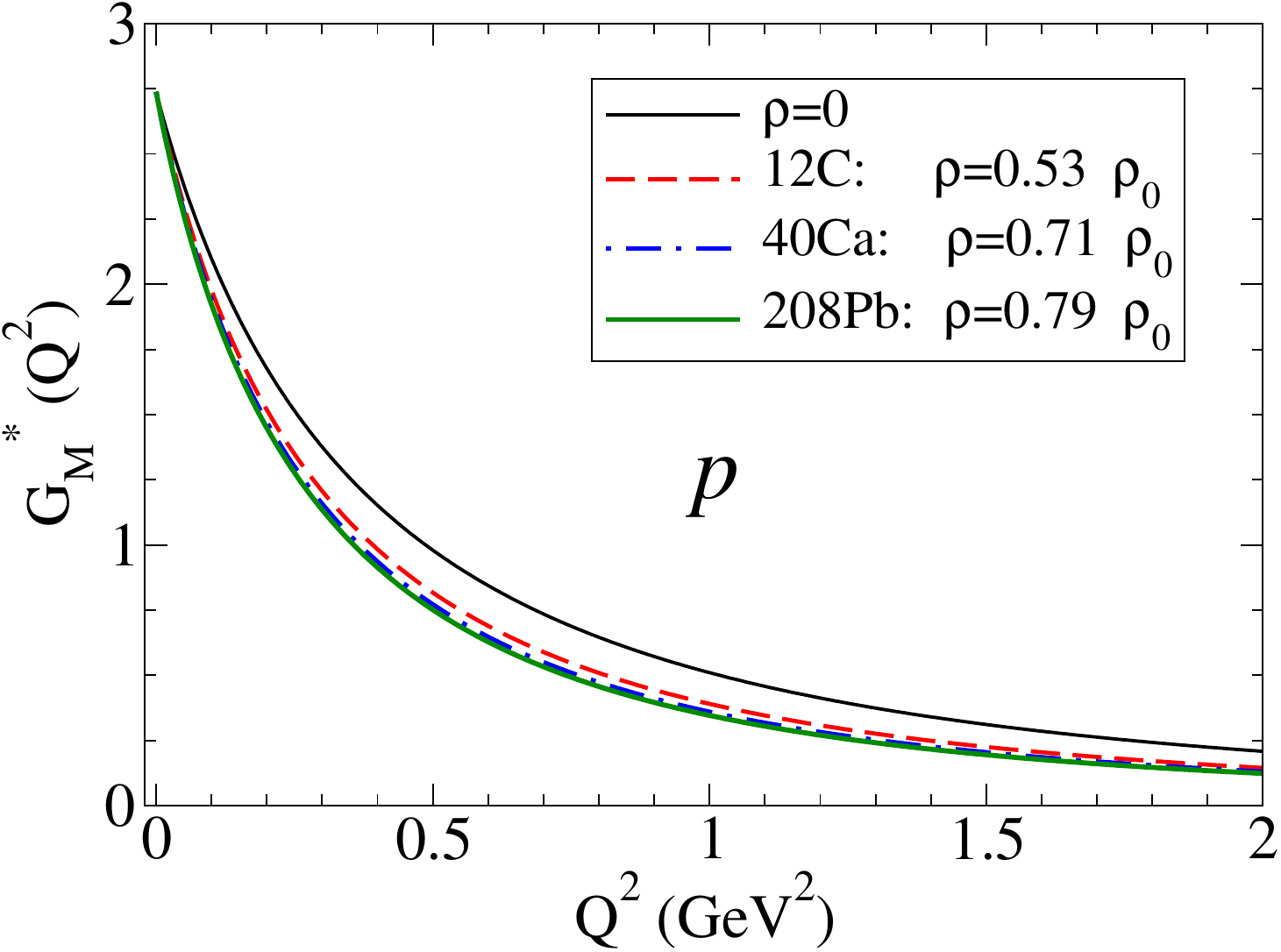}&
\includegraphics[width=1.7in]{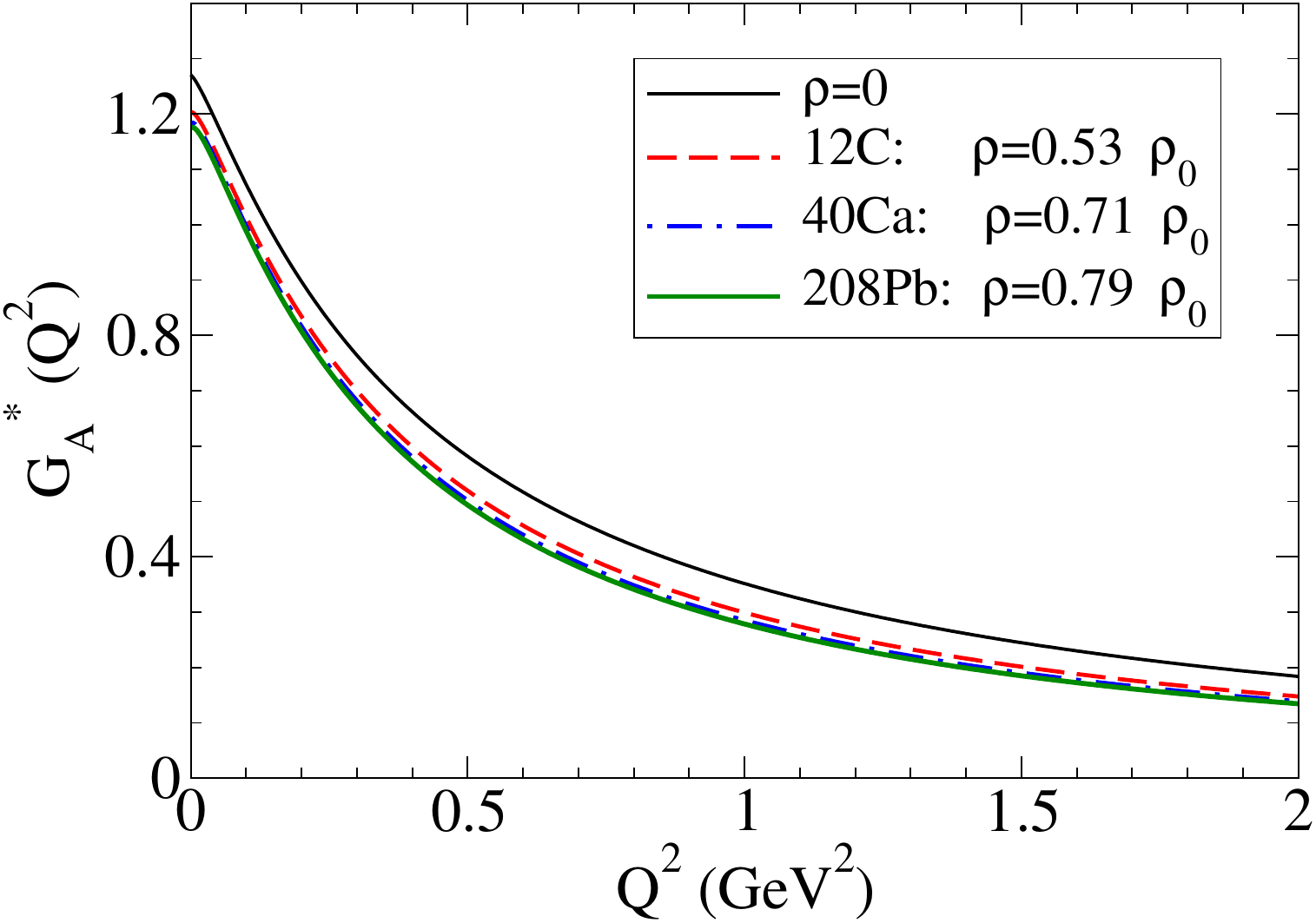}\\
\includegraphics[width=1.7in]{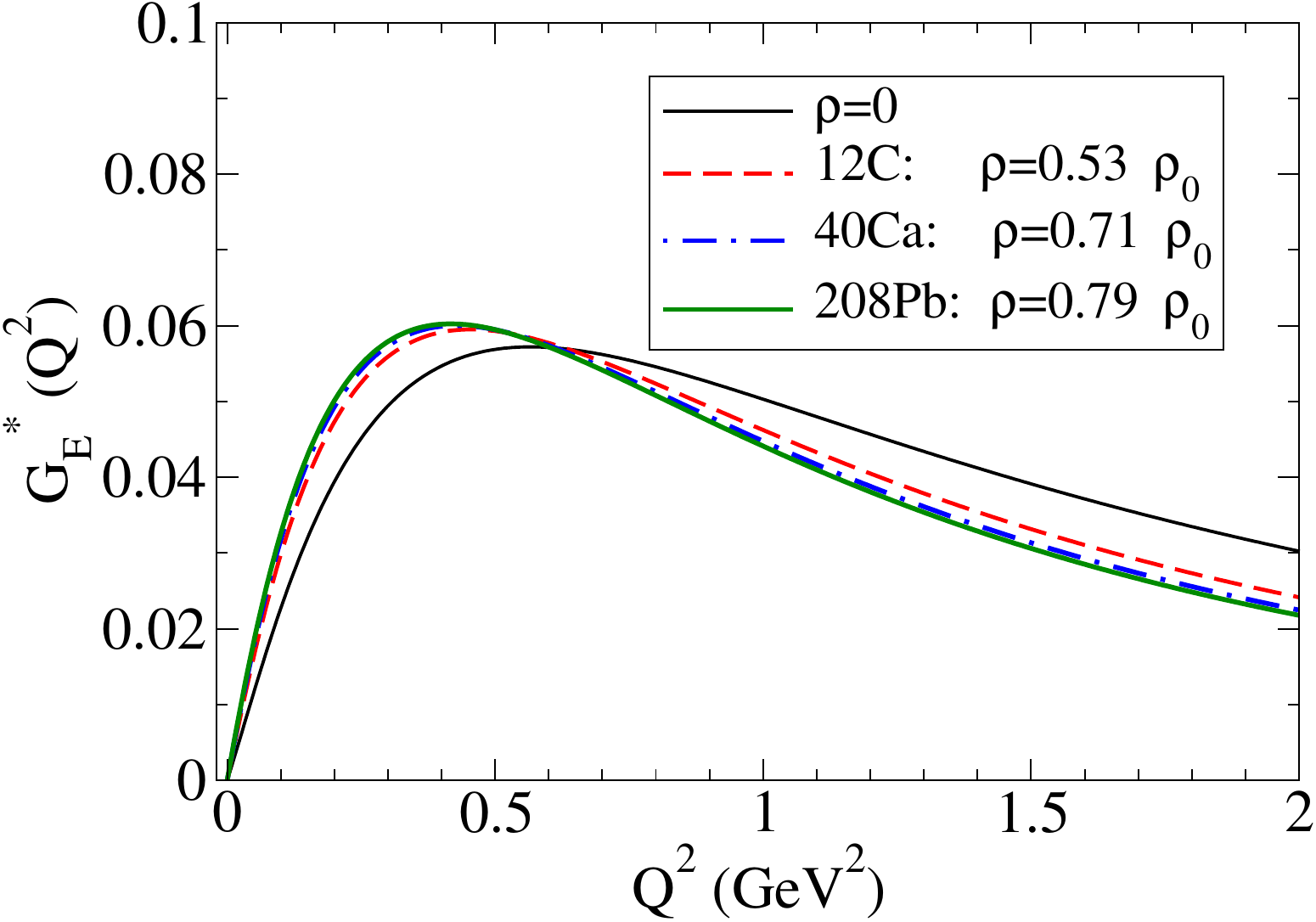} &
\includegraphics[width=1.7in]{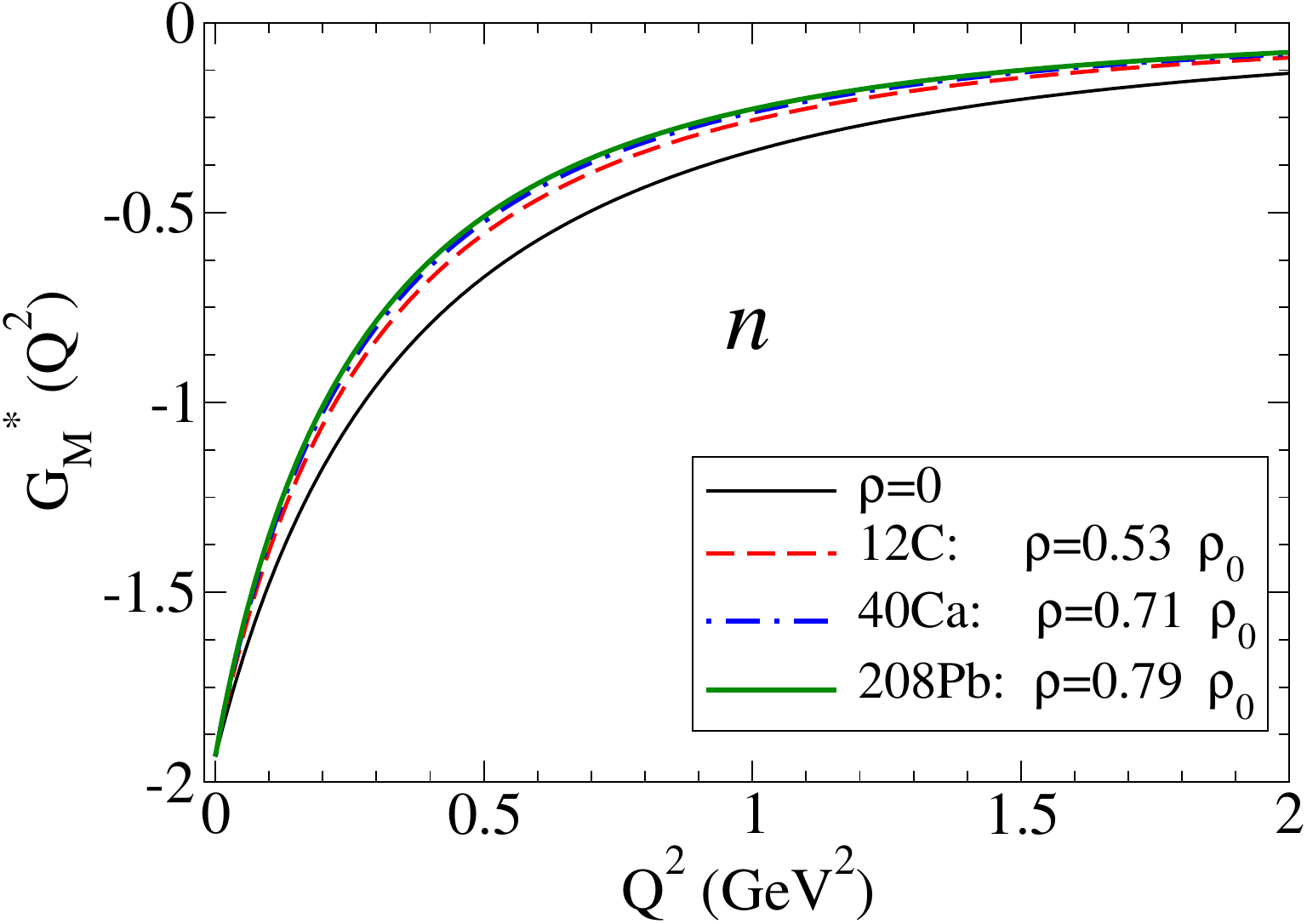}&
\includegraphics[width=1.7in]{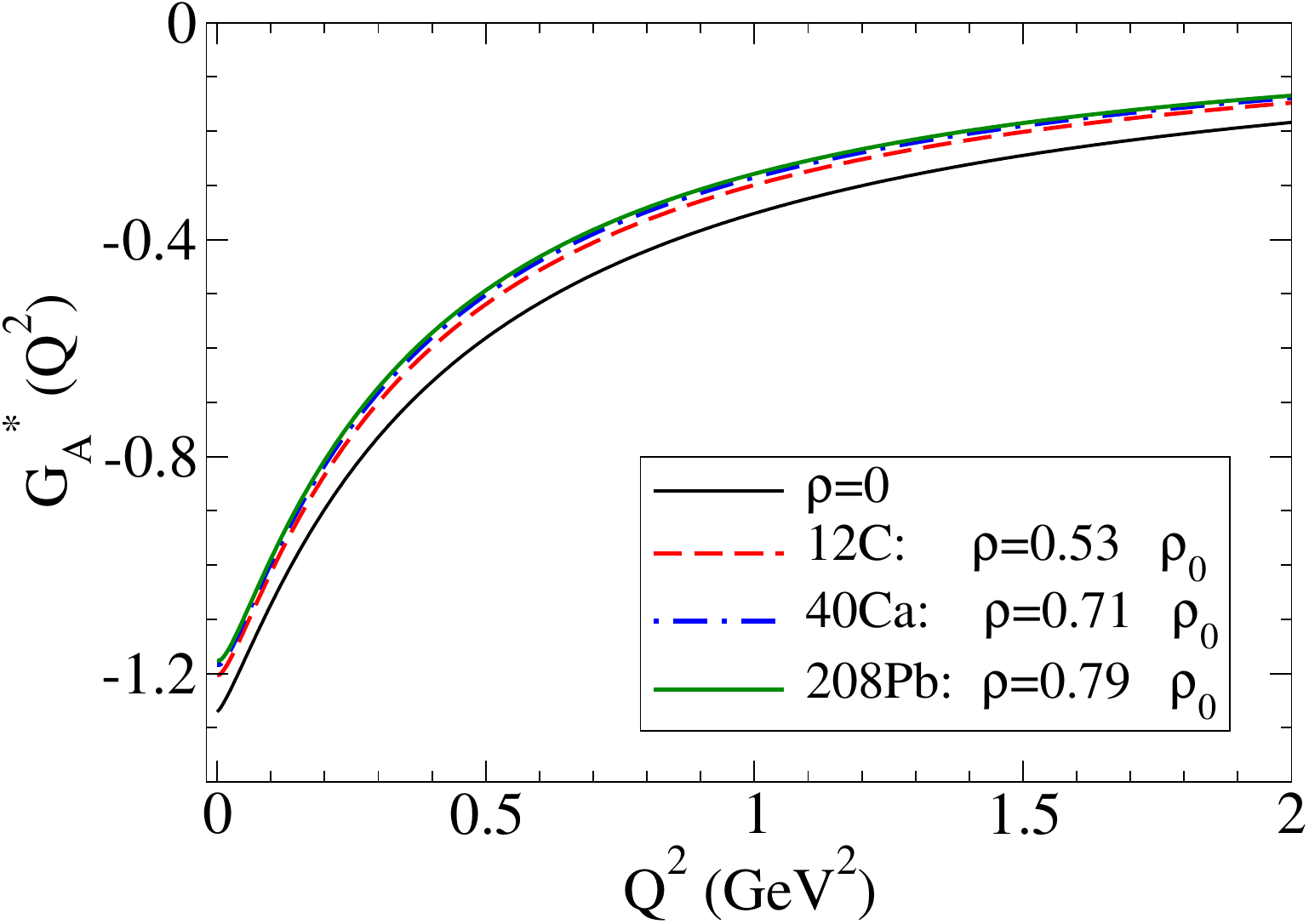}\\
    \end{tabular}
    \caption{$G_E$, $G_M$, and $G_A$ form factors for the nucleon $B=p,n$ bound to a nucleus.
\label{figNucleus}}
\end{figure}

In the discussion, we consider three typical cases: $^{12}$C (12C), $^{40}$Ca (40Ca),  and $^{208}$Pb (208Pb).
The form factors $G_E$, $G_M$,  and $G_A$ are presented in Figure~\ref{figNucleus}.
For the axial-vector form factors, we use the relations $G_A(p)= G_A( n \to p)$ and $G_A(n)= - G_A( n \to p)$, following Table~\ref{tab-Axial-Transitions}.

From the results in Figure~\ref{figNucleus}, we can conclude that the proton and the neutron electromagnetic and axial form factors are in general reduced in the nuclear medium compared to the free space ($\rho=0$).
The conclusion is exact for $Q^2 > 0$.
The exceptions are the results for $G_E$ and $G_M$ at $Q^2=0$.
The value of $G_E(0)$ is the same in medium and in vacuum because the charge is preserved in medium.
The result for $G_M(0)$ is approximately the same in medium and in vacuum due to the dominance of the valence quark contribution~\cite{Octet3}.
The explicit expressions for the bare contribution to $G_M(0)$ are presented in Table~\ref{tabGMB0}/Appendix~\ref{appBareEMFF}.


Our model calculations for $n$ and $p$ form factors for $\rho=0.5 \rho_0$ and $\rho= \rho_0$ have been used in the calculation of the $\nu p \to \nu p$, $\bar \nu p \to \bar \nu p$, $\bar \nu n \to \bar \nu n$ and $\bar \nu n \to \bar \nu n$ single-differential cross-sections, as well as the $\bar \nu p \to e^+ n$ and $\nu n \to e^- p$ single-differential cross-sections~\cite{GA-Medium1,Cheoun13b,Cheoun13a}. 
Here, $\nu$ represents the electron neutrino. 
From the calculations, one can conclude  that the in-medium single-differential cross-sections are suppressed compared with the free space,
and that the the suppression increases with the density $\rho$~\cite{GA-Medium1}.

\begin{figure}[H]

   \hspace*{-3mm}
    \begin{tabular}{ccc}
\includegraphics[width=1.7in]{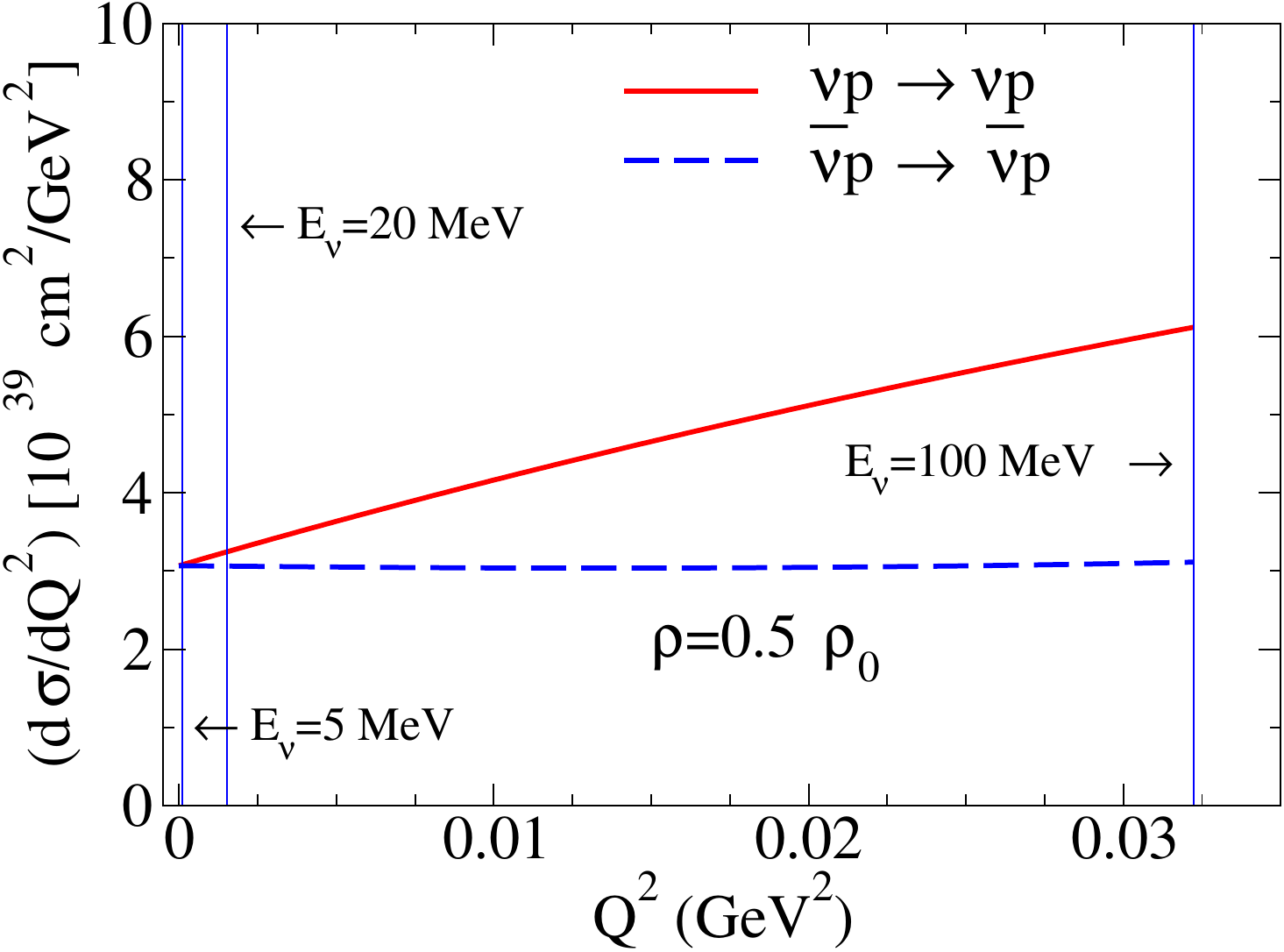} &
\includegraphics[width=1.7in]{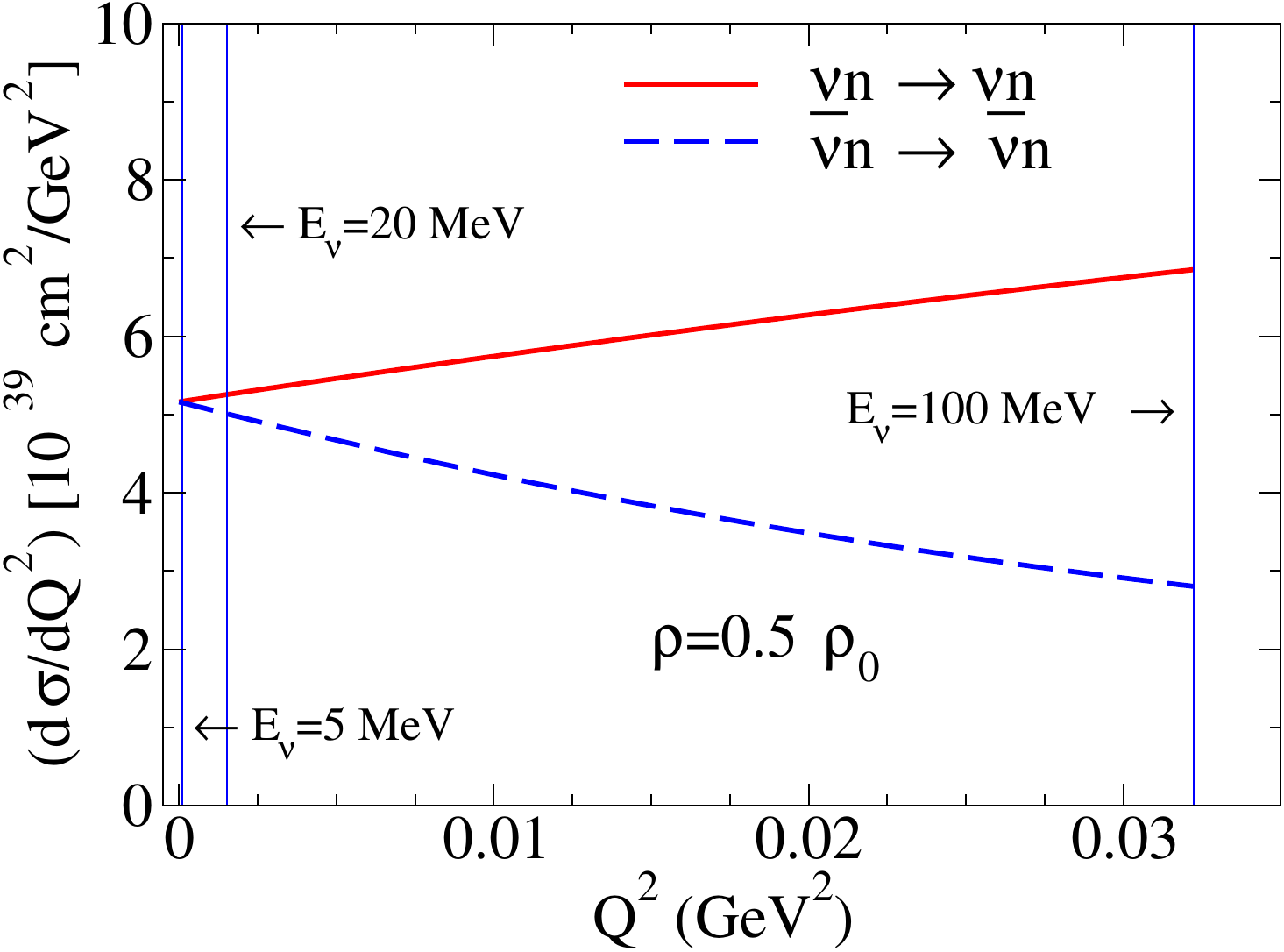}&
\includegraphics[width=1.7in]{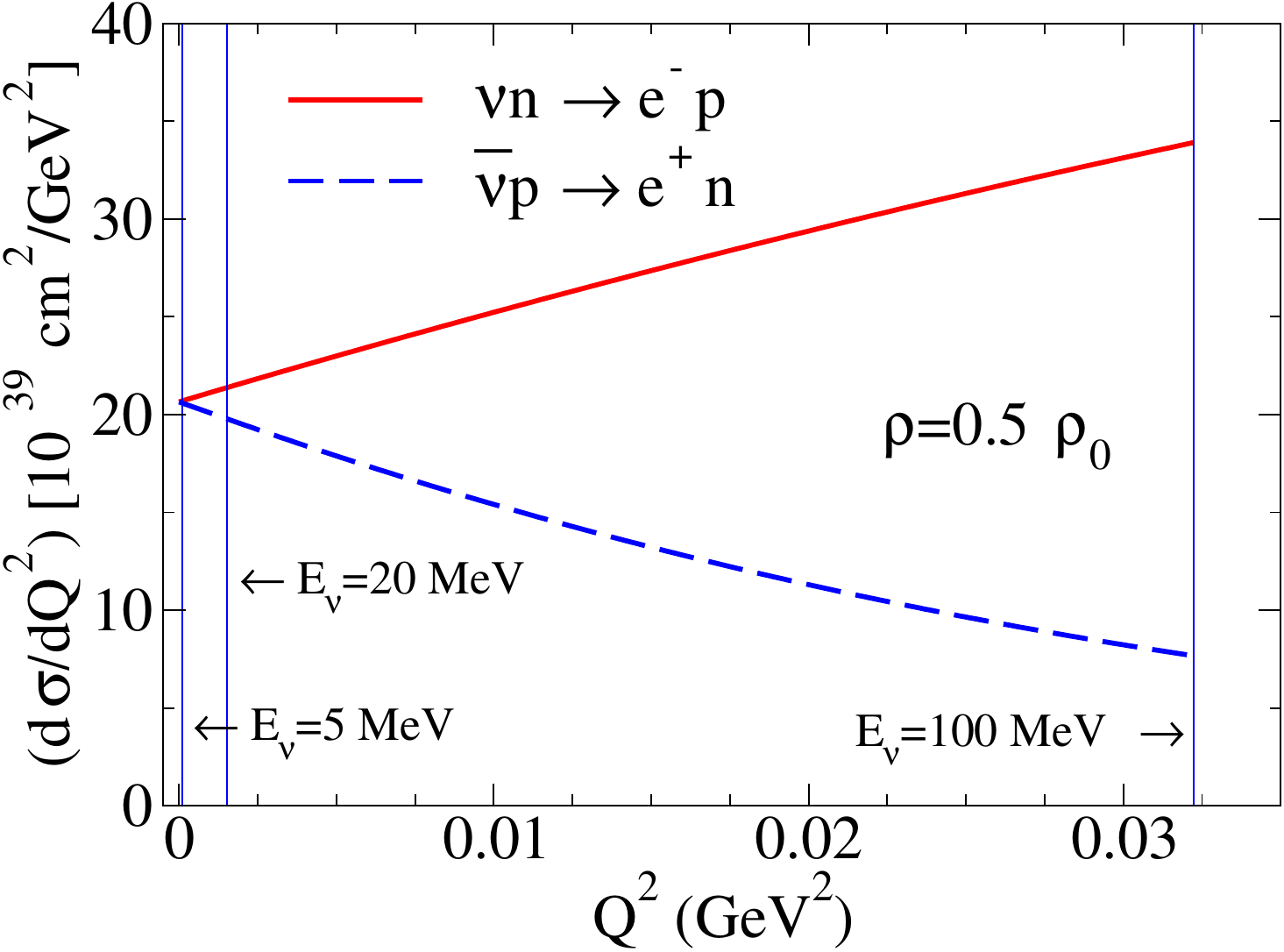}\\
   \end{tabular}
    \caption{Single-differential 
 cross-sections in terms of $Q^2$~\cite{GA-Medium1}.
    Comparison between $\nu p \to \nu p$ and  $\bar \nu p \to \bar \nu p$
    (left); $\nu n \to \nu n$ and  $\bar \nu n \to \bar \nu n$
    (center); and
    $\nu n \to e^- p$ and  $\bar \nu p \to e^+ n$ (right)
    for the neutrino/antineutrino energy $E_\nu = 100$ MeV.
    The nuclear matter density is $\rho= 0.5 \rho_0$.
    The vertical lines indicate the upper limit in $Q^2$ for $E_\nu =5$ MeV, $E_\nu= 20$ MeV, and $E_\nu= 100$ MeV.
\label{fig-cross-section}}
\end{figure}

The numerical calculations for the neutrino energy of $E_\nu =100$ MeV, a typical energy for neutrinos ($Q^2 \lesssim 0.035$ GeV$^2$), for a density $\rho=0.5 \rho_0$ are presented in Figure~\ref{fig-cross-section}.
Higher neutrino energies are achieved in experiments with KDAR neutrinos, muon neutrinos generated by kaon decay, where the neutrino energy is $E_\nu = 236$ MeV  ($Q^2 \lesssim 0.15$ GeV$^2$)~\cite{KDAR}.
Within our framework the electroweak form factor calculations can be extended to larger neutrino energies $E_\nu$, corresponding to larger $Q^2$ values (in the massless 
neutrino limit, the maximum value of $Q^2$ is determined by $Q^2= \frac{4M^2 E_\nu}{M + 2 E_\nu}$, where $M$ is the nucleon effective mass)~\cite{GA-Medium1}.

The comparison between $\nu N \to \nu N$ and $\bar \nu N \to \bar \nu N$ cross-sections
presented in Figure~\ref{fig-cross-section}
is consistent with the well-known dominance of the neutrino--nucleon cross-section dominance over the antineutrino--nucleon cross-sections observed in BNL and other experiments~\mbox{\cite{Ahrens87a,Cheoun13a,Cheoun13b,Athar-Review22,Alberico02a}.}

The numerical calculations of  nucleons $G_E$, $G_M$, and $G_A$ form factors can  also be used in the calculation of the coherent neutrino scattering on nucleus, in the low-$Q^2$ limit, defined by $ \sqrt{Q^2}  (2R) \lesssim  1$,
  where $R$ is radius of the nucleus~\cite{Snowmass22a,Akimov22a,Simons22a,Hoferichter20a,Abdullah20a,Athar-Review22,Gondolo02a,Cadeddu23a}.

\subsection{Hyperons in Dense Nuclear Matter}  


We now discuss  calculations for the electromagnetic and axial-vector form factors of the octet baryon members for densities larger than $\rho_0$.
We follow the formalism from Section~\ref{secMedium1}, with an alternative model for the pion-decay constant $f_\pi^\ast$, where the original expression derived from  perturbation theory~\cite{ChPT97} is modified for $\rho > \rho_0$,  in order to saturate at a certain value of $\rho$ near $3 \rho_0$~\cite{InPreparation}.
As an example, we consider calculations up to $\rho= 2 \rho_0$.

\begin{figure}[H]

   \hspace*{-3mm}
    \begin{tabular}{ccc}
\includegraphics[width=1.7in]{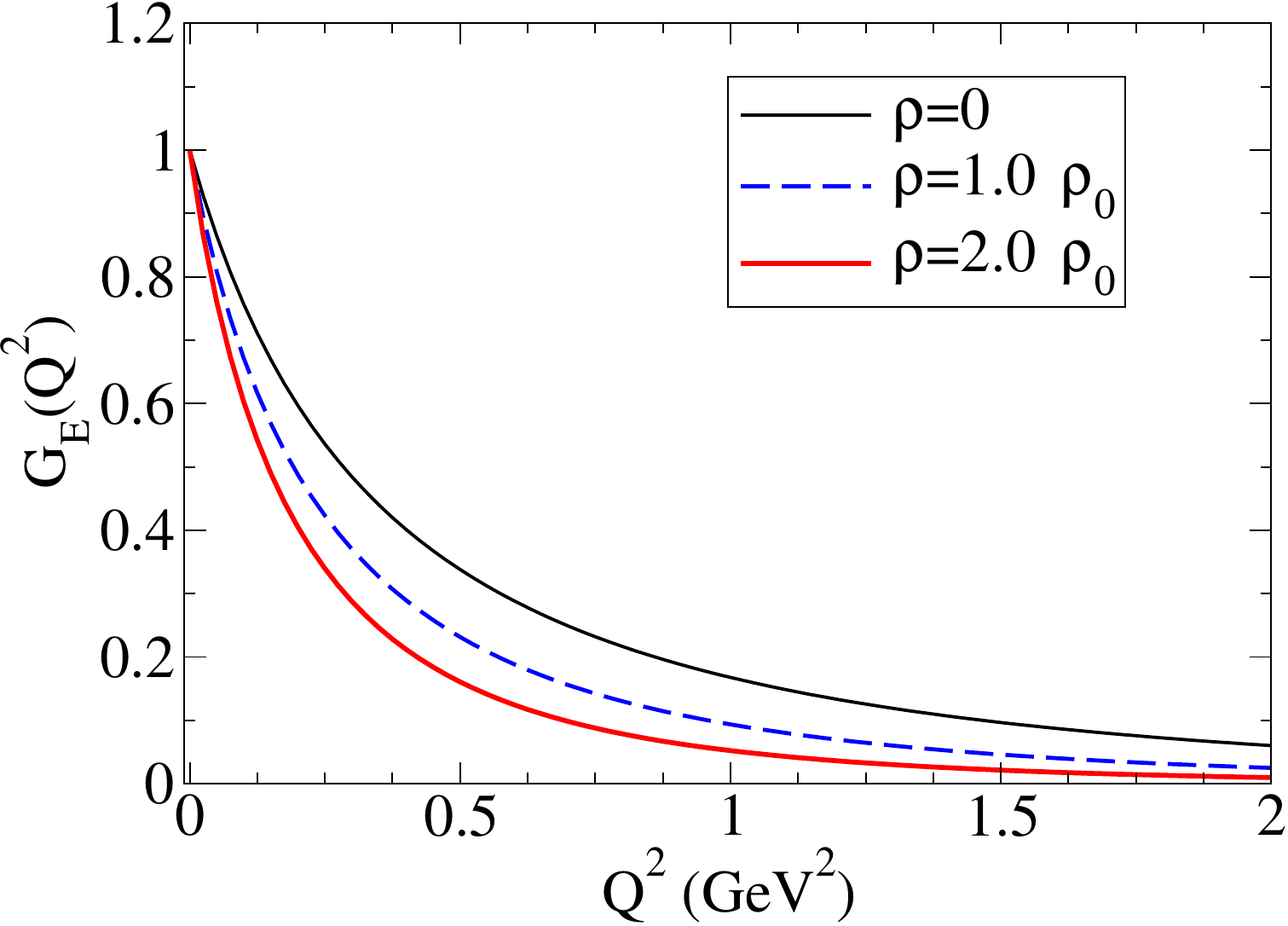}&
\includegraphics[width=1.7in]{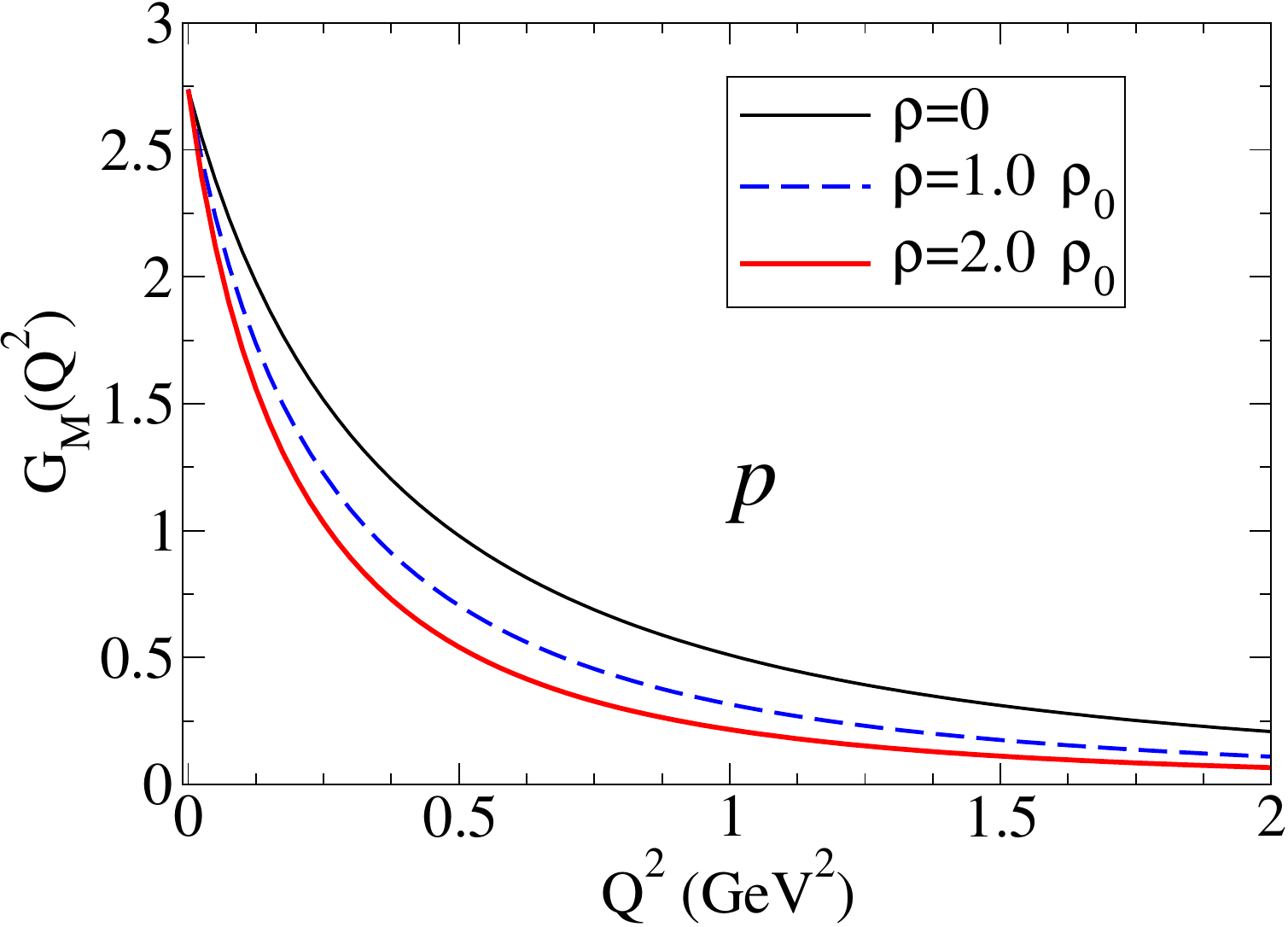}&
\includegraphics[width=1.7in]{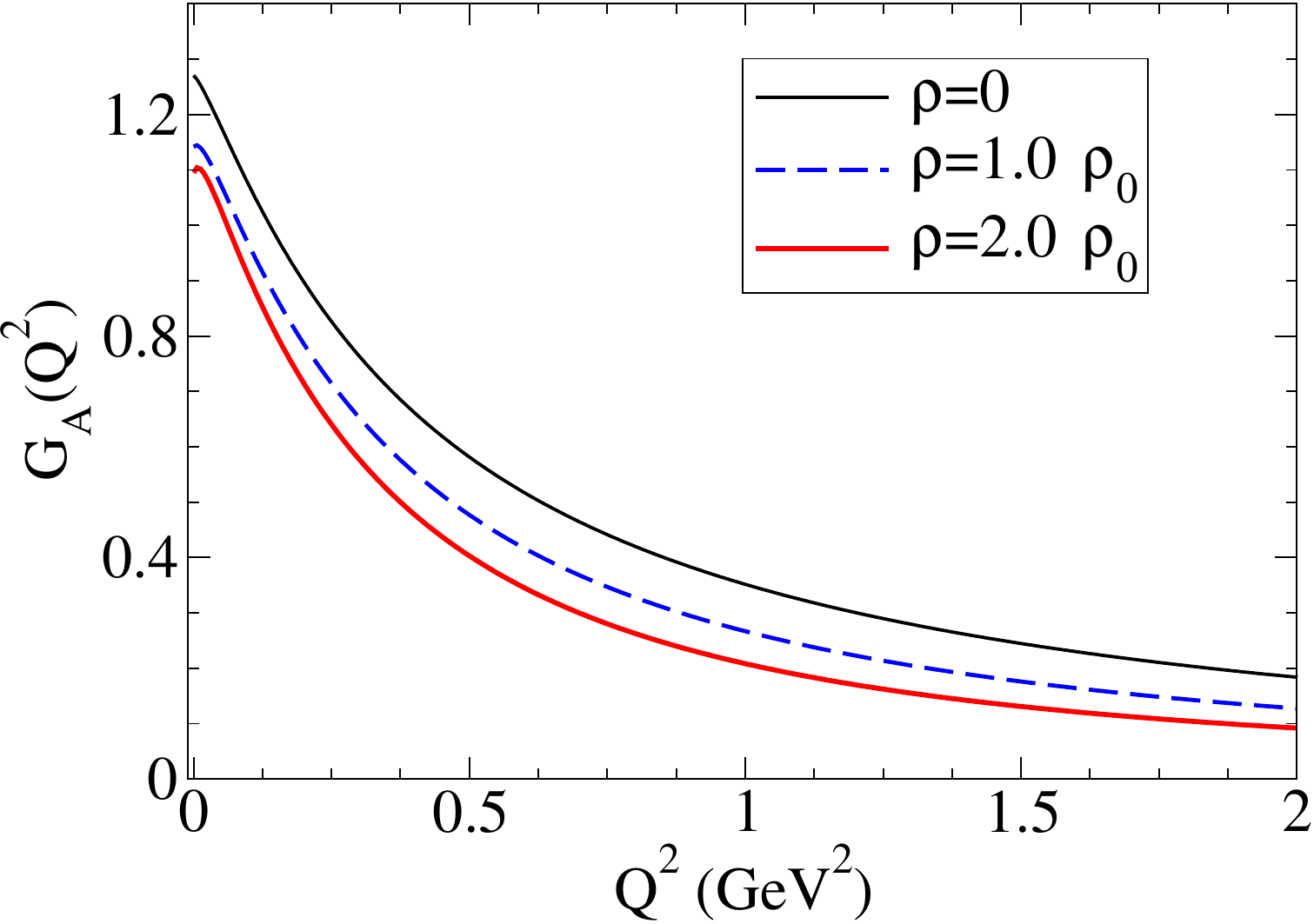} \\
\includegraphics[width=1.7in]{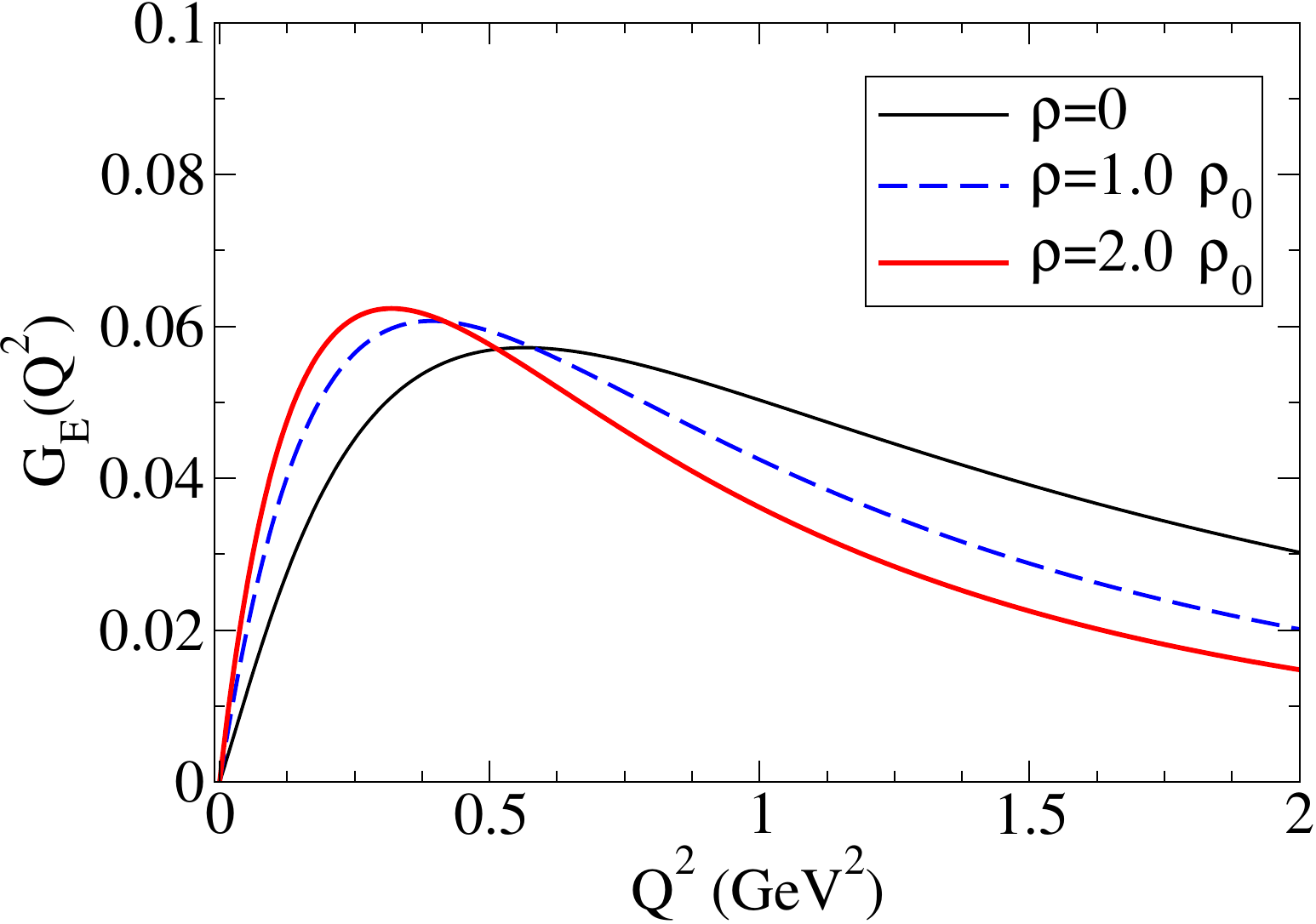} &
\includegraphics[width=1.7in]{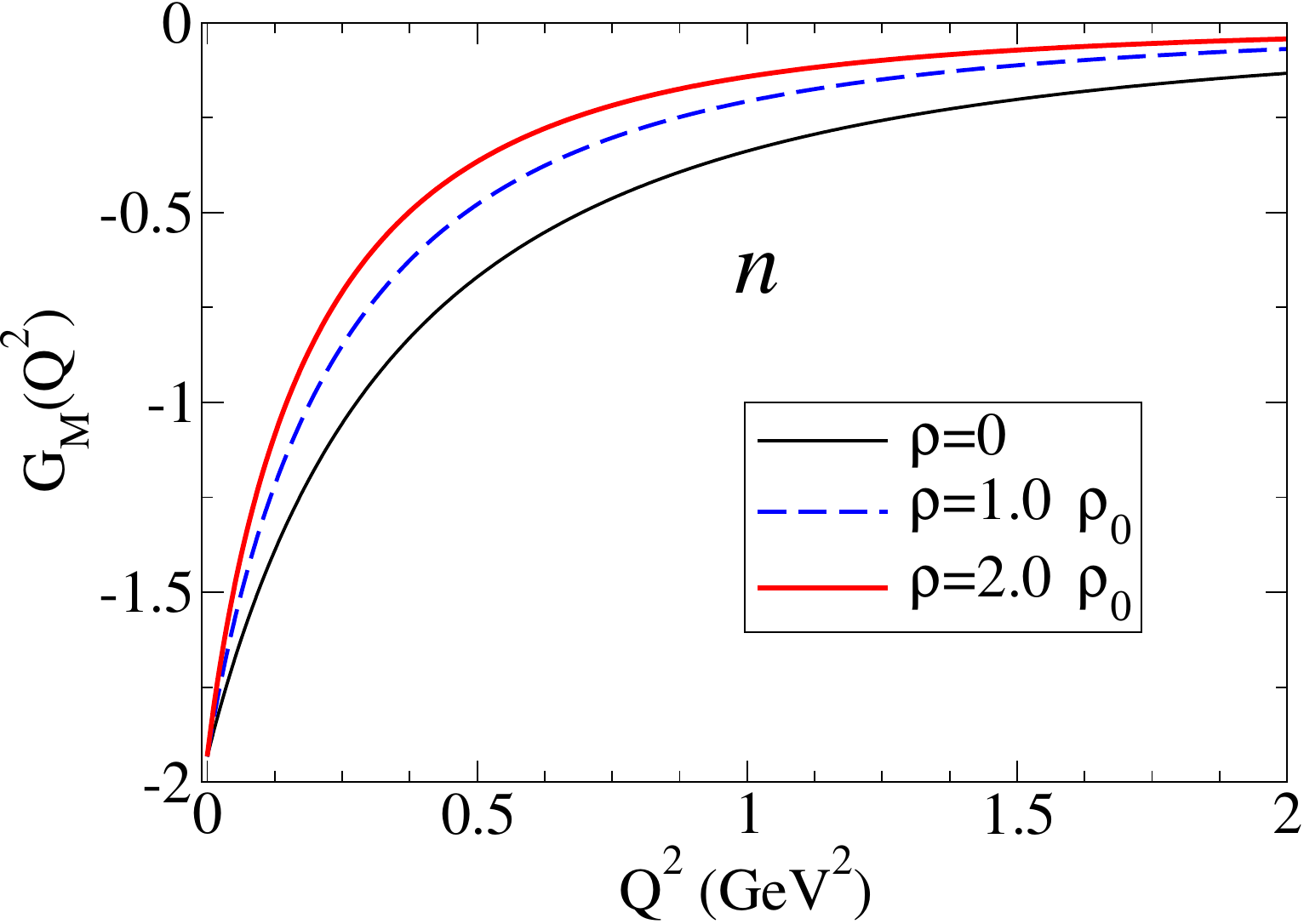}&
\includegraphics[width=1.7in]{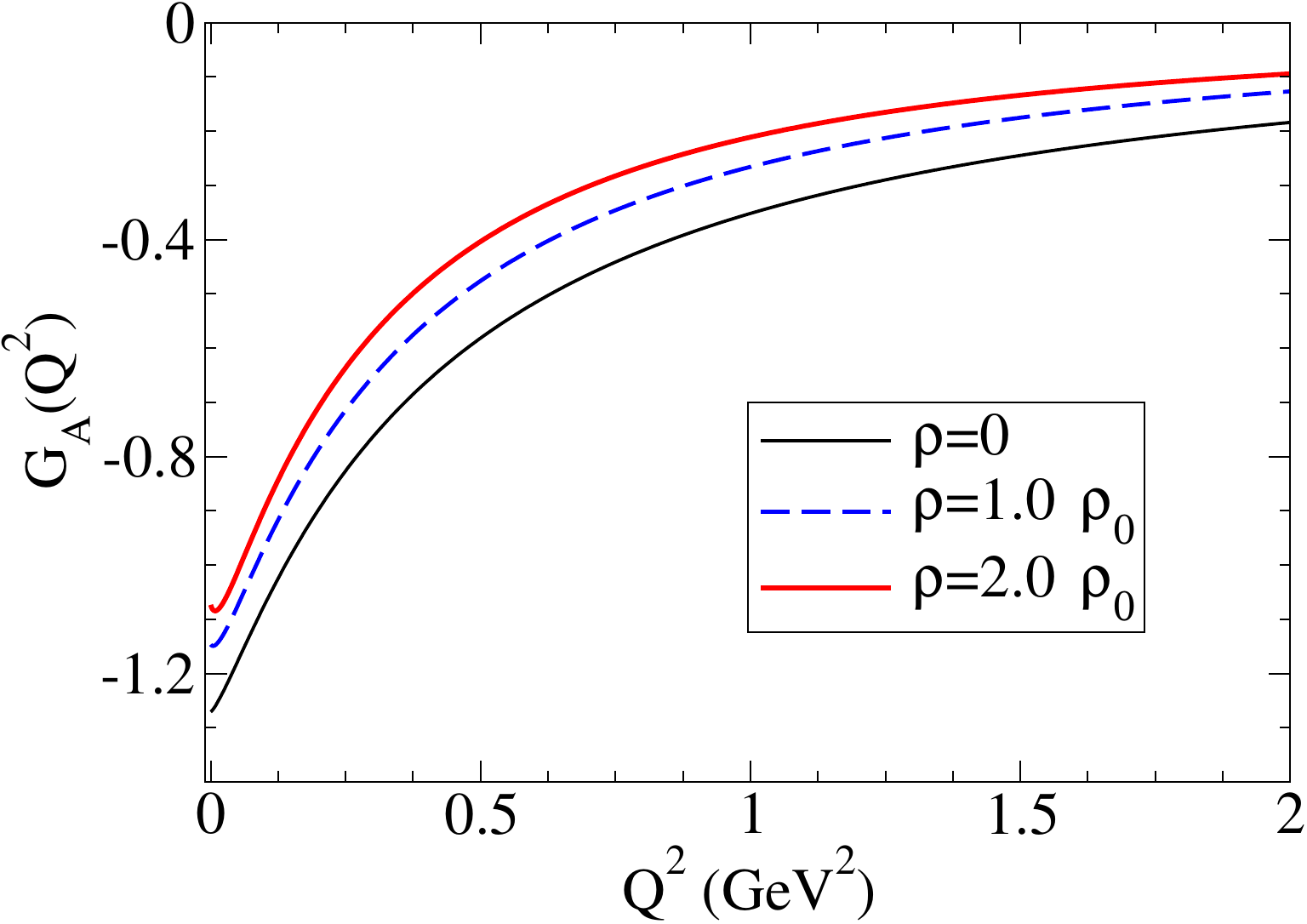}\\
\end{tabular}
\caption{$G_E$, $G_M$, and $G_A$ form factors
  for the nucleon bound to a nucleus ($B=n,p$).
\label{fig-Hyperons1}}
\end{figure}

\begin{figure}[H]

   \hspace*{-3mm}
    \begin{tabular}{ccc}
\includegraphics[width=1.7in]{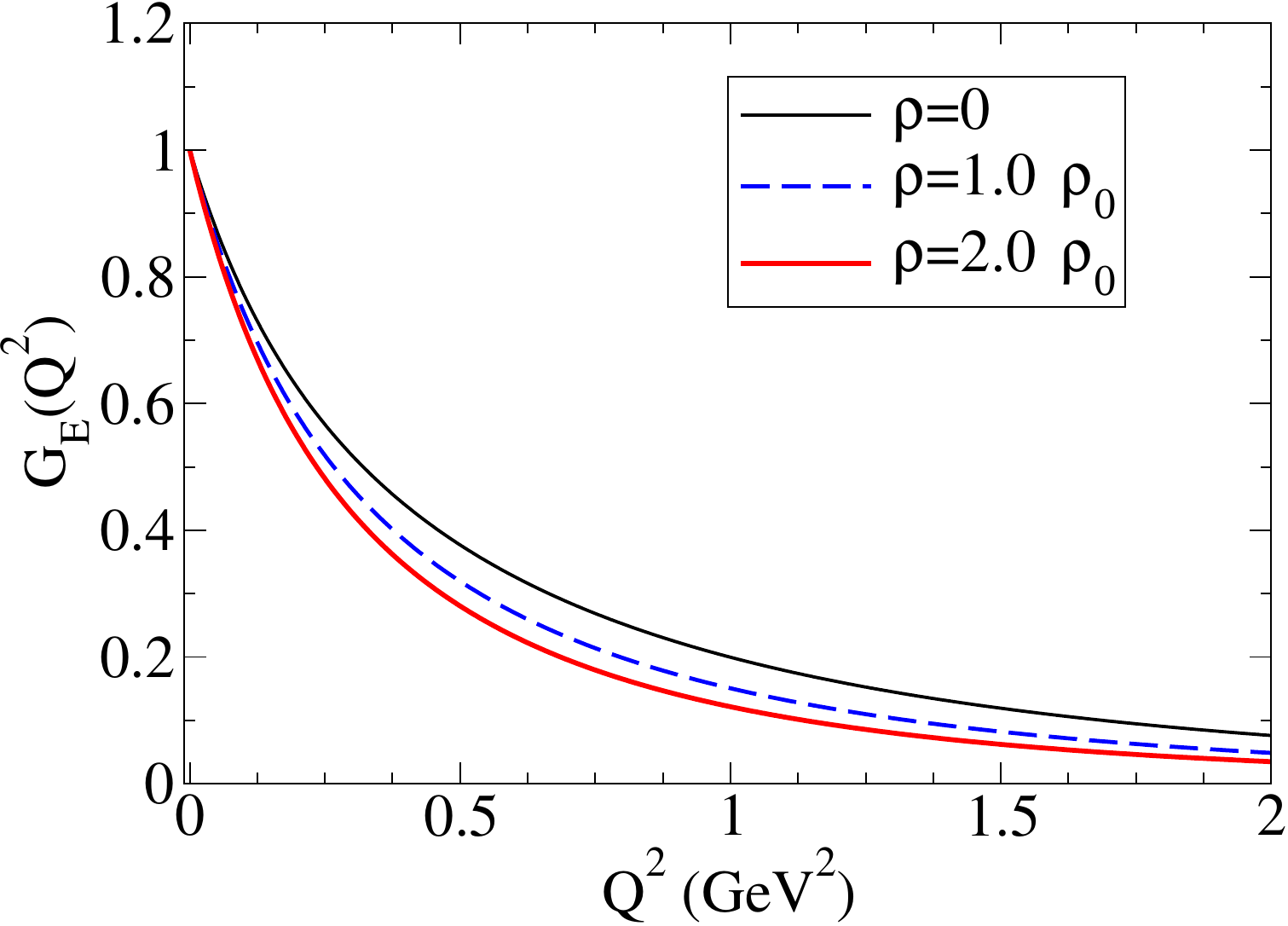} &
\includegraphics[width=1.7in]{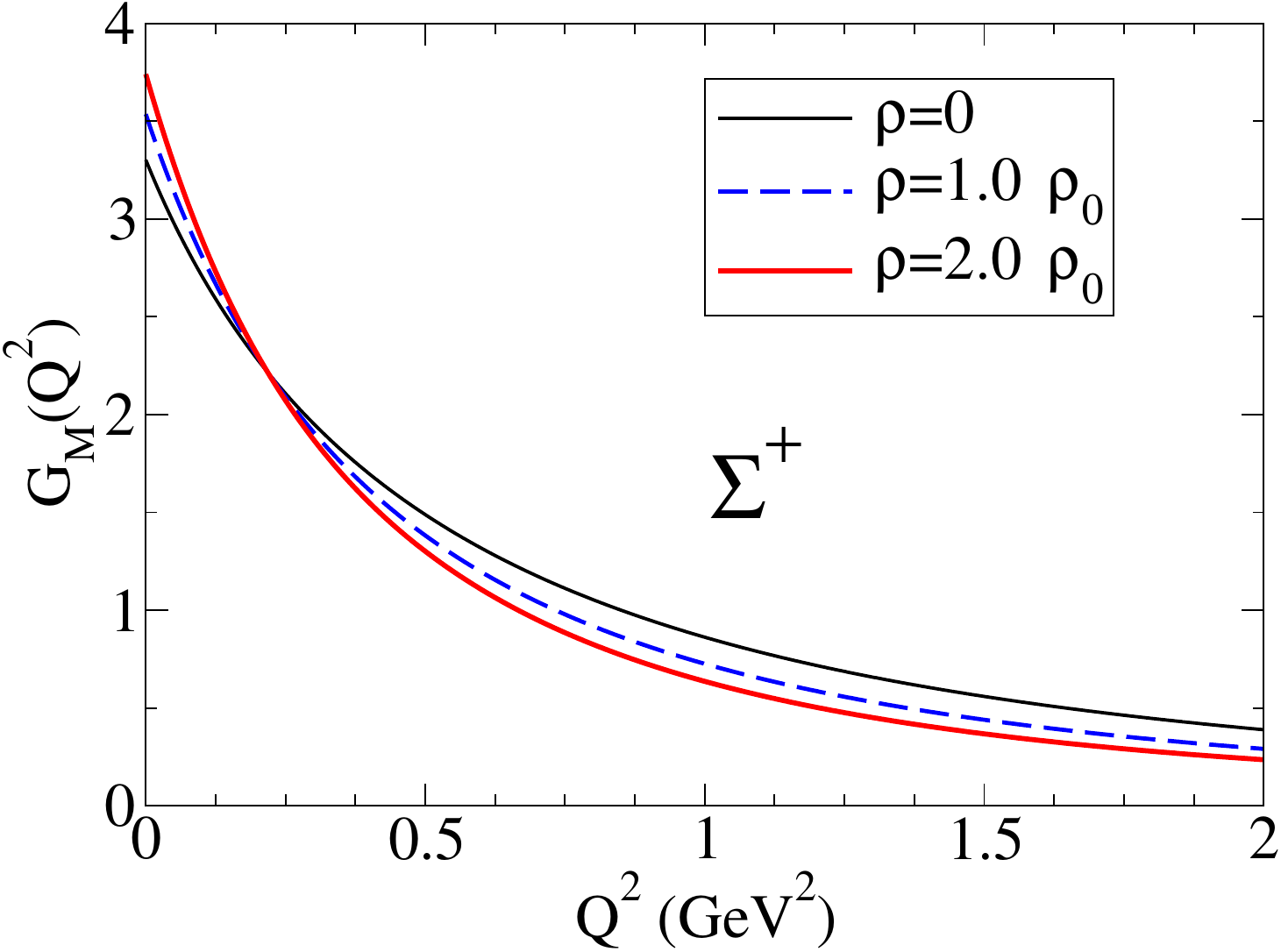} &
\includegraphics[width=1.7in]{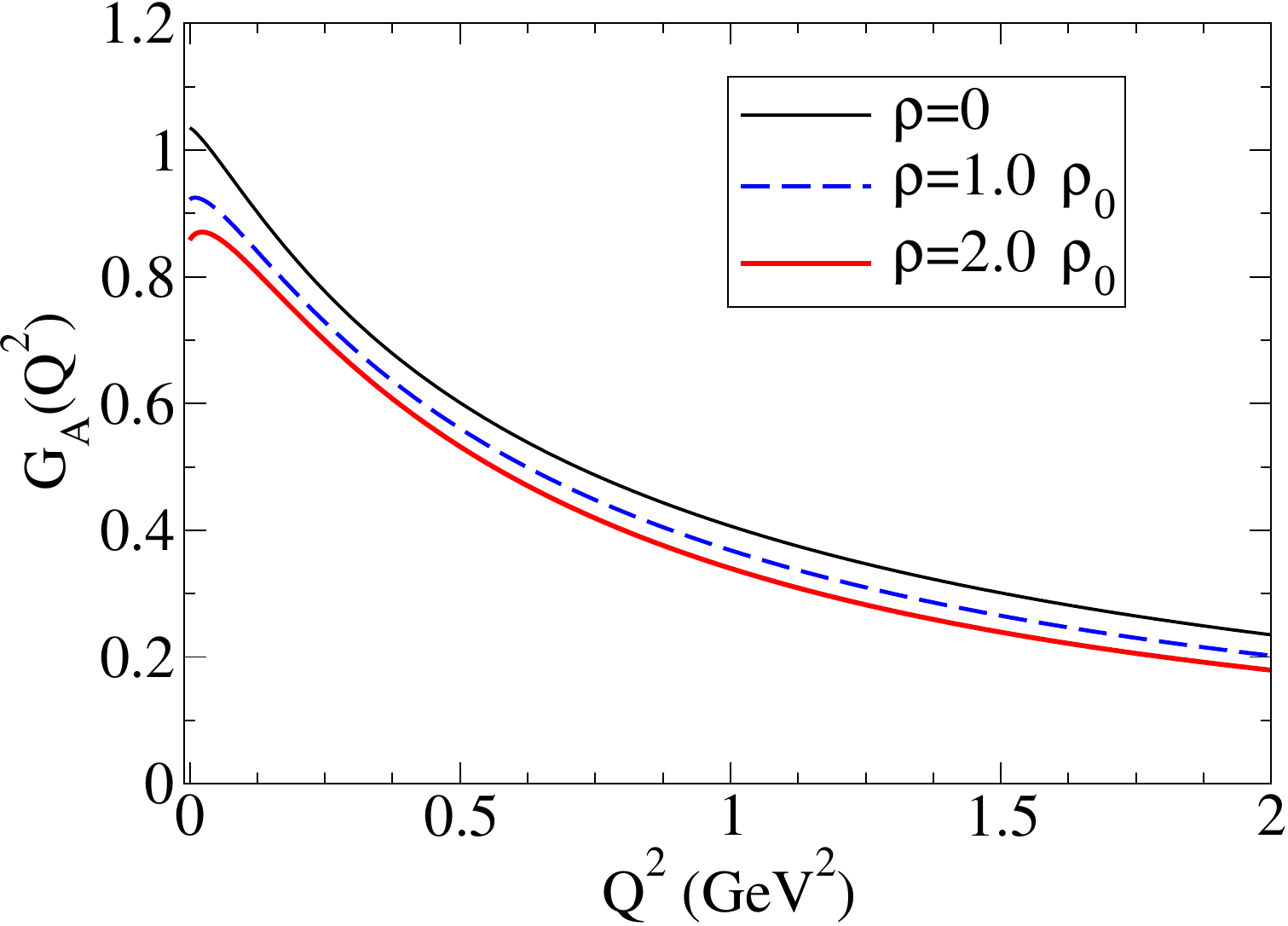} \\
\includegraphics[width=1.7in]{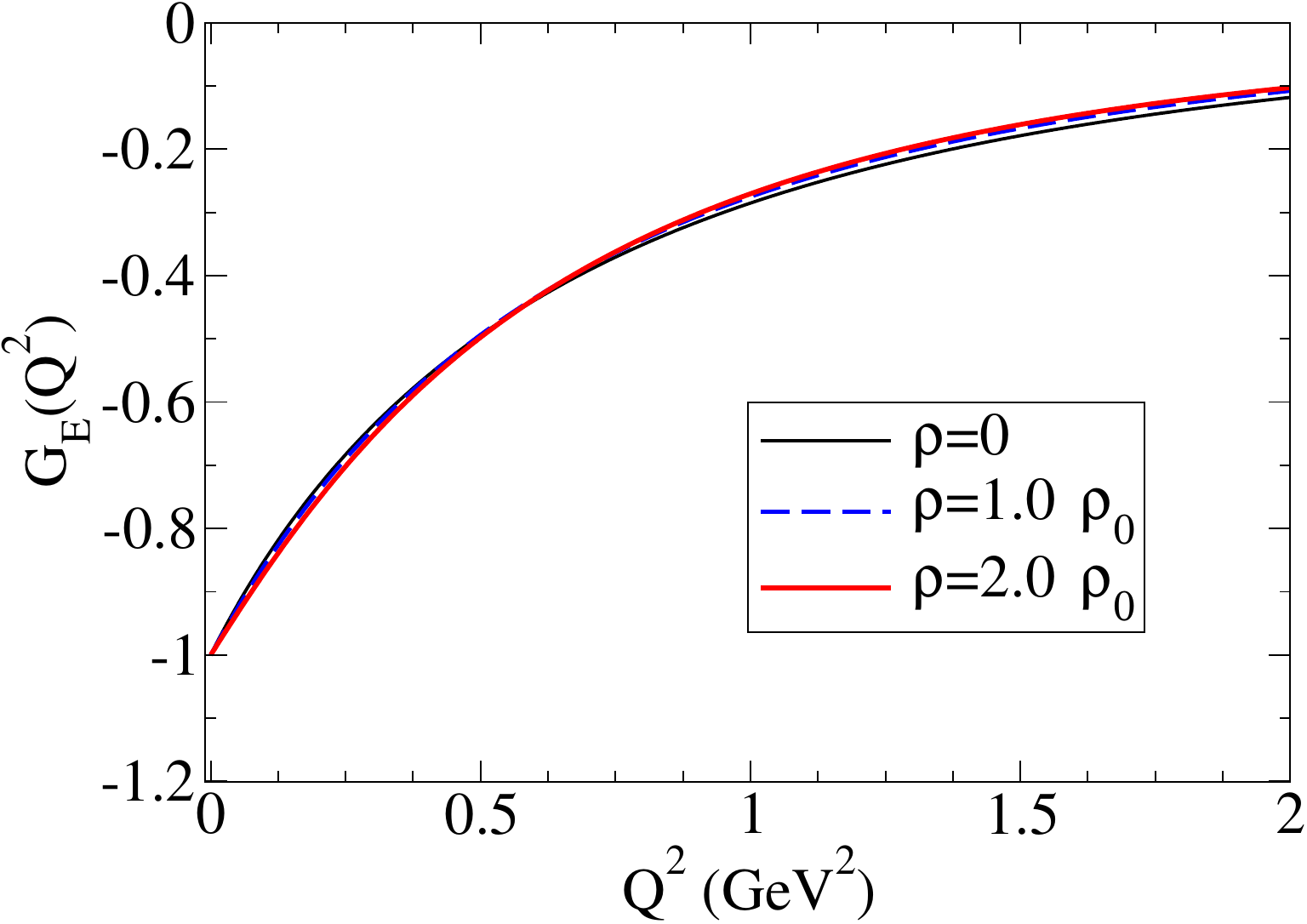} &
\includegraphics[width=1.7in]{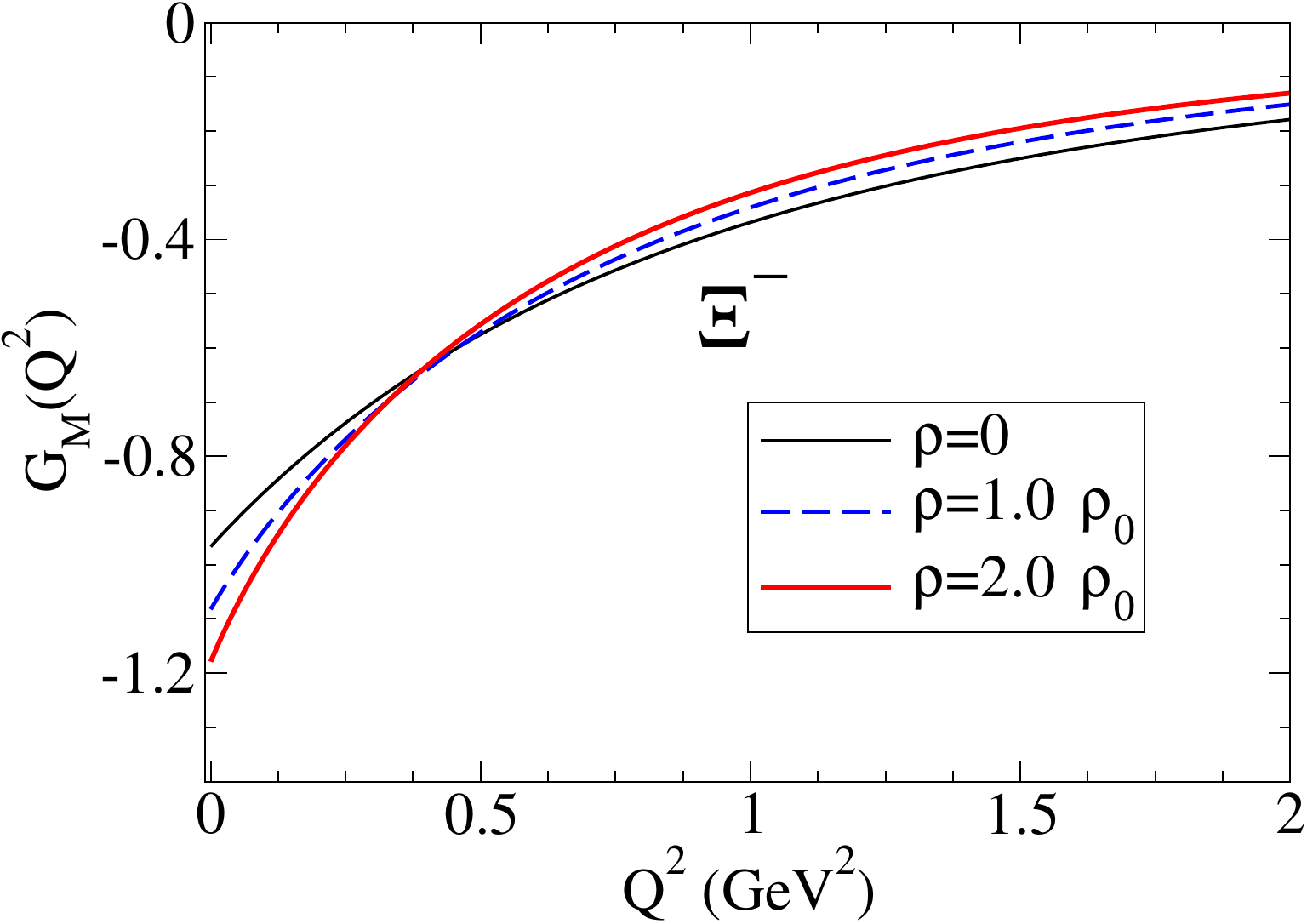} &
\includegraphics[width=1.7in]{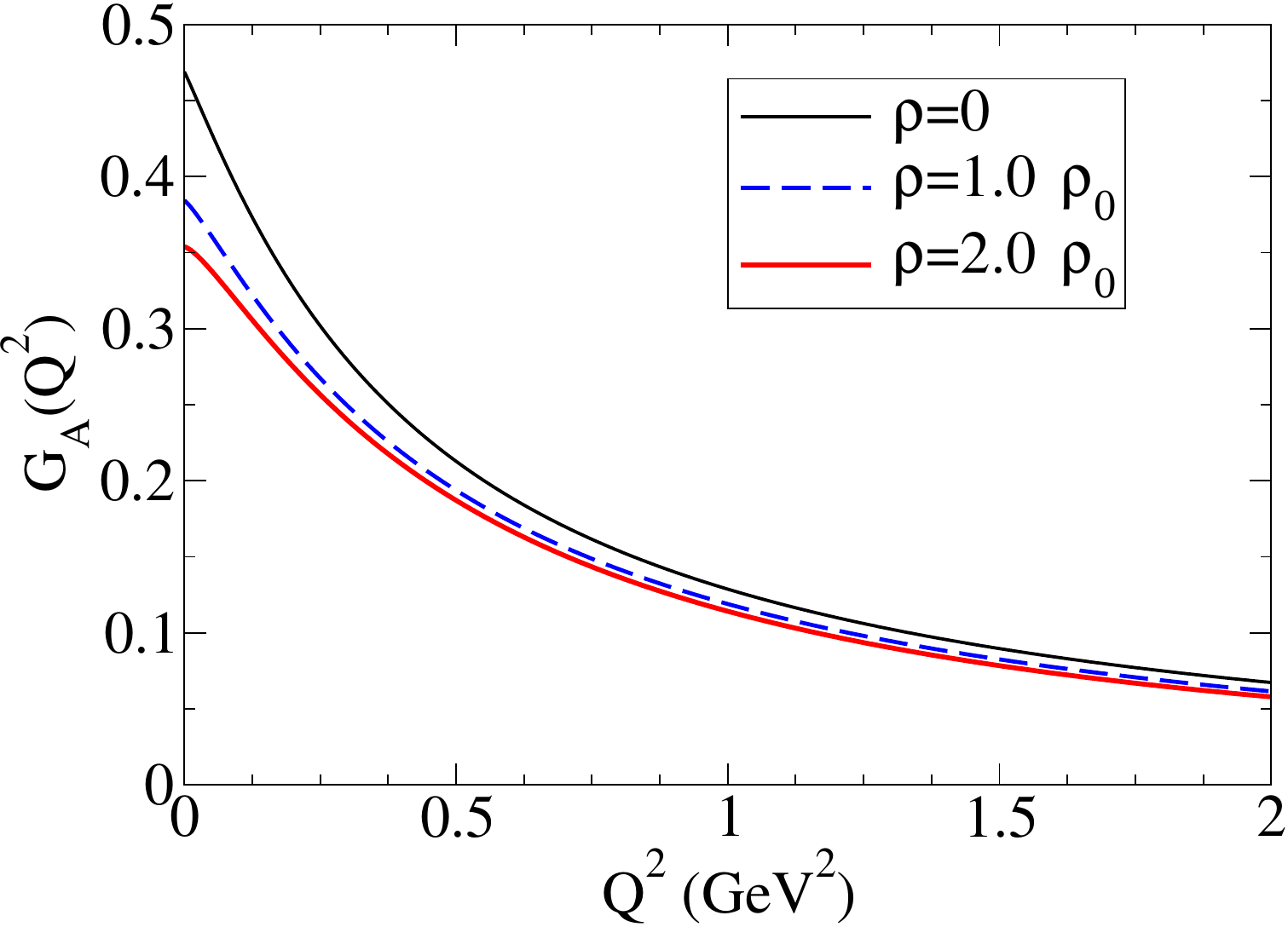}\\
\end{tabular}
\caption{$G_E$, $G_M$, and $G_A$ form factors 
  for $\Sigma^+$ and $\Xi^-$ bound to a nucleus.
\label{fig-Hyperons2}}
\end{figure}

The calculations for the nucleon ($p$ and $n$) are presented in Figure~\ref{fig-Hyperons1}, and the calculations for the $\Sigma^+$ and $\Xi^-$ are presented in Figure~\ref{fig-Hyperons2}.
For the last case, we used $G_A (\Sigma^+)= \sqrt{2} G_A( \Sigma^0 \to \Sigma^+)$ and $G_A(\Xi^-)= G_A(\Xi^- \to \Xi^0)$ (see Table~\ref{tab-Axial-Transitions}).
Similar relations can be derived for $G_A(\Sigma^-)$  and $G_A(\Xi^0)$.


From the results for the nucleon (Figure~\ref{fig-Hyperons1}), one can confirm the trend observed for $\rho < \rho_0$.
Except for the electric form factor of the neutron, the form factors
are reduced (quenched) in the nuclear medium.

{
  In Figure~\ref{fig-Hyperons1}, one can notice that the electric form factor of the neutron is suppressed for $Q^2 > 1$ GeV$^2$ when compared with the function in free space.
  The same happens to the proton $G_E$ and $G_M$ form factors and to the neutron $G_M$ form factor.
These results are a consequence of the dominance of the valence quark contributions on the nucleon electromagnetic form factors for large $Q^2$.
In these conditions, the electromagnetic form factors are determined by the quark form factors, parametrized in a vector meson dominance form in terms of vector meson poles in nuclear medium (see Appendix~\ref{appBareEMFF}).
The suppression of the form factors in medium is then the consequence of the reduction in the vector meson masses in medium.
The functions associated with lower masses (higher densities) are suppressed relative to the functions associated with larger masses (lower densities).}

As for the hyperons (Figure~\ref{fig-Hyperons2}), one has a similar reduction in form factors, except for the magnetic form factors, which are enhanced at low $Q^2$ and suppressed at large $Q^2$.
In the present model, the enhancement of $G_M$ is a consequence of the dominance of bare contribution to $G_M$ near $Q^2=0$.
Near $Q^2=0$, the bare contribution to $G_M(0)$ is enhanced if the mass
of the baryon $B$ in medium ($M_B^\ast$) is larger than the nucleon mass $M_N^\ast$ in medium ($\frac{M_B^\ast}{M_N^\ast} > 1$).
The explicit expressions for the bare contributions to 
$G_M^\ast (0)$ are presented in Table~\ref{tabGMB0}/Appendix~\ref{appBareEMFF}. 
The trend changes for larger values of $Q^2$, as can be observed in the figures.

The medium effects on $\Sigma^-$ and $\Xi^0$ for $G_M$ and $G_E$ can be inferred from the results for $\Sigma^+$ and $\Xi^-$ using the SU(3) symmetry, a good approximation in these cases.
Notice for $\Xi^-$ in Figure~\ref{fig-Hyperons2} the very mild dependence of the electric form factor in $G_E$.
Systems with more strange quarks tend to be less affected in nuclear medium.

The discussion of $G_M$ changes when we convert the results to nuclear magneton units in the free space.
In that case, $G_M$ is, in general, enhanced at low $Q^2$~\cite{Octet3}.

The present formalism, combined with model parametrizations of the neutrino/anti-neutrino flux, can be used to calculate differential cross-sections for neutrino/antineutrino--baryon scattering in nuclear medium in terms of $Q^2$, as well as the neutrino and antineutrino mean free paths in dense nuclear matter (densities above the normal nuclear matter)~\mbox{\cite{Cheoun13a,Cheoun13b,Athar-Review22,Alberico02a,Parada18a}.}

The conclusion of this section is that the neutrino/antineutrino--nucleon cross-sections are in general reduced in the nuclear medium.
Under study is the impact of the hyperon form factors on the neutrino/antineutrino--hyperon cross-sections, and other interactions with neutrinos in a nuclear medium~\cite{InPreparation}.


\section{Discussion \label{secDiscussion}}

The results presented in Sections~\ref{secResults1} and \ref{secResults2}  correspond to two different applications of the covariant spectator quark model to symmetric nuclear matter.
The model for the electromagnetic form factors~\cite{Octet2} is based on the simplest model for the quark--diquark system based on an $S$-state wave function.
In the study of the axial structure, it was concluded that the $S$-state model was insufficient to describe the observed experimental data as well as the lattice QCD data for the nucleon axial form factors, and the model was improved with the inclusion of a $P$-state~\cite{AxialFF,GA-Medium1}.
In the future, we should consider a combined analysis of
the electromagnetic and axial structure of octet baryons.
New lattice QCD simulations for electromagnetic and axial transitions with accurate data for  neutral baryons (neutron, $\Lambda$, and $\Sigma^0$) can also help to improve the description of the octet baryon form factor data in the free space.

The model description of the physics associated with  meson cloud dressing can be improved.
The contributions of the kaon cloud that have been proven to be relevant for some electromagnetic transitions should be taken into account in more detailed studies of the electromagnetic and axial structure of  octet baryons.
The inclusion of a kaon cloud component  automatically improves the description of the $\Xi^0 \to \Sigma^+$ axial-vector coupling due to the correction of the normalization constant $Z_\Xi$ [reduction in $G_A(0)$]~\cite{GA-Medium1}.
The effect of the kaon cloud will also improve  the
description of the baryon octet electromagnetic form factors
since the coupling with the kaon is expected to be stronger for the 
$\Sigma$ and $\Xi$ systems~\cite{Octet2}.
Concerning the meson cloud contributions to the axial form factors, there is also the opportunity of improving the model considering a more theoretically motivated parametrization of the meson cloud, taking into account the explicitly pion and kaon cloud contributions weighted by  explicit SU(3) baryon--meson coupling.

In our studies, we also provide calculations of the induced pseudoscalar form factor $G_P$, seldom discussed in the literature.
These calculations are predictions of the model since no free parameters are adjusted in the calculations, except for small bare contributions, determined by the analysis of the lattice QCD data for the nucleon.

From the experimental side, we expect that, in the near future,  the proton in-medium form factors will be measured for several nuclei. 
There is the hope that new experiments based on the polarization transfer method may provide information about the ratio $G_E/G_M$ for the neutron or hyperons bound to a nucleus.
Concerning the axial structure of baryons, we expect that new and accurate measurements of beta decays of heavy nuclei may improve our knowledge on the quenching of the nucleon axial-vector coupling in nuclear matter.

The calculations of the electromagnetic and axial-vector form factors are used to study neutrino--nucleon and antineutrino--nucleon single-differential cross-sections in terms of the neutrino energy $E_\nu$.
{
The model calculations can be used in the near future for the study of the neutrino--hyperon and antineutrino--hyperon scattering in dense nuclear matter, generalizing the studies on neutrino/antineutrino--nucleon scattering~\cite{Cheoun13a,Cheoun13b,Alberico02a}.
The extension of the model for higher densities expands the range of application of the model to studies with astrophysical implications, including neutrino scattering off nuclear targets, high-energy nucleus--nucleon collisions, and cores of compact stars.
The study of the neutrino/antineutrino mean free paths in dense nuclear matter is an example of a direct application of the present model.
For more realistic applications, we need to consider the case of asymmetric nuclear matter in order to study nuclear systems with different numbers of protons and neutrons.}

\section{Outlook and Conclusions \label{secConclusions}}

The literature regarding the properties of baryons in a nuclear medium is scarce.
Theoretical studies on the subject are very important for the study of  hadrons in extreme conditions, including nucleus--nucleus and hyperon--nucleons collisions, neutrino and antineutrino propagation in dense matter, neutrino--baryon and antineutrino--baryon in-medium interactions, and cores of compact stars, among other processes.

Models based on the degrees of freedom observed in free space measurements, involving valence quarks and the meson cloud excitations of baryon cores, are particularly useful because they can help in the interpretation of the results based on the physical processes associated with empirical data (vacuum/free space).

The formalism used in the in-medium calculations was tested in previous works in free space, including physical systems and lattice QCD simulations.
The latter are important to disentangle the physical mechanisms associated with valence quarks, within our model the interpretation of valence/meson cloud contributions.

We conclude that the electromagnetic and axial structure of octet baryons is modified in the nuclear medium.
The impact of the medium effects increases with  medium density.

For charged baryons, the absolute values of the electric and axial-vector form factors are in general suppressed in the nuclear medium.
As for the magnetic form factor, in general,
the absolute value of $G_M$ is reduced in medium above a certain value of $Q^2$
in natural units 
(in units of nuclear magneton in free space, $|G_M^\ast|$ is enhanced at low $Q^2$ and suppressed for larger values of $Q^2$).
%
For neutral baryons, $G_E$ and $G_A$ have small magnitudes and milder  in-medium modifications.
The magnetic form factors of neutral baryons have properties similar to  charged baryons
(suppression for finite $Q^2$).
A consequence of these medium modifications is the reduction in the single-differential neutrino/antineutrino cross-sections in medium in neutral-current and charged-\mbox{current reactions.}

The calculations presented here can be used in the study of reactions of neutrinos/antineutrinos with nuclei and reactions of neutrinos/antineutrinos with hyperons in dense nuclear matter (above the normal nuclear matter density).
The numerical results for $G_E$, $G_M$, and $G_A$ can be used to calculate the single-differential cross-sections for neutrino/antineutrino--baryon scattering in a nuclear medium, and using model parametrizations of neutrino flux
to calculate the neutrino/antineutrino--baryon differential cross-sections, as well as the free neutrino/antineutrino mean free paths in dense nuclear matter.

In the future, the methods discussed in the present work for octet baryons
can be extended to other baryon systems, like decuplet baryons
and transitions between octet baryons and decuplet baryons,
as well as for densities larger than  normal nuclear matter.
{
For applications to problems with astrophysical implications,
we also consider the generalization of the formalism to asymmetric nuclear matter.}

\vspace{6pt}

\authorcontributions{Writing---original draft preparation, G.R.;
  writing---review and editing, K.T.~and M.-K.C.  All authors have read and agreed to the published version of the manuscript.}

\funding{G.R.~and M.-K.C.~were supported by the National
Research Foundation of Korea (Grant  No.~RS-2021-NR060129). K.T.~was supported by Conselho
Nacional de Desenvolvimento 
Cient\'{i}fico e Tecnol\'ogico (CNPq, Brazil), Processes No.~313063/2018-4,
No.~426150/2018-0, and No.~014199/2022-2, 
and FAPESP Processes No.~2019/00763-0
and No.~2023/07313-6, and his work was also part of the project
Instituto Nacional de Ci\^{e}ncia e 
Tecnologia---Nuclear Physics and Applications 
(INCT-FNA), Brazil, Process No.~464898/2014-5.}

\dataavailability{The experimental data presented here are available
  in the literature (references included).
  The tables with numerical values used in the figures
  can be supplied \mbox{upon request.}}




\conflictsofinterest{The authors declare no conflicts of interest.}




\appendixtitles{yes} %
\appendixstart
\appendix

\section{Gell--Mann Matrices \label{appGM-matrices}}

The  Gell--Mann matrices are used in the definition
of the octet baryon axial transitions (Section~\ref{secDefs})
and on the quark electromagnetic and axial currents (Sections~\ref{secQEMcurr}
and \ref{sec-Axial}).
We use here the standard definition~\cite{Gaillard84}
{\small\ba
&&
\lambda_1= 
\left(
\begin{array}{ccc} 0 & 1 & 0 \cr 
                   1 & 0 & 0 \cr
                   0 & 0 & 0 \cr
\end{array}
\right),
\hspace{.5cm}
\lambda_2= 
\left(
\begin{array}{ccc} 0 & -i & 0 \cr 
                   i & 0 & 0 \cr
                   0 & 0 & 0 \cr
\end{array}
\right),
\hspace{.5cm}
\lambda_3= 
\left(
\begin{array}{ccc} 1 & 0 & 0 \cr 
                   0 & -1 & 0 \cr
                   0 & 0 & 0 \cr
\end{array}
\right), \nonumber \\
&& 
\lambda_4= 
\left(
\begin{array}{ccc} 0 & 0 & 1 \cr 
                   0 & 0 & 0 \cr
                   1 & 0 & 0 \cr
\end{array}
\right),
\hspace{.5cm}
\lambda_5= 
\left(
\begin{array}{ccc} 0 & 0 & -i \cr 
                   0 & 0 & 0 \cr
                   i & 0 & 0 \cr
\end{array}
\right), \\
&& 
\lambda_6= 
\left(
\begin{array}{ccc} 0 & 0 & 0 \cr 
                   0 & 0 & 1 \cr
                   0 & 1 & 0 \cr
\end{array}
\right),
\hspace{.5cm}
\lambda_7= 
\left(
\begin{array}{ccc} 0 & 0 & 0 \cr 
                   0 & 0 & -i \cr
                   0 & i & 0 \cr
\end{array}
\right),
\hspace{.5cm}
\lambda_8= 
\frac{1}{\sqrt{3}}
\left(
\begin{array}{ccc} 1 & 0 & 0 \cr 
                   0 & 1 & 0 \cr
                   0 & 0 & -2 \cr
\end{array}
\right). 
\nonumber
\ea}

We can also define $\lambda_0= {\rm diag}(1,1,1)$.

The neutral current transitions ($\Delta I=0$ and  $\Delta S=0$) are associated with the operator\vspace{12pt}
\ba
I_0 = \lambda_3.
\ea
The transitions that increase/decrease the isospin ($\Delta I=\pm 1$) are defined by 
\ba
I_\pm = \frac{1}{2}( \lambda_1 \pm i \lambda_2). 
\ea
The transitions associated with $s \rightleftarrows u$ transitions ($\Delta S =\pm 1$) are represented by the operator
\ba
V_\pm = \frac{1}{2}( \lambda_4 \pm i \lambda_5). 
\ea

{
For the discussion of the quark electromagnetic current, we also define} 
\ba
\lambda_s= 
\left(
\begin{array}{ccc} 0 & 0 & 0 \cr 
                   0 & 0 & 0 \cr
                   0 & 0 & -2 \cr
\end{array}
\right),
\ea
which can also be written as 
\ba
\lambda_s =  \sqrt{3} \lambda_8 - \tau_0,
\ea
where $\tau_0= {\rm diag}(1,1,0)$ is the SU(3) generalization of the SU(2) unitary matrix.

\section{Information About the Valence Quark Contributions
\label{appendixBare}}

We consider the octet baryon wave function $\Psi_B(P,k)$ given by Equations~(\ref{eqPsi-total})--(\ref{eq-PsiP}).
The flavor wave functions are determined by the SU(3) flavor symmetry as presented in Table~\ref{tabPHI}.
The spin wave functions are determined by the covariant spectator quark model~\cite{Nucleon,Omega,NDeltaSL1,NDeltaSL2}
\ba
& &
\phi_S^0 = u_B(P,s), \hspace{1cm}
\phi_S^1 = - (\varepsilon_{P \lambda}^\ast)_\alpha U_B^\alpha(P,s),  \nonumber \\
& & U_B^\alpha(P,s) = -\frac{1}{\sqrt{3}} \gamma_5 \left( \gamma^\alpha - \frac{P^\alpha}{M_B} \right)
u_B(P,s),
\ea
where $u_B(P,s)$ is the baryon $B$ Dirac spinor in terms of the momentum ($P$) and spin projection ($s$), and $(\varepsilon_{P \lambda})_\alpha$ is the diquark polarization in the fixed-axis representation discussed in Refs.~\cite{Nucleon,NDeltaSL2,FixedAxis}.
The $S$-state wave function (\ref{eq-PsiS}) is a generalization of the nonrelativistic wave function~\cite{Nucleon}.

\begin{table}[H]
  \caption{Flavor wave functions of the octet baryons.
\label{tabPHI}}
\footnotesize
\begin{adjustwidth}{-\extralength}{0cm}
\begin{tabularx}{\fulllength}{lCC}
\toprule
\boldmath{$B$}   & \boldmath{$\ket{M_A}$}  &  \boldmath{$\ket{M_S}$}  \\
\midrule
$p$  \vspace{2pt}   &  $\sfrac{1}{\sqrt{2}} (ud -du) u$ \vspace{2pt}& 
        $\sfrac{1}{\sqrt{6}} \left[
        (ud + du) u - 2 uu d \right]$ \vspace{2pt}\\
$n$\vspace{2pt}     & $\sfrac{1}{\sqrt{2}} (ud -du) d$  \vspace{2pt}& 
         $-\sfrac{1}{\sqrt{6}} \left[
         (ud + du) d - 2 ddu \right]$\vspace{2pt}  \\
& & \\
$\Lambda^0$\vspace{2pt} &
$\sfrac{1}{\sqrt{12}}
\left[
s (du-ud) - (dsu-usd) + 2(ud -du)s
\right]$\vspace{2pt}
& 
$\sfrac{1}{2}
\left[ (dsu-usd) - s (ud-du)
\right]$ \vspace{2pt}\\ 
& & \\
$\Sigma^+$ \vspace{2pt} &  $\sfrac{1}{\sqrt{2}} (us -su) u $ \vspace{2pt}&  
$\sfrac{1}{\sqrt{6}} \left[(us + su) u - 2 uu s \right]$\vspace{2pt}
\\
$\Sigma^0$ \vspace{2pt}&
$\sfrac{1}{2}
\left[ (dsu+usd) -s (ud+du)
\right]$\vspace{2pt}
& 
$\sfrac{1}{\sqrt{12}}
\left[
s (ud + du ) +(dsu+usd) -2(ud+du)s
\right]$ \vspace{2pt}\\
$\Sigma^-$ \vspace{2pt}& $\sfrac{1}{\sqrt{2}} (ds -sd) d$ \vspace{2pt}& 
$\sfrac{1}{\sqrt{6}}\left[ (sd + ds) d - 2 dd s \right]$ \vspace{2pt}
              \\
              & & \\
$\Xi^0$ \vspace{2pt}&
$\sfrac{1}{\sqrt{2}} (us -su) s$ \vspace{2pt}
& 
$-\sfrac{1}{\sqrt{6}} \left[(us + su) s - 2 ss u\right]$ \vspace{2pt}  \\
$\Xi^-$ & $\sfrac{1}{\sqrt{2}} (ds -sd) s$
& 
$-\sfrac{1}{\sqrt{6}} \left[(ds + sd) s - 2 ss d\right]$  \\
\bottomrule
\end{tabularx}
\end{adjustwidth}
\end{table}


The octet baryon flavor states $\left| M_{A,S} \right>$ used in the wave functions (\ref{eq-PsiS}) and (\ref{eq-PsiP}) are included in Table~\ref{tabPHI}.

\subsection{Quark Electromagnetic Form Factors
\label{appBareEMFF}}

Motivated by the vector meson dominance mechanism, we use the following parametrizations for the quark form factors 
$f_{i0}$ and $f_{i\pm}$ ($i=1,2$)
\ba
&&
\hspace{-1.2cm}
f_{1 \pm}(Q^2) = \lambda_q + (1 - \lambda_q) \frac{m_\rho^2}{m_\rho^2 + Q^2} 
+ c_\pm \frac{M_h^2 Q^2}{(M_h^2 + Q^2)^2}, 
\label{eqF1p}\\
& &
\hspace{-1.2cm}
f_{10} (Q^2) = \lambda_q + (1 - \lambda_q) \frac{m_\phi^2}{m_\phi^2 + Q^2} 
+ c_0 \frac{M_h^2 Q^2}{(M_h^2 + Q^2)^2}, 
\label{eqf10}\\
&&
\hspace{-1.2cm}
f_{2\pm}(Q^2) =   \kappa_\pm \left\{ 
d_\pm 
 \frac{m_\rho^2}{m_\rho^2 + Q^2} 
 + (1- d_\pm)  \frac{M_h^2}{M_h^2 + Q^2}  \right\},
 \label{eqF2p} \\
&&
\hspace{-1.2cm}
f_{20}(Q^2) =   \kappa_0 \left\{ 
d_0 
 \frac{m_\phi^2}{m_\phi^2 + Q^2} 
+ (1- d_0)  \frac{M_h^2}{M_h^2 + Q^2}  \right\}, 
\label{eqf20}
\ea 
where $m_\rho$ and $m_\phi$  represent the masses of  mesons  $\rho$ and $\phi$, respectively.
In this representation, we consider for simplicity the approximation $m_\omega = m_\rho$ for the isoscalar channels ($\ell =+$). 
The terms with $M_h$ correspond to an effective heavy vector meson that parametrizes the short-range effects.
The value of $M_h$ is fixed as $M_h = 2 M_N$~\cite{Nucleon,Lattice}.
Numerically, we use $m_\rho= 0.7753$ GeV 
and $m_\phi= 1.020$ GeV.

\begin{table}[H]
\caption{Mixed symmetric ($j_i^S$)
and anti-symmetric ($j_i^A$)
coefficients for the octet baryons
defined by Equation~(\ref{eqjiAS}).
The coefficients are used in the calculation
of the valence quark contributions to the
electromagnetic form factors (\ref{eqF10}) and (\ref{eqF20}).
\label{tableJI}}
\begin{tabularx}{\textwidth}{lCC}
\toprule
\boldmath{$B$}   & \boldmath{$j_i^S$}  &   \boldmath{$j_i^A$}  \\
\midrule
$p$   \vspace{2pt}  & $\sfrac{1}{6} (f_{i+}-f_{i-}) $ \vspace{2pt}&
        $\sfrac{1}{6} (f_{i+}+3 f_{i-})  $ \vspace{2pt}\\
$n$   \vspace{2pt}  & $\sfrac{1}{6} (f_{i+}+  f_{i-}) $\vspace{2pt} &
        $\sfrac{1}{6} (f_{i+}-  3 f_{i-})$ \vspace{2pt}\\
& & \\
$\Lambda$\vspace{2pt} & $\sfrac{1}{6}f_{i+}$ \vspace{2pt}&
 $\sfrac{1}{18} (f_{i+}-  4 f_{i0})$ \vspace{2pt}\\
& & \\
$\Sigma^+$ \vspace{2pt} & $\sfrac{1}{18}
(f_{i+} + 3 f_{i-} -4 f_{i0}) $ \vspace{2pt}&
 $\sfrac{1}{6} (f_{i+}+3 f_{i-})  $ \vspace{2pt}\\
$\Sigma^0$\vspace{2pt} &  $\sfrac{1}{36} (2 f_{i+}-  8 f_{i0})$\vspace{2pt}
& $\sfrac{1}{6}f_{i+}$\vspace{2pt} \\
$\Sigma^-$ \vspace{2pt}& $\sfrac{1}{18}
(f_{i+} - 3 f_{i-} -4 f_{i0}) $\vspace{2pt} &
        $\sfrac{1}{6} (f_{i+}-  3 f_{i-})           $\vspace{2pt} \\
& & \\
$\Xi^0$ \vspace{2pt}& $\sfrac{1}{18} (2 f_{i+} + 6  f_{i-} -2 f_{i0}) $ \vspace{2pt}&
$-\sfrac{1}{3}f_{i0}$\vspace{2pt} \\
$\Xi^-$ & $\sfrac{1}{18} (2 f_{i+} -6  f_{i-} -2 f_{i0}) $ &
$-\sfrac{1}{3}f_{i0}$ \\
\bottomrule
\end{tabularx}
\end{table}

In the calculations, we use the quark anomalous
magnetic moments
\ba
\kappa_u = 1.711, \hspace{.5cm}  \kappa_d =1.987,  \hspace{.5cm}  \kappa_s =1.462,
\ea
and the parameters
\ba
& &
\lambda_q=1.21, \hspace{.5cm}
c_0=4.427, \hspace{.5cm} d_0=-1.860, \nonumber \\
& &
 c_+= 4.160, \hspace{.35cm} c_- = 1.160, \hspace{.35cm}
  d_+ = d_- = -0.686.
\ea
The quark anomalous magnetic moments $\kappa_u$ and $\kappa_d$ were determined by the fit to the nucleon and octet baryon electromagnetic data in Ref.~\cite{Octet2}, while $\kappa_s$ was determined in the study of the electromagnetic form factors of the decuplet baryons, taking into account the $\Omega^-$ magnetic moment~\cite{Omega}.
The renaming parameters were  determined in the study of the nucleon and decuplet baryon form factors~\cite{Nucleon,Omega}.

The form factors $f_{i\ell}$ ($i=1,2$) with $\ell=0,\pm$ are related to the flavor form factors  $f_{i q}$ ($q=u,d,s$):
\ba
 Q_s f_{is} = f_{i0}, \hspace{.5cm}
Q_u f_{iu} = \frac{1}{6} f_{1+} +  \frac{1}{2} f_{1-}, \hspace{.5cm}
     Q_d f_{id} = \frac{1}{6} f_{1+} -  \frac{1}{2} f_{1-},  
\ea
 where $Q_u = +\frac{2}{3}$ and $Q_d = Q_s= -\frac{1}{3}$ are the quark charges.

From the previous relations and the normalization of $f_{2\ell}(0)$, one can write
 $\kappa_+ = 2 \kappa_u - \kappa_d$ and
  $\kappa_- = \frac{1}{3} (2 \kappa_u +  \kappa_d)$.

The coefficients $j_i^{A,S}$ for the octet baryon defined by Equation~(\ref{eqjiAS}) are presented in Table~\ref{tableJI}.

For the discussion of the magnetic form factors
near $Q^2=0$, we include in Table~\ref{tabGMB0}
the valence quark contributions to $G_{MB}^\ast(0)$,
in natural units, 
in terms of in-medium baryon masses.

\begin{table}[H]
\caption{Explicit expressions for the bare contribution to $G_{MB}^\ast(0)$ in terms 
of the octet baryon mass in medium ($M_B^\ast$)
and the free space nucleon mass.
Values in natural units.
\label{tabGMB0}}
\begin{tabularx}{\textwidth}{CC}
\toprule
   \boldmath{$B$} & \boldmath{$G_{MB}^\ast(0)$ $\left(\frac{e}{2 M_B^\ast} \right)$} \\
\midrule
 $p$   \vspace{2pt}     & 
$ 1  + \left(\frac{8}{9} \kappa_u   +  \frac{1}{9}  \kappa_d \right)    $ \vspace{2pt} \\
 $n$        \vspace{2pt} & 
$ - \frac{2}{3}  - \left(\frac{2}{9} \kappa_u   +  \frac{4}{9}  \kappa_d \right)   $  
\vspace{10pt}
\\
$\Lambda$  \vspace{2pt}  & 
$ - \frac{1}{3}   - \frac{1}{3} \kappa_s \frac{M_\Lambda^\ast}{M_N^\ast}$   \vspace{10pt}\\
 $\Sigma^+$  \vspace{2pt}&  $ 1 + \left(\frac{8}{9}\kappa_u + \frac{1}{9}\kappa_s  \right) 
\frac{M_\Sigma^\ast}{M_N^\ast}$ \vspace{2pt} \\
%
 $\Sigma^0$  \vspace{2pt}& $ \frac{1}{3}  + 
\left(\frac{4}{9}\kappa_u - \frac{2}{9}\kappa_d + \frac{1}{9}\kappa_s  \right) 
\frac{M_\Sigma^\ast}{M_N^\ast}$  \vspace{2pt}\\
 $\Sigma^-$  \vspace{2pt}& $ -\frac{1}{3}  - 
\left(\frac{4}{9}\kappa_d - \frac{1}{9}\kappa_s \right) 
\frac{M_\Sigma^\ast}{M_N^\ast}$ \vspace{10pt} \\
%
 $\Xi^0$   \vspace{2pt}  &   $ -\frac{2}{3}  - 
\left(\frac{2}{9}\kappa_u + \frac{4}{9}\kappa_s \right) 
\frac{M_\Xi^\ast}{M_N^\ast}$ \vspace{2pt} \\
 $\Xi^-$    &    $ -\frac{1}{3}  + 
\left(\frac{1}{9}\kappa_d - \frac{4}{9}\kappa_s \right) 
\frac{M_\Xi^\ast}{M_N^\ast}$ \\
\bottomrule
\end{tabularx}
\end{table}

\subsection{Quark Axial Form Factors \label{appBareAxial}}

As discussed in Refs.~\cite{AxialFF,GA-Medium1}, we  consider
the following parametrizations for the
quark axial form factors
\ba
& &
\hspace{-1cm}
g_A^q (Q^2)= \lambda_q + (1-\lambda_q)\frac{m_\rho^2}{m_\rho^2+ Q^2}
+ c_-  \frac{ Q^2 M_h^2}{(M_h^2+ Q^2)^2}, \label{gqA}\\
& &
\hspace{-1cm}
g_P^q (Q^2)= \alpha  \frac{m_\rho^2}{m_\rho^2+ Q^2} +
\beta  \frac{M_h^2}{M_h^2+ Q^2},
\label{eqQuarkGP}
\ea
where the hadron masses have the same meaning as in the previous section, and
\ba
\alpha = -3.9011, \hspace{.5cm} \beta= 0.3297.
\ea

The calculation of the valence quark contributions to the axial form
factors $G_A^{\rm B}$ and $G_P^{\rm B}$ based on Equations~(\ref{eqGA-bare})
and (\ref{eqGP-bare}) requires the knowledge of the coefficient ${\cal F}$.
These coefficients are determined by the values of $f_X^{A,S}$,
given by Equation~(\ref{eq-calF}), and presented in Table~\ref{tabFAS}.

\begin{table}[H]
\caption{Axial transitions.
Coefficients $f_I^{S,A}$ and $f_{V}^{S,A}$ for the possible axial
transitions and the global factor ${\cal F}$.
The column SU(3) is the estimate based on the functions
$F'$ and $D'$ determined by an SU(3) baryon--meson coupling.
For the neutral current transitions:
the operators act on the isospin states
$p=\Xi^0 = (1\; 0)^T$ and $n= \Xi^- = (0\; 1)^T$
for isospin $1/2$ states and
$I_\Sigma = \mbox{diag} (1, 0, -1)$ act on isospin \mbox{1 states} (charge operator).
\label{tabFAS}}

\begin{adjustwidth}{-\extralength}{0cm}

\begin{tabularx}{\fulllength}{CCCCCCC}
\toprule
  &  \boldmath{$B \to B'$} &   \boldmath{$f_{X}^A$} & \boldmath{$f_{X}^S$} &  \boldmath{${\cal F}$} &  \boldmath{${\cal G}_{\rm SU(3)}$} \\
\midrule
$|\Delta I| = 1 $ 
& $n \to p$                  
      &$1$ &$-\frac{1}{3}$ &  $\frac{10}{9}$ &  $F^\prime+ D^\prime$\\
         &$\Sigma^{\pm} \to \Lambda$  
&$\pm\frac{1}{\sqrt{6}}$ &$\mp\frac{1}{\sqrt{6}}$
&    $\pm  \frac{2\sqrt{2}}{3\sqrt{3}}$ &  $\pm \sqrt{\frac{2}{3}} D^\prime$\\
      &    $\Sigma^- \to \Sigma^0$
&$\frac{1}{\sqrt{2}}$ &$\frac{1}{3\sqrt{2}}$
&  $\frac{4\sqrt{2}}{9} $ & $\sqrt{2} F'$   \\
& $\Sigma^0 \to \Sigma^+$
&$\frac{1}{\sqrt{2}}$ &$\frac{1}{3\sqrt{2}}$
&  $\frac{4\sqrt{2}}{9} $& $\sqrt{2} F'$ \\
&$\Xi^- \to \Xi^0$ 
&$0$   &$\frac{2}{3}$ &  $- \frac{2}{9}$   &  $F'-D'$  \\
& & \\
    $|\Delta S| = 1 $ 
&$\Lambda \to p$ 
&  $-\frac{2}{\sqrt{6}}$ & $0$  & $- \sqrt{\frac{2}{3}}$ &
$-\sqrt{\frac{3}{2}}  \left(F' + \frac{1}{3}D'\right)$  \\
  & $\Sigma^{-} \to n$ 
          &$0$ &  $-\frac{2}{3}$ & $\frac{2}{9}$ & $-F'+D'$  \\
          & $\Sigma^0 \to p$ 
          &$0$ &  $-\frac{\sqrt{2}}{3}$  &  $\frac{\sqrt{2}}{9}$
&  $-\frac{1}{ \sqrt{2}} (F'-D')$   \\
           &$\Xi^- \to \Lambda$ 
&$-\frac{1}{\sqrt{6}}$ &  $-\frac{1}{\sqrt{6}}$
&   $- \frac{\sqrt{2}}{3\sqrt{3}}$   &  $- \sqrt{\frac{3}{2}} \left(F'-\frac{1}{3}D'\right)$ \\
   &$\Xi^- \to \Sigma^0$ 
           &$\frac{1}{\sqrt{2}}$ &$-\frac{1}{3\sqrt{2}}$  &   $\frac{5 \sqrt{2}}{9}$  &
           $\frac{1}{\sqrt{2}} (F' + D')$ \\
           &$\Xi^0 \to \Sigma^+$ 
&$1$ &$-\frac{1}{3}$  & $\frac{10}{9}$  &$F'+D'$  \\
 & & \\
    $\Delta I = 0$,  &  $N \to N$ &  $\tau_3$ &  $- \frac{1}{3} \tau_3$ &
  $\frac{10}{9} \tau_3$  &      $(F'+D')\tau_3$  \\ 
    $\Delta S = 0$  
 &    $\Sigma \to \Sigma $   &   $I_\Sigma$   &
$\frac{1}{3} I_\Sigma$  &  $\frac{8}{9} I_\Sigma$   & $2F' I_\Sigma$ \\
 &  $\Xi \to \Xi$   &   $0$ &
 $\frac{2}{3} \tau_3$  &  $-\frac{2}{9} \tau_3$ & $ (F' - D')\tau_3$ \\
\bottomrule
\end{tabularx}
\end{adjustwidth}
\end{table}


In the calculation of the overlap integrals, we use the variables
\ba
\bar P = \frac{1}{2}(P' + P), \hspace{1cm}
\tilde k^\prime = k - \frac{\bar P \cdot k}{\bar P^2} \bar P,
\ea
where $\bar P^2 = M_B^2 ( 1 + \tau)$.

The overlap integrals $B_i(Q^2)$ ($i=0,\ldots,4$) are defined as
\ba
& &
B_0 = \int_k \psi_S(P',k) \psi_S(P,k), \hspace{.5cm}
B_1 = \int_k \frac{\bar P \cdot k}{M_B}\psi_P(P',k) \psi_S(P,k), \nonumber \\
& &
B_2 = \int_k \frac{(\bar P \cdot k)^2}{\bar P^2}\psi_P(P',k) \psi_P(P,k), \hspace{.5cm}
B_3 = \int_k (- \tilde k^{\prime 2}) \psi_P(P',k) \psi_P(P,k), \nonumber \\
& &
B_4 = \int_k \frac{(q \cdot k)^2}{Q^2} \psi_P(P',k) \psi_P(P,k), \hspace{.5cm}
B_5 = (B_3 - 3 B_4)/\tau. 
\ea
The integrals take a simpler form at the Breit frame.
In the Breit frame,
\ba
\frac{\bar P \cdot k}{M_B} = \sqrt{1 + \tau} E_D, \hspace{.5cm}
 \frac{(\bar P \cdot k)^2}{\bar P^2} = E_D^2, \hspace{.5cm}
 \tilde k^{\prime 2} = - {\bf k}^2, \hspace{.5cm}
 \frac{(q \cdot k)^2}{Q^2} = - k_z^2,
\ea
where $E_D$ is the diquark energy $E_D = \sqrt{m_D^2 + {\bf k}^2}$.

The normalization of the radial wave functions,
defined in Section~\ref{secBare}, lead to
\mbox{$B_0(0)=1$} and  $3 B_4(0)=B_0(0)=1$. 
For simplicity, we redefined the integral $B_5$ compared to the previous works~\cite{AxialFF,GA-Medium1}.
Notice, however, that the integrals $B_3$ and $B_4$
are correlated in the limit $Q^2=0$ or $\tau=0$
according to  $B_3(Q^2) \simeq 3 B_4 (Q^2)$.
The last relation ensures that $B_5(0)$ is finite.

\section{Information About the Meson Cloud Contributions \label{appendixMesonCloud}}

\subsection{Electromagnetic Transitions \label{appMesonCloud1}}

The pion cloud contributions to the octet baryon form actors are calculated
based on Equations~(\ref{eqF1B}) and (\ref{eqF2B}).
The coefficients $a_j$ ($j=1,2,3$), are written
in Table~\ref{tab-F1F2-pion} in terms of the couplings
$\beta_B$ and the bare functions
$\tilde e_{0B}$ and $\tilde \kappa_{0B}$.

\begin{table}[H]
  \caption{Pion cloud coefficients for the functions $F_{1B}$ and $F_{2B}$~\cite{Octet,Octet2,Octet4}.
\label{tab-F1F2-pion}}
\begin{tabularx}{\textwidth}{Lccc}
\toprule
\boldmath{$B$}   & \boldmath{$a_1$}  &   \boldmath{$a_2$}  &  \boldmath{$a_3$} \\
\midrule
$p$ & $2 \beta_N$ & $\beta_N(\tilde e_{0p} + 2 \tilde e_{0 n})$ &
$\beta_N(\tilde \kappa_{0p} + 2 \tilde \kappa_{0 n})$ \\
$n$ & $- 2 \beta_N$ & $\beta_N(2 \tilde e_{0p} + \tilde e_{0 n})$ &
$\beta_N(2\tilde \kappa_{0p} + \tilde \kappa_{0 n})$ \\
  & & \\
 $\Lambda$ &  0&  $\beta_\Lambda (\tilde e_{0\Sigma^+} + \tilde e_{0 \Sigma^0} + \tilde e_{0 \Sigma^-} \!)$ &
$\beta_\Lambda (\tilde \kappa_{0\Sigma^+} + \tilde \kappa_{0 \Sigma^0} + \tilde \kappa_{0 \Sigma^-} \!)$ \\
 & & \\
    $\Sigma^+$ & $(\beta_\Lambda + \beta_\Sigma)$ &
    $\beta_\Lambda \tilde e_{0\Lambda} + \beta_\Sigma ( \tilde e_{0\Sigma^+}  + \tilde e_{0\Sigma^0} \!)$
    &  $\beta_\Lambda \tilde \kappa_{0\Lambda} + \beta_\Sigma ( \tilde \kappa_{0\Sigma^+}  + \tilde \kappa_{0\Sigma^0} \!) $\\
  $\Sigma^0$ & 0 &  $\beta_\Lambda \tilde e_{0\Lambda} + \beta_\Sigma ( \tilde e_{0\Sigma^+}  + \tilde e_{0\Sigma^-} \!)$&
   $\beta_\Lambda \tilde \kappa_{0\Lambda} + \beta_\Sigma ( \tilde \kappa_{0\Sigma^+}  + \tilde \kappa_{0\Sigma^-} \!)$ \\
       $\Sigma^-$ & $-(\beta_\Lambda + \beta_\Sigma)$ &
    $\beta_\Lambda \tilde e_{0\Lambda} + \beta_\Sigma ( \tilde e_{0\Sigma^-}  + \tilde e_{0\Sigma^-} \!)$
    &  $\beta_\Lambda \tilde \kappa_{0\Lambda} + \beta_\Sigma ( \tilde \kappa_{0\Sigma^0}  + \tilde \kappa_{0\Sigma^-} \!) $\\
& & \\
 $\Xi^0$     & $2 \beta_\Xi$ &  $\beta_\Xi (\tilde e_{0\Xi^0} + 2\tilde e_{0\Xi^-} \!) $ &
                               $\beta_\Xi (\tilde \kappa_{0\Xi^0} + 2\tilde \kappa_{0\Xi^-} \!) $\\
 $\Xi^-$      &  $-2 \beta_\Xi$ & $\beta_\Xi (2 \tilde e_{0\Xi^0} + \tilde e_{0\Xi^-} \!) $ &
                                $\beta_\Xi (2\tilde \kappa_{0\Xi^0} + \tilde \kappa_{0\Xi^-} \!) $  \\
\bottomrule
\end{tabularx}
\end{table}

For the electromagnetic couplings with the meson cloud, we use the values of $\beta_B$ determined by the SU(6) quark model, defined by  $\alpha=0.6$~\cite{Octet2}: 
\ba
\beta_N= 1, \hspace{.5cm} \beta_\Lambda = \frac{12}{25},  \hspace{.5cm} \beta_\Sigma= \frac{16}{25},  \hspace{.5cm} \beta_\Xi = \frac{1}{25}. 
\ea

The functions $b_i$, $c_i$, and $d_i$ ($i=1,2$)
are parametrized in terms two generic cutoffs $\Lambda_i$
associated with the Dirac ($i=1$) and Pauli ($i=2$)
and the values of $b_1(0)$, $b_2(0)$,
$c_2(0)$, $d_2(0)$, and the constant
$d_1^\prime$, based on
\ba
c_1= c_1(0) \left( \frac{\Lambda_1^2}{\Lambda_1^2  + Q^2} \right)^2, \hspace{1cm}
c_2= c_2(0) \left( \frac{\Lambda_2^2}{\Lambda_2^2  + Q^2} \right)^3, \\
d_1= d_1^\prime \frac{Q^2}{\Lambda_1^2} \left( \frac{\Lambda_1^2}{\Lambda_1^2  + Q^2} \right)^3, \hspace{1cm}
d_2= d_2(0) \left( \frac{\Lambda_2^2}{\Lambda_2^2  + Q^2} \right)^3,
\ea
and
\ba
& &
b_1= b_1(0) \left( \frac{\Lambda_1^2}{\Lambda_1^2  + Q^2} \right)^3
\left[1 + \frac{1}{Z_N b_1(0)} \left(\frac{1}{24} T_1 \log m_\pi + R_1  \right) Q^2\right], \\
& &
b_2= b_2(0) \left( \frac{\Lambda_2^2}{\Lambda_1^2  + Q^2} \right)^4
\left[1 + \frac{1}{Z_N b_2(0)} \left(- \frac{1}{24} T_2 \frac{M}{m_\pi} + R_2  \right) Q^2\right]. 
\ea
In the last relations, $T_1 = 13.27$ GeV$^{-2}$,
$T_2= 7.42$ GeV$^{-2}$, $R_1 = 1.036$ GeV$^{-2}$, and
\mbox{$R_2= -1.987$ GeV$^{-2}$.}
The values of $T_1$ and $T_2$ are calculated
in terms of the pion decay constant  $F_\pi \simeq 0.093$ GeV
and nucleon axial coupling $g_A \simeq 1.27$
in order to describe the asymptotic properties of the  nucleon wave functions
at the physical pion mass.

Notice that $d_1(0)=0$ and that the shape of the function is determined by the constant $d_1^\prime$.
The parametrizations for $b_1$ and $b_2$ reproduce
the expected chiral expressions
for the isovector nucleon charge radius, both Dirac and Pauli,
according to chiral perturbation theory~\cite{Bernard02}.


The relevant parameters associated with the pion cloud parametrization are in \mbox{Table~\ref{table-PionCloud}.}

\begin{table}[H]
\caption{Pion cloud parameters. 
\label{table-PionCloud} }
\begin{tabularx}{\textwidth}{CCCCCcc}
\toprule
\boldmath{$b_1(0)$}   & \boldmath{$d_1^\prime$}  &   \boldmath{$b_2(0)$}  &  \boldmath{$c_2(0)$} & \boldmath{$d_2(0)$} & \boldmath{$\Lambda_1^2$} \textbf{(GeV}\boldmath{$^2$}\textbf{)} &  \boldmath{$\Lambda_2^2$} \textbf{(GeV\boldmath{$^2$})}\\
\midrule
0.0510 & $-0.148$ & 0.216 & 0.00286 & 0.0821 & 0.618 & 1.281 \\
\bottomrule
\end{tabularx}
\end{table}

\subsection{Axial Transitions \label{appMesonCloud2}}

As discussed in Section~\ref{sec-Axial-MC}, we consider a general parametrization of the meson cloud based on the form of the nucleon axial-vector form factor (\ref{eqGAN-MC}) with~\cite{GA-Medium1}
\ba
G_{AN}^{\rm MC0} = 0.6059, \hspace{1cm} \Lambda =1.05 \; \mbox{GeV}.
\ea

The coefficients $\eta_{BB'}$ are determined by the coefficients from Table~\ref{tabFAS} for ${\cal G}_{\rm SU(3)}$  according to
\ba
\eta_{BB'} = \frac{ {\cal G}_{\rm SU(3)}^0}{G_A^{\rm MC0}},
\ea
where $ {\cal G}_{\rm SU(3)}^0$ represents the value of the function for $Q^2=0$
expressed in terms of $F^{\prime \ast} (0)$ and  $D^{\prime \ast} (0)$.
The coefficients from  $G_A^{\rm MC}$ are calculated in
terms of
\ba
F^\prime (0)= 0.1775, \hspace{1cm} D^\prime (0)=0.4284.
\ea
The value of $G_A^{\rm MC0}$ corresponds to the result for the nucleon
$G_A^{\rm MC0} = F^\prime (0) + D^\prime (0)$.

In the extension to the nuclear medium,
the denominator ($G_A^{\rm MC0}$) is preserved,
while $F^\prime (0)$ and $D^\prime (0)$  are modified by the factor $\left(\frac{g_{\pi NN}^\ast}{g_{\pi NN}}\right)^2$.

The normalization of the baryon wave function (\ref{eqDressedWF})
follows the formalism discussed for the electromagnetic transitions
based on Equations~(\ref{eqF1B})--(\ref{eqZBs}),
but the value of $b_1(0)$ is readjusted according to the
asymptotic value of $G_{AN}(Q^2) \simeq Z_N G_{AN}^{\rm B}(Q^2)$.
We then use $b_1(0)= 0.121$~\cite{GA-Medium1}.


\begin{adjustwidth}{-\extralength}{0cm}

\reftitle{References}



\begin{thebibliography}{999}



\bibitem{BrownRho}
Brown, G.E.; Rho, M. 
Scaling effective Lagrangians in a dense medium.
\emph{Phys. Rev. Lett.} \textbf{1991}, \emph{66}, 2720.



\bibitem{JLabbook}
Brooks, W.K.; Strauch, S.; Tsushima, K. 
Medium Modifications of Hadron Properties and Partonic Processes. 
\emph{J. Phys. Conf. Ser.} \textbf{2011}, \emph{299}, 012011. 




\bibitem{QMCReview}
Saito, K.; Tsushima, K.; Thomas, A.W.
Nucleon and hadron structure changes in the nuclear medium and impact on
observables. 
\emph{Prog. Part. Nucl. Phys.} \textbf{2007}, \emph{58}, 1--167.




\bibitem{Miyatsu15a}
Miyatsu, T.; Cheoun, M.K.; Saito, K.
Equation of State for Neutron Stars With Hyperons and Quarks in the Relativistic Hartree--fock Approximation.
\emph{Astrophys. J.} \textbf{2015}, \emph{813}, 135. 




\bibitem{GA-Medium1}
Ramalho, G.; Tsushima, K.; Cheoun, M.K.
Weak interaction axial form factors of the octet baryons in nuclear medium. 
\emph{Phys. Rev. D} \textbf{2025}, \emph{111}, 013002.


\bibitem{EMC1}
Aubert, J.J.; Bassompierre, G.; Becks, K.H.; Best, C.; Bohm, E.; de Bouard, X.; Brasse, F.W.; Broll, C.; Brown, S.; Carr, J.; {et~al.} 
The ratio of the nucleon structure functions $F_{2n}$ for iron and deuterium. 
\emph{Phys. Lett. B} \textbf{1983}, \emph{123}, 275.

\bibitem{EMC2}
Arnold, R.G.; Bosted, P.E.; Chang, C.C.; Gomez, J.; Katramatou, A.T.; Petratos, G.; Rahbar, A.A.; Rock, S.; Sill, A.F.; Szalata, Z.M.; {et~al.}
Measurements of the a-Dependence of Deep Inelastic electron Scattering from Nuclei.
\emph{Phys. Rev. Lett.} \textbf{1984}, \emph{52}, 727.






\bibitem{Hen17a}
Hen, O.; Miller, G.A.; Piasetzky, E.; Weinstein, L.B.
Nucleon-Nucleon Correlations, Short-lived Excitations, and the Quarks Within.
\emph{Rev. Mod. Phys.} \textbf{2017}, \emph{89}, 045002. 







\bibitem{Dieterich01}
Dieterich, S.; Bartsch, P.; Baumann, D.; Bermuth, J.; Bohinc, K.; Bohm, R.; Bosnar, D.; Derber, S.; Ding, M.; Distler, M.; {et~al.} 
Polarization transfer in the
$^4$He$(\vec{e},e' \vec{p})^3$H reaction.
\emph{Phys. Lett.} 2001, B500, 47.


\bibitem{Strauch03a}
Strauch, S.;  Dieterich, S.; Aniol, K.A.; Ann, J.R.; Baker, O.K.; Bertozzi, W.; Boswell, M.; Brash, E.J.; Chai, Z.; Chen, J.P.; {et~al.} 
Polarization transfer in the
$^4$He$(\vec{e},e' \vec{p})^3$H
reaction up to $Q^2 =$ 2.6 (GeV/c)$^2$. 
\emph{Phys. Rev. Lett.} \textbf{2003}, \emph{91}, 052301. 



\bibitem{JLab-data} 
Strauch, S. [E93-049 Collaboration].
Medium modification of the proton form-factor. 
\emph{{Eur. Phys. J. A}} \textbf{2004}, \emph{19}, 153.




\bibitem{Paolone10}
Paolone, M.; Malace, S.P.; Strauch, S.; Albayrak, I.; Arrington, J.; Berman, B.L.; Brash, E.J.; Briscoe, B.; {et~al.} 
Polarization Transfer in the $^4$He$(e,e'p)^3$H at $Q^2 = 0.8$ and 1.3 (GeV/c)$^2$. 
\emph{Phys. Rev. Lett.} \textbf{2010}, \emph{105}, 072001. 






\bibitem{Malace11a}
Malace, S.P.; Paolone, M.; Strauch, S.; Albayrak, I.; Arrington, J.; Berman, B.L.; Brash, E.J.; Briscoe, W.J.; Camsonne, A.; Chen, J.P.; {et~al.}
A precise extraction of the induced polarization in the $^4$He$(e,e'p)^3$H reaction.  
%
\emph{Phys. Rev. Lett.} \textbf{2011}, \emph{106}, 052501.



\bibitem{Malov00}
Malov, S.; Wijesooriya, K.; Baker, F.T.; Bimbot, L.; Brash, E.J.; Chang, C.C.; Finn, J.M.; Fissum, K.G.; Gao, J.; Gilman, R.; {et~al.}
Polarization transfer in the $^{16}$O$(\vec{e}, e' \vec{p})^{15}$N reaction. 
\emph{Phys. Rev. C} \textbf{2000}, \emph{62}, 057302.




\bibitem{MAMI2021}
Izraeli, D. et al.  [A1 Collaboration].
Measurement of polarization-transfer to bound protons in carbon and its virtuality dependence.  
\emph{Phys. Lett. B} \textbf{2018}, \emph{781}, 95--98.


\bibitem{Kolar23a}
Kolar, T. et al.  [A1 Collaboration]. 
Measurement of polarization transfer in the quasi-elastic
$^{40}$Ca$(\vec{e},e' \vec{p})$ process. 
\emph{Phys. Lett. B} \textbf{2023}, \emph{847}, 138309.






\bibitem{Octet2} 
Ramalho, G.; Tsushima, K.; Thomas, A.W.
Octet Baryon Electromagnetic form Factors in Nuclear Medium.  
\emph{J. Phys. G} \textbf{2013}, \emph{40}, 015102.



\bibitem{Octet3}
Ramalho, G.; de Melo, J.P.B.C.; Tsushima, K.
Octet baryon electromagnetic form factor double ratios $(G_E^\ast/G_M^\ast)/(G_E/G_M)$ in a nuclear medium.
\emph{Phys. Rev. D} \textbf{2019}, \emph{100}, 014030.





\bibitem{Frank96} 
Frank, M.R.; Jennings, B.K.; Miller, G.A.
The Role of color neutrality in nuclear physics: Modifications of nucleonic wave functions.
\emph{Phys. Rev. C} \textbf{1996}, \emph{54}, 920.



\bibitem{Lu99a} 
Lu, D.H.; Tsushima, K.; Thomas, A.W.; Williams, A.G.; Saito, K.
Electromagnetic form-factors of the bound nucleon.  
\emph{Phys. Rev. C} \textbf{1999}, \emph{60}, 068201.



\bibitem{Smith04} 
Smith, J.R.; Miller, G.A.
Chiral solitons in nuclei: Electromagnetic form-factors.  
\emph{Phys. Rev. C} \textbf{2004}, \emph{70}, 065205. 


\bibitem{Cloet09a}
Cloet, I.C.; Miller, G.A.; Piasetzky, E.; Ron, G.
Neutron Properties in the Medium.  
\emph{Phys. Rev. Lett.} \textbf{2009}, \emph{103}, 082301.


\bibitem{deAraujo18} 
de Araújo, W.R.B.; de Melo, J.P.B.C.; Tsushima, K.
Study of the in-medium nucleon electromagnetic form factors using a light-front nucleon wave function combined with the quark-meson coupling model. 
\emph{Nucl. Phys. A} \textbf{2018}, \emph{970}, 325.




\bibitem{Miller02} 
Miller, G.A.
Light front cloudy bag model: Nucleon electromagnetic form-factors. 
\emph{Phys. Rev. C} \textbf{2002}, \emph{66}, 032201.


\bibitem{Yakhshiev03} 
Yakhshiev, U.T.; Meissner, U.G.; Wirzba, A.
Electromagnetic form-factors of bound nucleons revisited.  
\emph{Eur. Phys. J. A} \textbf{2003}, \emph{16}, 569.



\bibitem{QMCEMFFMedium2}
Lu, D.H.; Thomas, A.W.; Tsushima, K.; Williams, A.G.; Saito, K.
In-medium electron-nucleon scattering.  
\emph{Phys. Lett. B} \textbf{1998}, \emph{417}, 217.




\bibitem{QMCEMFFMedium3} 
Lu, D.H.; Tsushima, K.; Thomas, A.W.; Williams, A.G.; Saito, K.
The Neutron charge form-factor in helium-3. 
\emph{Phys. Lett. B} \textbf{1998}, \emph{441}, 27. 



\bibitem{QMCEMFFMedium4}
Tsushima, K.; Kim, H.; Saito, K.
Effect of the bound nucleon form-factors on charged current neutrino nucleus scattering. 
\emph{Phys. Rev. C} \textbf{2004}, \emph{70}, 038501.





\bibitem{Christov95a}
Christov, C.V.; Gorski, A.Z.; Goeke, K.; Pobylitsa, P.V.
Electromagnetic form-factors of the nucleon in the chiral quark soliton model.
\emph{Nucl. Phys. A} \textbf{1995}, \emph{592}, 513.




\bibitem{PAC35}
Letter of Intent to JLab PAC 35. In  
{\it Neutron Properties in the Nuclear Medium Studied 
by Polarization Measurements};  
Gilman, R., Higinbotham, D.W., Lichtenstadt, J., Ron, G., Strauch, S.
\url{https://hallaweb.jlab.org/collab/PAC/PAC35/LOI-10-007-Neutron-Modification.pdf}






\bibitem{Brown85a}
Brown, B.A.; Wildenthal, B.H.
Experimental and Theoretical Gamow-Teller Beta-Decay Observables for the sd-Shell Nuclei. 
\emph{Atom. Data Nucl. Data Tabl.} \textbf{1985}, \emph{33}, 347.





\bibitem{Gysbers19a}
Gysbers, P.; Hagen, G.; Holt, J.D.; Jansen, G.R.; Morris, T.D.; Navr\'atil, P.; Papenbrock, T.; Quaglioni, S.; Schwenk, A.; Stroberg, S.R.; {et~al.}
Discrepancy between experimental and theoretical $\beta$-decay rates resolved from first principles.  
\emph{Nat. Phys.} \textbf{2019}, \emph{15}, 428.











\bibitem{Thomas84}
Thomas, A.W.
Chiral Symmetry And The Bag Model: A New Starting Point For Nuclear
Physics.
\emph{Adv. Nucl. Phys.} \textbf{1984}, \emph{13}, 1--137.

\bibitem{Lu01a}
Lu, D.H.; Thomas, A.W.; Tsushima, K.
Medium modification of the nucleon axial form-factor.  \emph{arXiv} \textbf{2001},  arXiv:nucl-th/0112001.










\bibitem{Christov96a}
Christov, C.V.; Blotz, A.; Kim, H.-C.; Pobylitsa, P.; Watabe, T.; Meissner, T.; Arriola, E.R.; Goeke, K. 
Baryons as non-topological chiral solitons.
\emph{Prog. Part. Nucl. Phys.} \textbf{1996}, \emph{37}, 91--191.







\bibitem{Rakhimov98a}
Rakhimov, A.M.; Khanna, F.C.; Yakhshiev, U.T.; Musakhanov, M.M.
Density dependence of meson nucleon vertices in nuclear matter.
\emph{Nucl. Phys. A} \textbf{1998}, \emph{643}, 383.













\bibitem{Meier97a}
Meier, F.; Walliser, H.
Quantum corrections to baryon properties in chiral soliton models. 
\emph{Phys. Rept.} \textbf{1997}, \emph{289}, 383.























\bibitem{Meissner87a}
Meissner, U.G.; Kaiser, N.; Weise, W.
Nucleons as Skyrme Solitons with Vector Mesons: Electromagnetic and Axial Properties.
\emph{Nucl. Phys. A} \textbf{1987}, \emph{466}, 685.



\bibitem{Meissner89a}
Meissner, U.G.
Chiral Symmetry and Medium Modifications of Nucleon Properties. 
\emph{Phys. Lett. B} \textbf{1989}, \emph{220}, 1.














\bibitem{Ramalho-Pena23}
Ramalho, G.; Pe\~na, M.T.
Electromagnetic transition form factors of baryon resonances. 
\emph{Prog. Part. Nucl. Phys.} \textbf{2024}, \emph{136}, 104097.





\bibitem{NSTAR} 
Aznauryan, I.G.; Bashir, A.; Braun, V.; Brodsky, S.J.; Burkert, V.D.; Chang, L.; Chen, C.h.; El-Bennich, B.; Cloet, I.C.; Cole, P.L.; {et~al.}
Studies of Nucleon Resonance Structure in Exclusive Meson Electroproduction. 
\emph{Int. J. Mod. Phys. E} \textbf{2013}, \emph{22}, 1330015.



\bibitem{Aznauryan12} 
Aznauryan, I.G.; Burkert, V.D.
Electroexcitation of nucleon resonances.  
\emph{Prog. Part. Nucl. Phys.} \textbf{2012}, \emph{67}, 1.



\bibitem{Compton}
Eichmann, G.;  Ramalho, G.
Nucleon resonances in Compton scattering. 
\emph{Phys. Rev. D} \textbf{2018}, \emph{98}, 093007.





\bibitem{AxialFF}
Ramalho, G.; Tsushima, K.
Axial form factors of the octet baryons in a covariant quark model. 
\emph{Phys. Rev. D} \textbf{2016}, \emph{94}, 014001.




\bibitem{Gaillard84}
Gaillard, J.M.;  Sauvage, G.
Hyperon Beta Decays.
\emph{Ann. Rev. Nucl. Part. Sci.} \textbf{1984}, \emph{34}, 351.


\bibitem{Bernard02}
Bernard, V.; Elouadrhiri, L.;  Meissner, U.G.
Axial structure of the nucleon: Topical Review. 
\emph{J. Phys. G} 2002, 28, R1.




\bibitem{Gorringe04} 
Gorringe, T.; Fearing, H.W.
Induced pseudoscalar coupling of the proton weak interaction. 
\emph{Rev. Mod. Phys.} \textbf{2004}, \emph{76}, 31.  




\bibitem{Schindler07a} 
Schindler, M.R.;  Scherer, S.
Nucleon Form Factors of the Isovector Axial-Vector Current:
Situation of Experiments and Theory.  
\emph{Eur. Phys. J. A} \textbf{2007}, \emph{32}, 429.




\bibitem{Spectator-Review} 
Ramalho, G.
$N^\ast$ Form Factors based on a Covariant Quark Model. 
\emph{Few Body Syst.} \textbf{2018}, \emph{59}, 92.






\bibitem{NDeltaSL1}
Ramalho, G.; Pe\~na, M.T.;  Gross, F.
$D$-state effects in the electromagnetic $N \Delta$ transition. 
\emph{Phys. Rev. D} \textbf{2008}, \emph{78}, 114017. 



\bibitem{Nucleon}
Gross, F.; Ramalho, G.; Pe\~na, M.T.
A Pure $S$-wave covariant model for the nucleon. 
\emph{Phys. Rev. C} \textbf{2008}, \emph{77}, 015202. 




\bibitem{Nucleon2}
Gross, F.; Ramalho, G.; Pe\~na, M.T.
Covariant nucleon wave function with $S$, $D$, and $P$-state components. 
\emph{Phys. Rev. D} \textbf{2012}, \emph{85}, 093005.



\bibitem{Omega} 
Ramalho, G.; Tsushima, K.; Gross, F.
A Relativistic quark model for the $\Omega^-$ electromagnetic form factors. 
\emph{Phys. Rev. D} \textbf{2009}, \emph{80}, 033004. 




\bibitem{Octet}
Ramalho, G.; Tsushima, K.
Octet baryon electromagnetic form factors in a relativistic quark model. 
\emph{Phys. Rev. D} \textbf{2011}, \emph{84}, 054014.






\bibitem{Octet4}
Gross, F.; Ramalho, G.;  Tsushima, K.
Using baryon octet magnetic moments and masses to fix the pion cloud contribution. 
\emph{Phys. Lett. B} \textbf{2010}, \emph{690}, 183. 



\bibitem{OctetDecuplet1} 
Ramalho, G.;  Tsushima, K.
Octet to decuplet electromagnetic transition in a relativistic quark model.
\emph{Phys. Rev. D} \textbf{2013}, \emph{87}, 093011.



\bibitem{DeltaFF}
Ramalho, G.; Pe\~na, M.T.
Electromagnetic form factors of the Delta in a S-wave approach.
\emph{J. Phys. G} \textbf{2009}, \emph{36}, 085004.


\bibitem{DeltaFF2}
Ramalho, G.; Pe\~na, M.T.; Gross, F.
Electromagnetic form factors of the Delta with D-waves. 
\emph{Phys. Rev. D} \textbf{2010}, \emph{81}, 113011.



\bibitem{DeltaFF3}
Ramalho, G.; Pe\~na, M.T.;  Gross, F.
Electric quadrupole and magnetic octupole moments of the Delta. 
\emph{Phys. Lett. B} \textbf{2009}, \emph{678}, 355.




\bibitem{Omega2}
Ramalho, G.;  Pe\~na, M.T.
Extracting the $\Omega^-$ electric quadrupole moment from lattice QCD data. 
\emph{Phys. Rev. D} \textbf{2011}, \emph{83}, 054011. 



\bibitem{Delta-shape}
Ramalho, G.; Pe\~na, M.T.; Stadler, A.
The shape of the $\Delta$ baryon in a covariant spectator quark model.  
\emph{Phys. Rev. D} \textbf{2012}, \emph{86}, 093022.



\bibitem{Sigma0Lambda}
Ramalho, G.; Tsushima, K.
Covariant spectator quark model description of the $\gamma^\ast \Lambda \to \Sigma^0$ transition. 
\emph{Phys. Rev. D} \textbf{2012}, \emph{86}, 114030.




\bibitem{OctetDecuplet2}
Ramalho, G.; Tsushima, K.
What is the role of the meson cloud in the $\Sigma^{*0} \to \gamma \Lambda$ and $\Sigma^\ast \to \gamma \Sigma$ decays?  
\emph{Phys. Rev. D} \textbf{2013}, \emph{88}, 053002.





\bibitem{HyperonFF1}
Ramalho, G.; Pe\~na, M.T.; Tsushima, K.
Hyperon electromagnetic timelike elastic form factors at large $q^2$. 
\emph{Phys. Rev. D} \textbf{2020}, \emph{101}, 014014.




\bibitem{HyperonFF3}
Ramalho, G.
Electromagnetic form factors of the $\Omega^-$ baryon in the spacelike and timelike regions. 
\emph{Phys. Rev. D} \textbf{2021}, \emph{103}, 074018.


\bibitem{HyperonFF2}
Ramalho, G.; Pe\~na, M.T.; Tsushima, K.; Cheoun, M.K.
Electromagnetic $|G_E/G_M|$ ratios of hyperons at large timelike $q^2$. 
\emph{Phys. Lett. B} \textbf{2024}, \emph{858}, 139060.






\bibitem{OctetDecupletD1}
Ramalho, G.
Ovariant model for Dalitz decays of decuplet baryons to octet baryons. 
\emph{Phys. Rev. D} \textbf{2020}, \emph{102}, 054016. 




\bibitem{OctetDecupletD2} 
Ramalho, G.; Tsushima, K.
Meson cloud contributions to the Dalitz decays of decuplet to octet baryons. 
\emph{Phys. Rev. D} \textbf{2023}, \emph{108}, 074019.








\bibitem{QMCEMFFMedium5}
Saito, K.; Tsushima, K.; Thomas, A.W.
Selfconsistent description of finite nuclei based on a relativistic quark model.
\emph{Nucl. Phys. A} \textbf{1996}, \emph{609}, 339.








\bibitem{Lu01b}
Lu, D.H.; Yang, S.N.; Thomas, A.W.
On the role of the pion cloud in nucleon electromagnetic form-factors. 
\emph{Nucl. Phys. A} \textbf{2001}, \emph{684}, 296.






\bibitem{Thomas83}
Theberge, S.; Thomas, A.W.
Magnetic Moments Of The Nucleon Octet Calculated In The Cloudy Bag Model. 
\emph{Nucl. Phys. A} \textbf{1983}, \emph{393}, 252.




\bibitem{Lu98x}
Lu, D.H.; Thomas, A.W.; Williams, A.G.
Electromagnetic form-factors of the nucleon in an improved quark model. 
\emph{Phys. Rev. C} \textbf{1998}, \emph{57}, 2628--2637.




\bibitem{Kubodera85}
Kubodera, K.; Kohyama, Y.; Oikawa, K.; Kim, C.W.
Weak Interactions Form-factors of the Octet Baryons in the Cloudy Bag Model.
\emph{Nucl. Phys. A} \textbf{1985}, \emph{439}, 695.




\bibitem{Yamaguchi89}
Yamaguchi, T.; Tsushima, K.; Kohyama, Y.; Kubodera, K.
Semileptonic Beta Decay Form-factors and Magnetic Moments of Octet Baryons: Recoil Effects and Center-of-mass Corrections in the Cloudy Bag Model Including Gluonic Effects. 
\emph{Nucl. Phys. A} \textbf{1989}, \emph{500}, 429.

\bibitem{Tsushima88}
Tsushima, K.; Yamaguchi, T.; Kohyama, Y.; Kubodera, K.
Weak Interaction Form-factors and Magnetic Moments of Octet Baryons: Chiral Bag Model With Gluonic Effects.
\emph{Nucl. Phys. A} \textbf{1988}, \emph{489}, 557.









\bibitem{Eichmann16a}
Eichmann, G.; Sanchis-Alepuz, H.; Williams, R.; Alkofer, R.; Fischer, C.S.
Baryons as relativistic three-quark bound states. 
\emph{Prog. Part. Nucl. Phys.} \textbf{2016}, \emph{91}, 100.



\bibitem{Cloet14a}
Cloet, I.C.; Roberts, C.D.
Explanation and Prediction of Observables using Continuum Strong QCD. 
\emph{Prog. Part. Nucl. Phys.} \textbf{2014}, \emph{77}, 1. 




\bibitem{Maris03a}
Maris, P.; Roberts, C.D. 
Dyson-Schwinger equations: A Tool for hadron physics. 
\emph{Int. J. Mod. Phys. E} \textbf{2003}, \emph{12}, 297.

































\bibitem{Bernard98a}
Bernard, V.; Fearing, H.W.; Hemmert, T.R.; Meissner, U.G.
The form-factors of the nucleon at small momentum transfer. 
\emph{Nucl. Phys. A} \textbf{1998}, \emph{635}, 121. Erratum in  \emph{Nucl. Phys. A} \textbf{1998}, \emph{642}, 563.





\bibitem{Hammer04a}
Hammer, H.W.; Drechsel, D.;  Meissner, U.G.
On the pion cloud of the nucleon. 
\emph{Phys. Lett. B} \textbf{2004}, \emph{586}, 291. 


\bibitem{Meissner07a}
Meissner, U.G.
The Pion cloud of the nucleon: Facts and popular fantasies. 
\emph{AIP Conf. Proc.} \textbf{2007}, \emph{904}, 142.



\bibitem{Ronchenn13a}
Ronchen, D.; Doring, M.; Huang, F.; Haberzettl, H.; Haidenbauer, J.; Hanhart, C.; Krewald, S.; Meissner, U.G.; Nakayama, K.
Coupled-channel dynamics in the reactions
$\pi N \to \pi N$, $\eta N$, $K \Lambda$, $K \Sigma$. 
\emph{Eur. Phys. J. A} \textbf{2013}, \emph{49}, 44.




\bibitem{Pascalutsa07a}
Pascalutsa, V.; Vanderhaeghen, M.;  Yang, S.N.
Electromagnetic excitation of the $\Delta(1232)$-resonance. 
\emph{Phys. Rept.} \textbf{2007}, \emph{437}, 125.



\bibitem{Alexandrou08a}
Alexandrou, C.; Koutsou, G.; Neff, H.; Negele, J.W.; Schroers, W.;  Tsapalis, A. 
Nucleon to delta electromagnetic transition form factors in lattice QCD. 
\emph{Phys. Rev. D} \textbf{2008}, \emph{77}, 085012.





\bibitem{NDeltaSL2}
Ramalho, G.; Pe\~na, M.T.; Gross, F.
A Covariant model for the nucleon and the Delta. 
\emph{Eur. Phys. J. A} \textbf{2008}, \emph{36}, 329.





\bibitem{LatticeD}
Ramalho, G.; Pe\~na, M.T.
Valence quark contribution for the
$\gamma N \to \Delta$ quadrupole transition extracted from lattice QCD. 
\emph{Phys. Rev. D} \textbf{2009}, \emph{80}, 013008. 






\bibitem{Lattice}
Ramalho, G.; Pe\~na, M.T.
Nucleon and $\gamma N \to \Delta$ lattice form factors in a constituent quark model. 
\emph{J. Phys. G} \textbf{2009}, \emph{36}, 115011.




\bibitem{Siegert1}
Ramalho, G.
Parametrizations of the $\gamma^\ast N \to \Delta(1232)$ quadrupole form factors and Siegert's theorem. 
\emph{Phys. Rev. D} \textbf{2016}, \emph{94}, 114001. 




\bibitem{Siegert2}
Ramalho, G.
New low-$Q^2$ measurements of the $\gamma^\ast N \to \Delta(1232)$ Coulomb quadrupole form factor, pion cloud parametrizations and Siegert's theorem.
\emph{Eur. Phys. J. A} \textbf{2018}, \emph{54}, 75.




\bibitem{Roper1}
Ramalho, G.; Tsushima, K.
Valence quark contributions for the
$\gamma N \to P_{11}(1440)$ form factors. 
\emph{Phys. Rev. D} \textbf{2010}, \emph{81}, 074020.



\bibitem{Roper2}
Ramalho, G.; Tsushima, K.
$\gamma^\ast N \to N(1710)$ transition at high momentum transfer. 
\emph{Phys. Rev. D} \textbf{2014}, \emph{89}, 073010.




\bibitem{SemiRel}
Ramalho, G.
Semirelativistic approximation to the $\gamma^\ast N \to N(1520)$ and $\gamma^\ast N \to N(1535)$ transition form factors. 
\emph{Phys. Rev. D} \textbf{2017}, \emph{95}, 054008.




\bibitem{N1535SL1}
Ramalho, G.;  Pe\~na, M.T.
A covariant model for the
$\gamma N \to N(1535)$ transition at high momentum transfer. 
\emph{Phys. Rev. D} \textbf{2011}, \emph{84}, 033007.



\bibitem{N1535SL2}
Ramalho, G.; Tsushima, K.
A simple relation between the $\gamma N \to N(1535)$ helicity amplitudes. 
\emph{Phys. Rev. D} \textbf{2011}, \emph{84}, 051301. 



\bibitem{N1520SL}
Ramalho, G.; Pe\~na, M.T.
$\gamma^\ast N \to N^\ast(1520)$ form factors in the spacelike region. 
\emph{Phys. Rev. D} \textbf{2014}, \emph{89}, 094016.





\bibitem{SQTM}
Ramalho, G.
Using the Single Quark Transition Model to predict nucleon resonance amplitudes. 
\emph{Phys. Rev. D} \textbf{2014}, \emph{90}, 033010.






\bibitem{LambdaStar}
Ramalho, G.; Jido, D.; Tsushima, K.
Valence quark and meson cloud contributions for the
$\gamma^* \Lambda \to \Lambda^*$ and $\gamma^* \Sigma^0 \to \Lambda^*$ reactions. 
\emph{Phys. Rev. D} \textbf{2012}, \emph{85}, 093014. 




\bibitem{Delta1600}
Ramalho, G.; Tsushima, K.
A Model for the $\Delta(1600)$ resonance and $\gamma N \to \Delta(1600)$ transition. 
\emph{Phys. Rev. D} \textbf{2010}, \emph{82}, 073007.


\bibitem{N1535TL}
Ramalho, G.; Pe\~na, M.T.
Covariant model for the Dalitz decay of the $N(1535)$ resonance. 
\emph{Phys. Rev. D} \textbf{2020}, \emph{101}, 114008. 



\bibitem{N1520TL}
Ramalho, G.;  Pe\~na, M.T.
$\gamma^\ast N \to N^\ast(1520)$ form factors in the timelike regime. 
\emph{Phys. Rev. D} \textbf{2017}, \emph{95}, 014003.





\bibitem{NDeltaTL3}
Ramalho, G.; Pe\~na, M.T.; Weil, J.; van Hees, H.; Mosel, U.
Role of the pion electromagnetic form factor in the $\Delta(1232) \to \gamma^\ast N$ timelike transition. 
\emph{Phys. Rev. D} \textbf{2016}, \emph{93}, 033004.



\bibitem{NDeltaTL4}
Ramalho, G.; Pe\~na, M.T.
Timelike $\gamma^* N \to \Delta$ form factors and Delta Dalitz decay.
\emph{Phys. Rev. D} \textbf{2012}, \emph{85}, 113014.




\bibitem{DIS}
Gross, F.; Ramalho, G.; Pe\~na, M.T.
Spin and angular momentum in the nucleon. 
\emph{Phys. Rev. D} \textbf{2012}, \emph{85}, 093006.















\bibitem{Gross69}
Gross, F.
Three-Dimensional Covariant Integral Equations For Low-Energy Systems. 
\emph{Phys. Rev.} \textbf{1969}, \emph{186}, 1448.

\bibitem{Gross97}
Stadler, A.; Gross, F.; Frank, M.
Covariant equations for the three-body bound state. 
\emph{Phys. Rev. C} \textbf{1997}, \emph{56}, 2396.





\bibitem{FixedAxis}
Gross, F.; Ramalho, G.; Pe\~na, M.T.
Fixed-axis polarization states: Covariance and comparisons. 
\emph{Phys. Rev. C} \textbf{2008}, \emph{77}, 035203.





\bibitem{Savkli01a}
Savkli, C.;  Gross, F.
Quark-antiquark bound states in the relativistic spectator formalism. 
\emph{Phys. Rev. C} \textbf{2001}, \emph{63}, 035208.


\bibitem{Gross06a}
Gross, F.; Agbakpe, P.
The Shape of the nucleon.
\emph{Phys. Rev. C} \textbf{2006}, \emph{73}, 015203.








\bibitem{Lin09}
Lin, H.W.; Orginos, K.
Strange Baryon Electromagnetic Form Factors and SU(3) Flavor Symmetry Breaking.  
\emph{Phys. Rev. D} \textbf{2009}, \emph{79}, 074507.
%





\bibitem{Alexandrou11a}
Alexandrou, C.  {et~al.} [ETM Collaboration].
Axial Nucleon form factors from lattice QCD. 
\emph{Phys. Rev. D} \textbf{2011}, \emph{83}, 045010.






\bibitem{Alexandrou13a}
Alexandrou, C.; Constantinou, M.; Dinter, S.; Drach, V.; Jansen, K.; Kallidonis, C.; Koutsou, G. 
Nucleon form factors and moments of generalized parton distributions using $N_f=2+1+1$ twisted mass fermions. 
\emph{Phys. Rev. D} \textbf{2013}, \emph{88}, 014509.




\bibitem{ARehim15a}
Abdel-Rehim, A.; Alexandrou, C.; Constantinou, M.; Dimopoulos, P.; Frezzotti, R.; Hadjiyiannakou, K.; Jansen, K.; Kallidonis, C.; Kostrzewa, B.;  Koutsou, G.; {et~al.}
Nucleon and pion structure with lattice QCD simulations at physical value of the pion mass.
\emph{Phys. Rev. D} \textbf{2015}, \emph{92}, 114513. Erratum in \emph{Phys. Rev. D} \textbf{2016}, \emph{93}, 039904.





\bibitem{Jang20a}
Jang, Y.C.; Gupta, R.; Yoon, B.; Bhattacharya, T.
Axial Vector Form Factors from Lattice QCD that Satisfy the PCAC Relation. 
\emph{Phys. Rev. Lett.} \textbf{2020}, \emph{124}, 072002. 






\bibitem{Djukanovic22a}
Djukanovic, D.;~von Hippel, G.; Koponen, J.; Meyer, H.B.; Ottnad, K.; Schulz, T.; Wittig, H. 
Isovector axial form factor of the nucleon from lattice QCD. 
\emph{Phys. Rev. D} \textbf{2022}, \emph{106}, 074503.


\bibitem{Bali23a}
Bali, G.S.;  Collins, S.; Heybrock, S.; Löffler, M.; Rödl, R.; Söldner, W.; Weishäupl, S.
Octet baryon isovector charges from Nf=2+1 lattice QCD. 
\emph{Phys. Rev. D} \textbf{2023}, \emph{108}, 034512.



\bibitem{Chang18a}
Chang, C.C.; Nicholson, A.N.; Rinaldi, E.; Berkowitz, E.; Garron, N.; Brantley, D.A.; Monge-Camacho, H.; Monahan, C.J.; Bouchard, C.;  Clark, M.A.; {et~al.}
A per-cent-level determination of the nucleon axial coupling from quantum chromodynamics.  
\emph{Nature} \textbf{2018}, \emph{558}, 91.




\bibitem{Meyer22a}
Meyer, A.S.; Walker-Loud, A.; Wilkinson, C.
Status of Lattice QCD Determination of Nucleon Form Factors and their Relevance for the Few-GeV Neutrino Program. 
\emph{Ann. Rev. Nucl. Part. Sci.} \textbf{2022}, \emph{72}, 205.






\bibitem{GAholography} 
Ramalho, G.
Holographic estimate of the meson cloud contribution to nucleon axial form factor.
\emph{Phys. Rev. D} \textbf{2018}, \emph{97}, 073002.


























\bibitem{Jenkins93}
Jenkins, E.E.; Luke, M.E.; Manohar, A.V.; Savage, M.J. 
Chiral perturbation theory analysis of the baryon magnetic moments. 
\emph{Phys. Lett. B} \textbf{1993}, \emph{302}, 482





\bibitem{Meissner97}
Meissner, U.G.;  Steininger, S.
Baryon magnetic moments in chiral perturbation theory.  
\emph{Nucl. Phys. B} \textbf{1997}, \emph{499}, 349.





\bibitem{Kubis99}
Kubis, B.; Hemmert, T.R.; Meissner, U.G.
Baryon form-factors. 
\emph{Phys. Lett. B} \textbf{1999}, \emph{456}, 240. 






\bibitem{Franklin02}
Franklin, J.
Phenomenological quark model for baryon magnetic moments and beta decay ratios $(G_A/G_V)$. 
\emph{Phys. Rev. D} \textbf{2002}, \emph{66}, 033010.





\bibitem{Cheedket04}
Cheedket, S.; Lyubovitskij, V.E.; Gutsche, T.; Faessler, A.;
Pumsa-ard, K.; Yan, Y.
Electromagnetic form factors of the baryon octet in the perturbative chiral
quark modeluark model.  
\emph{Eur. Phys. J. A} \textbf{2004}, \emph{20}, 317. 


\bibitem{Leinweber05}
Leinweber, D.B.; Boinepalli, S.; Cloet, I.C.; Thomas, A.W.; Williams, A.G.; Young, R.D.; Zanotti, J.M.; Zhang, J.B.  
Precise determination of the strangeness magnetic moment of the nucleon. 
\emph{Phys. Rev. Lett.} \textbf{2005}, \emph{94}, 212001.





\bibitem{Boinepalli06} 
Boinepalli, S.; Leinweber, D.B.; Williams, A.G.; Zanotti, J.M.; Zhang, J.B.  
Precision electromagnetic structure of octet baryons in the chiral regime.
\emph{Phys. Rev. D} \textbf{2006}, \emph{74}, 093005. 



\bibitem{Wang09}
Wang, P.; Leinweber, D.B.; Thomas, A.W.; Young, R.D.
Chiral extrapolation of octet-baryon charge radii.  
\emph{Phys. Rev. D} \textbf{2009}, \emph{79}, 094001.










\bibitem{Bernard95a}
Bernard, V.; Kaiser, N.;  Meissner, U.G.
Chiral dynamics in nucleons and nuclei.  
\emph{Int. J. Mod. Phys. E} \textbf{1995}, \emph{4}, 193.


\bibitem{Perdrisat07a}
Perdrisat, C.F.; Punjabi, V.; Vanderhaeghen, M.
Nucleon Electromagnetic Form Factors.  
\emph{Prog. Part. Nucl. Phys.} \textbf{2007}, \emph{59}, 694.



\bibitem{JLab00a}
Jones, M.K.; Aniol, K.A.; Baker, F.T.; Berthot, J.; Bertin, P.Y.; Bertozzi, W.; Besson, A.; Bimbot, L.; Boeglin, W.U.; Brash, E.J.; {et~al.} 
$G_{Ep}/G_{Mp}$ ratio by polarization transfer in $\vec ep\rightarrow e\vec p$.  
\emph{Phys. Rev. Lett.} \textbf{2000}, \emph{84}, 1398--1402.





\bibitem{JLab02a}
Gayou, O.;  Benmokhtar, F.; Bertozzi, W.; Bimbot, L.; Brash, E.J.; Calarco, J.R.; Cavata, C.; Chai, Z.; Chang, C.C.; Chang, T.; {et~al.} 
Measurement of $G_{Ep}/G_{Mp}$ in
$\vec{e} p \to e \vec{p}$ to $Q^2=5.6$ GeV$^2$.  
\emph{Phys. Rev. Lett.} \textbf{2002}, \emph{88}, 092301.





\bibitem{JLab10a}
Puckett, A.J.R.; Brash, E.J.; Jones, M.K.; Luo, W.; Meziane, M.; Pentchev, L.; Perdrisat, C.F.; Punjabi, V.; Wesselmann, F.R. and Ahmidouch, A.; {et~al.} 
Recoil Polarization Measurements of the Proton Electromagnetic Form Factor
Ratio to $Q^2$ = 8.5 GeV$^2$.  
\emph{Phys. Rev. Lett.} \textbf{2010}, \emph{104}, 242301.





\bibitem{Zhan11}
Zhan, X.; Allada, K.; Armstrong, D.S.; Arrington, J.; Bertozzi, W.; Boeglin, W.; Chen, J.-P.; Chirapatpimol, K.; Choi, S.; Chudakov, E.; {et~al.}
High Precision Measurement of the Proton Elastic Form Factor Ratio $\mu_pG_E/G_M$ at low $Q^2$.  
\emph{Phys. Lett. B} \textbf{2011}, \emph{705}, 59.



\bibitem{Arrington07a}
Arrington, J.; Melnitchouk, W.;  Tjon, J.A.
Global analysis of proton elastic form factor data with two-photon exchange corrections.
\emph{Phys. Rev. C} \textbf{2007}, \emph{76}, 035205.







\bibitem{Ostrick99} 
Ostrick, M.; Herberg, C.; Andresen, H.G.; Annand, J.R.M.; Aulenbacher, K.; Becker, J.; Drescher, P.; Eyl, D.; Frey, A.; Grabmayr, P.; {et~al.} 
Measurement of the neutron electric form factor $G_{E,n}$ in the quasifree
$^2$H$(\vec{e}, e' \vec{p})p$ reaction.  
\emph{Phys. Rev. Lett.} \textbf{1999}, \emph{83}, 276.

\bibitem{Herberg99} 
Herberg, C.; Ostrick, M.; Andresen, H.G.; Annand, J.R.M.; Aulenbacher, K.; Becker, J.; Drescher, P.; Eyl, D.; Frey, A.; Grabmayr, P.; {et~al.}
Determination of the neutron electric form factor in the $D(e,e' n)p$
reaction and the influence of nuclear binding.
\emph{Eur. Phys. J. A} \textbf{1999}, \emph{5}, 131.


\bibitem{Glazier05} 
Glazier, D.I.; Seimetz, M.; Annand, J.R.M.; Arenhövel, H.; Antelo, M.A.; Ayerbe, C.; Bartsch, P.; Baumann, D.; Bermuth, J.; Böhm, R.; {et~al.}
Measurement of the Electric Form Factor of the Neutron at $Q^2 = 0.3$--0.8 (GeV/c)$^2$.  
\emph{Eur. Phys. J. A} \textbf{2005}, \emph{24}, 101. 

\bibitem{Passchier99} 
Passchier, I.; Alarcon, R.; Bauer, T.S.; Boersma, D.; Brand, J.F.J.v.; Buuren, L.D.v.; Bulten, H.J.; Ferro-Luzzi, M.; Heimberg, P.; Higinbotham, D.W.; {et~al.}
The charge form factor of the neutron from the reaction $^2$H$(\vec{e},e' n)p$.  
\emph{Phys. Rev. Lett.} \textbf{1999}, \emph{82}, 4988. 


\bibitem{Eden94} 
Eden, T.;  Madey, R.; Zhang, W.-M.; Anderson, B.D.; Arenhövel, H.; Baldwin, A.R.; Barkhuff, D.; Beard, K.B.; Bertozzi, W.; Cameron, J.M.;  {et~al.}
Electric form-factor of the neutron from the $^2$H$(\vec{e},e' n)^1$H reaction
at $Q^2=0.255$ (GeV/c)$^2$.  
\emph{Phys. Rev. C} \textbf{1994}, \emph{50}, 1749.

\bibitem{Zhu01} 
Zhu, H.  {et~al.} [E93026 Collaboration]. 
A measurement of the electric form-factor of the neutron   through
$d(\vec{e},e' n)p$ at $Q^2 = 0.5$ (GeV/c)$^2$.
\emph{Phys. Rev. Lett.} \textbf{2001}, \emph{87}, 081801.  

\bibitem{Warren03} 
Warren, G.  {et~al.} [Jefferson Lab E93-026 Collaboration].
Measurement of the electric form factor of the neutron
$Q^2=0.5$ (GeV/c)$^2$ and 1.0 (GeV/c)$^2$. 
\emph{Phys. Rev. Lett.} \textbf{2004}, \emph{92}, 042301. 

\bibitem{Madey03}
Madey, R.    {et~al.} [E93-038 Collaboration].
Measurements of $G_{En}/G_{Mn}$ from $^2$H$(\vec{e},e' n)^1$H
reaction to $Q^2 = 1.45$ (GeV/c)$^2$.  
\emph{Phys. Rev. Lett.} \textbf{2003}, \emph{91}, 122002. 


\bibitem{Riordan10}
Beck, A.; Beck, S.M.-T.; Puckett, A.J.R.; Qiang, Y.; Sirca, S.; Zhu, X.
Measurements of the Electric Form Factor of the Neutron
up to $Q^2=3.4$ GeV$^2$ using the Reaction $^3$He$(\vec{e},e'n)pp$.
\emph{Phys. Rev. Lett.} \textbf{2010}, \emph{105}, 262302.



\bibitem{Schiavilla01}
Schiavilla, R.; Sick, I.
Neutron charge form factor at large $q^2$.
\emph{Phys. Rev. C} \textbf{2001}, \emph{64}, 041002. 



\bibitem{Bosted95} 
Bosted, P.E.
An Empirical fit to the nucleon electromagnetic form-factors. 
\emph{Phys. Rev. C} \textbf{1995}, \emph{51}, 409.

\bibitem{Kubon02} 
Kubon, G.; Anklin, H.; Bartsch, P.; Baumann, D.; Boeglin, W.U.; Bohinc, K.; Böhm, R.; Distler, M.O.; Ewald, I.; Friedrich, J.; {et~al.}
Precise neutron magnetic form factors. 
\emph{Phys. Lett. B} \textbf{2002}, \emph{524}, 26.  


\bibitem{Anklin98} 
Anklin, H.;  deBever, L.J.; Blomqvist, K.I.; Boeglin, W.U.; Böhm, R.; Distler, M.; Edelhoff, R.; Friedrich, J.; Fritschi, D.; Geiges, R.; {et~al.}
Precise measurements of the neutron magnetic form factor.
\emph{Phys. Lett. B} \textbf{1998}, \emph{428}, 248.


\bibitem{Anklin94} 
Anklin, H.;  Bruins, E.E.W.; Day, D.; Fritschi, D.; Groft, B.; Joosse, F.C.P.; Jourdan, J.; Lichtenstadt, J.; Loppacher, M.; Masson, G.; {et~al.}
Precision measurement of the neutron magnetic form-factor.
\emph{Phys. Lett. B} \textbf{1994}, \emph{336}, 313.


\bibitem{Lachniet09} 
Lachniet, J.  {et~al.} [CLAS Collaboration].
A Precise Measurement of the Neutron Magnetic Form Factor $G_{Mn}$ in the
Few-GeV$^2$ Region.
\emph{Phys. Rev. Lett.} \textbf{2009}, \emph{102}, 192001.





\bibitem{PDG2010}
Nakamura, K.; Particle Data Group.
Review of particle physics.
\emph{J. Phys. G} \textbf{2010}, \emph{37}, 075021.



\bibitem{Eschrich01}
Gough; Eschrich, I.M. {et~al.} [SELEX Collaboration].
Measurement of the $\Sigma^-$ charge radius by $\Sigma^-$ electron elastic
scattering. 
\emph{Phys. Lett. B} \textbf{2001}, \emph{522}, 233. 







\bibitem{Park12} 
Park, K. {et~al.} [CLAS Collaboration].
Measurement of the generalized form factors near threshold via $\gamma^* p \to n\pi^+$ at high $Q^2$.
\emph{Phys. Rev. C} \textbf{2012}, \emph{85}, 035208.






\bibitem{PDG2022}
 Particle Data Group;  Workman, R.L.; Burkert, V.D.; Crede, V.; Klempt, E.; Thoma, U.; Tiator, L.; Agashe, K.; Aielli, G.; Allanach, B.C.; {et~al.}  
Review of Particle Physics. 
\emph{Prog. Theor. Exp. Phys.} \textbf{2022}, \emph{2022}, 083C01.




\bibitem{Choi93}
Choi, S.;  Duval, M.A.; Elouadrhiri, L.; Estenne, V.; Bardin, G.; Berthot, J.; Bertin, P.Y.; Botton, N.D.; Didelez, J.P.; Fonvieille, H.;  {et~al.} 
Axial and pseudoscalar nucleon form-factors from low-energy pion electroproduction.  
\emph{Phys. Rev. Lett.} \textbf{1993}, \emph{71}, 3927.












\bibitem{Tsushima22a}
Tsushima, K.
Magnetic moments of the octet, decuplet, low-lying charm, and low-lying bottom baryons in a nuclear medium.  
\emph{Prog. Theor. Exp. Phys.} \textbf{2022}, \emph{2022}, 043D02.





\bibitem{CMartinez17a}
Cobos-Martínez, J.J.; Tsushima, K.; Krein, G.; Thomas, A.W.
$\phi$ meson mass and decay width in nuclear matter and nuclei.
\emph{Phys. Lett. B} \textbf{2017}, \emph{771}, 113.














\bibitem{GTrelation}
Goldberger, M.L.; Treiman, S.B.
Decay of the pi meson.
\emph{Phys. Rev.} \textbf{1958}, \emph{110}, 1178.



\bibitem{ChPT97}
Kirchbach, M.; Wirzba, A.
In-medium chiral perturbation theory and pion weak decay in the presence of
background matter.
\emph{Nucl. Phys. A} \textbf{1997}, \emph{616}, 648. 










\bibitem{JLab04a}
Christy, M.E. et al. [E94110 Collaboration].
Measurements of electron proton elastic cross-sections for
$0.4 < Q^2 < 5.5$ (GeV/c)$^2$.  
\emph{Phys. Rev. C} \textbf{2004}, \emph{70}, 015206.



\bibitem{JLab05a}
Punjabi, V.; Perdrisat, C.F.; Aniol, K.A.; Baker, F.T.; Berthot, J.; Bertin, P.Y.; Bertozzi, W.; Besson, A.; Bimbot, L.;  Boeglin, W.U.; {et~al.}
Proton elastic form-factor ratios to $Q^2= 3.5$ GeV$^2$ by polarization transfer.  
\emph{Phys. Rev. C} \textbf{2005}, \emph{71}, 055202. Erratum in \emph{Phys. Rev. C} \textbf{2005}, \emph{71}, 069902.





\bibitem{JLab12a}
Puckett, A.J.R.; Brash, E.J.; Gayou, O.; Jones, M.K.; Pentchev, L.; Perdrisat, C.F.; Punjabi, V.; Aniol, K.A.; Averett, T.; Benmokhtar, F.; {et~al.}
Final Analysis of Proton Form Factor Ratio Data at $\mathbf{Q^2 = 4.0}$, 4.8 and 5.6 GeV$\mathbf{^2}$.  
\emph{Phys. Rev. C} \textbf{2012}, \emph{85}, 045203.



\bibitem{JLab17a}
Puckett, A.J.R.; Brash, E.J.; Jones, M.K.; Luo, W.; Meziane, M.; Pentchev, L.; Perdrisat, C.F.; Punjabi, V.; Wesselmann, F.R.;  Afanasev, A.; {et~al.}
Polarization Transfer Observables in Elastic Electron Proton Scattering at
$Q^2 = $2.5, 5.2, 6.8, and 8.5 GeV$^2$.  
\emph{Phys. Rev. C} \textbf{2017}, \emph{96}, 055203.
Erratum in \emph{Phys. Rev. C} \textbf{2018}, \emph{98}, 019907.








\bibitem{Cheoun13b}
Cheoun, M.K.; Choi, K.S.; Kim, K.S.; Saito, K.;
Kajino, T.; Tsushima, K.; Maruyama, T.
Effects of the density-dependent weak form factors on the neutrino reaction via neutral current for the nucleon in nuclear medium and $^{12}$C. 
\emph{Phys. Rev. C} \textbf{2013}, \emph{87}, 065502.




\bibitem{Cheoun13a}
Cheoun, M.K.; Choi, K.; Kim, K.S.; Saito, K.; Kajino, T.;
Tsushima, K.; Maruyama, T.
Asymmetry in the neutrino and anti-neutrino reactions in a nuclear medium. 
\emph{Phys. Lett. B} \textbf{2013}, \emph{723}, 464. 





\bibitem{Cai23a}
Cai, T.;  Moore, M.L.; Olivier, A.; Akhter, S.; Dar, Z.A.; Ansari, V.; Ascencio, M.V.; Bashyal, A.; Bercellie, A.; Betancourt, M.; {et~al.} 
Measurement of the axial vector form factor from antineutrino-proton scattering. 
\emph{Nature} \textbf{2023}, \emph{614},  48.





\bibitem{KDAR}
Abud, A.A.   {et~al.}   [The DUNE collaboration].
Searching for solar KDAR with DUNE.
\emph{JCAP} \textbf{2021}, \emph{10}, 065. 















\bibitem{Alberico02a}
Alberico, W.M.; Bilenky, S.M.; Maieron, C.
Strangeness in the nucleon: Neutrino-nucleon and polarized electron-nucleon scattering.
\emph{Phys. Rept.} \textbf{2002}, \emph{358}, 227.





\bibitem{Ahrens87a}
Ahrens, L.A.; Aronson, S.H.; Connolly, P.L.; Gibbard, B.G.; Murtagh, M.J.; Murtagh, S.J.; Terada, S.; White, D.H.; Callas, J.L.; Cutts, D.; {et~al.}
Measurement of Neutrino---Proton and anti-neutrino-Proton Elastic Scattering.
\emph{Phys. Rev. D} \textbf{1987}, \emph{35}, 785.











\bibitem{Athar-Review22}
Athar, M.S.; Fatima, A.; Singh, S.K.
Neutrinos and their interactions with matter. 
\emph{Prog. Part. Nucl. Phys.} \textbf{2023}, \emph{129}, 104019.













\bibitem{Snowmass22a}
Ruso, L.A.; Ankowski, A.M.; Bacca, S.; Balantekin, A.B.; Carlson, J.; Gardiner, S.; Gonz\'alez-Jim\'enez, R.; Gupta, R.; Hobbs, T.J.; Hoferichter, M.; {et~al.}
Theoretical tools for neutrino scattering: Interplay between lattice QCD, EFTs, nuclear physics, phenomenology, and neutrino event generators.   \emph{arXiv} \textbf{2022}, arXiv:2203.09030.







\bibitem{Akimov22a}
Akimov, D.; Alawabdeh, S.; An, P.; Awe, C.; Barbeau, P.S.; Barry, C.; Becker, B.; Belov, V.; Bernardi, I.; Bock, C.; {et~al.}  
The COHERENT Experimental Program.  \emph{arXiv} \textbf{2022},	arXiv:2204.04575.








\bibitem{Simons22a}
Simons, D.; Steinberg, N.; Lovato, A.; Meurice, Y.; Rocco, N.; Wagman, M.
Form factor and model dependence in neutrino-nucleus cross section predictions.
\emph{arXiv} \textbf{2022},  arXiv:2210.02455.







\bibitem{Hoferichter20a}
Hoferichter, M.; Men\'endez, J.; Schwenk, A.
Coherent elastic neutrino-nucleus scattering: EFT analysis and nuclear responses.
\emph{Phys. Rev. D} \textbf{2020}, \emph{102}, 074018.









\bibitem{Abdullah20a}
Abdullah, M.; Aristizabal Sierra, D.; Dutta, B.; Strigari, L.E.
Coherent Elastic Neutrino-Nucleus Scattering with directional detectors.
\emph{Phys. Rev. D} \textbf{2020}, \emph{102}, 015009.



\bibitem{Cadeddu23a}
Cadeddu, M.; Dordei, F.; Giunti, C.
A view of coherent elastic neutrino-nucleus scattering.  
\emph{EPL} \textbf{2023}, \emph{143}, 34001. 







\bibitem{Gondolo02a}
Gondolo, P.
Recoil momentum spectrum in directional dark matter detectors.  
\emph{Phys. Rev. D} \textbf{2002}, \emph{66}, 103513.






\bibitem{InPreparation}
Ramalho, G.; Tsushima, K.; Cheoun, M.K.
Manuscript in preparation; to be submitted.
  







\bibitem{Parada18a}
Hutauruk, P.T.P.; Oh, Y.; Tsushima, K.
Impact of medium modifications of the nucleon weak and electromagnetic form factors on the neutrino mean free path in dense matter.
\emph{Phys. Rev. D} \textbf{2018}, \emph{98}, 013009.




\end{thebibliography}





\PublishersNote{}
\end{adjustwidth}
\end{document}